\documentclass[fleqn,usenatbib]{mnras}

\usepackage{newtxtext,newtxmath}
\usepackage{float}
\usepackage[caption = false]{subfig}

\usepackage[T1]{fontenc}
\hypersetup{linkcolor=red}          

\DeclareRobustCommand{\VAN}[3]{#2}
\let\VANthebibliography\thebibliography
\def\thebibliography{\DeclareRobustCommand{\VAN}[3]{##3}\VANthebibliography}

\newcommand{\dd}{{\,\mathop{\kern0pt\mathrm{d}}\!{}}}

\usepackage{graphicx}	
\usepackage{amsmath}	
\usepackage{scalerel}
\usepackage{tikz}
\usepackage{caption}
\usetikzlibrary{svg.path}

\definecolor{orcidlogocol}{HTML}{A6CE39}
\tikzset{
	orcidlogo/.pic={
		\fill[orcidlogocol] svg{M256,128c0,70.7-57.3,128-128,128C57.3,256,0,198.7,0,128C0,57.3,57.3,0,128,0C198.7,0,256,57.3,256,128z};
		\fill[white] svg{M86.3,186.2H70.9V79.1h15.4v48.4V186.2z}
		svg{M108.9,79.1h41.6c39.6,0,57,28.3,57,53.6c0,27.5-21.5,53.6-56.8,53.6h-41.8V79.1z M124.3,172.4h24.5c34.9,0,42.9-26.5,42.9-39.7c0-21.5-13.7-39.7-43.7-39.7h-23.7V172.4z}
		svg{M88.7,56.8c0,5.5-4.5,10.1-10.1,10.1c-5.6,0-10.1-4.6-10.1-10.1c0-5.6,4.5-10.1,10.1-10.1C84.2,46.7,88.7,51.3,88.7,56.8z};
	}
}

\newcommand\orcidlink[1]{\href{https://orcid.org/#1}{\mbox{\scalerel*{
				\begin{tikzpicture}[yscale=-1,transform shape]
					\pic{orcidlogo};
				\end{tikzpicture}
			}{|}}}}

\usepackage{hyperref}

\title[M87 Black Hole: Systematic Uncertainty]{Supermassive black hole mass in the massive elliptical galaxy M87 from integral-field stellar dynamics using OASIS and MUSE with adaptive optics: assessing systematic uncertainties}
\author[D. A. Simon  et al.]{
	David A. Simon$^{\orcidlink{0000-0001-5742-2982}}$,$^{1}$\thanks{E-mail: david.simon@physics.ox.ac.uk}
	Michele Cappellari$^{\orcidlink{0000-0002-1283-8420}}$,$^{1}$
	Johanna Hartke$^{\orcidlink{0000-0002-8745-689X}}$$^{1,2,3}$
	\\
	$^{1}$Sub-department of Astrophysics, Department of Physics, University of Oxford, Denys Wilkinson Building, Keble Road, Oxford OX1 3RH\\
	$^{2}$European Southern Observatory, Alonso de C\'ordova 3107, Casilla 19001, Santiago, Chile\\
	$^{3}$Finnish Centre for Astronomy with ESO (FINCA), FI-20014 University of Turku, Finland
}

\date{Accepted 2023 October 23. Received 2023 September 22; in original form 2023 March 22}

\pubyear{2015}

\begin{document}
	\label{firstpage}
	\pagerange{\pageref{firstpage}--\pageref{lastpage}}
	\maketitle
	
	\begin{abstract}
		The massive elliptical galaxy M87 has been the subject of several supermassive black hole mass measurements from stellar dynamics, gas dynamics, and recently the black hole shadow by the Event Horizon Telescope (EHT). This uniquely positions M87 as a benchmark for alternative black hole mass determination methods. Here we use stellar kinematics extracted from integral-field spectroscopy observations with Adaptive Optics (AO) using MUSE and OASIS. We exploit our high-resolution integral field spectroscopy to spectrally decompose the central AGN from the stars. We derive an accurate inner stellar-density profile and find it is flatter than previously assumed. We also use the spectrally-extracted AGN as a reference to accurately determine the observed MUSE and OASIS AO PSF. We then perform Jeans Anisotropic Modelling (JAM), with a new flexible spatially-variable anisotropy, and measure the anisotropy profile, stellar mass-to-light variations, inner dark matter fraction, and black hole mass. Our preferred black hole mass is $M_{\rm BH}=(8.7\pm1.2 [\text{random}] \pm1.3 [\text{systematic}])  \times 10^9 \ M_\odot $. However, using the inner stellar density from previous studies, we find a preferred black hole mass of $M_{\rm BH} = (5.5^{+0.5}_{-0.3}) \times 10^9 \ M_\odot $, consistent with previous work. We find that this is the primary cause of the difference between our results and previous work, in addition to smaller contributions due to kinematics and modelling method. We conduct numerous systematic tests of the kinematics and model assumptions and conclude that uncertainties in the black hole mass of M87 from previous determinations may have been underestimated and further analyses are needed.
	\end{abstract}
	
	
	
	
	\section{Introduction}
	Supermassive black holes play an important role in galaxy evolution. This is shown through empirical relations between black hole mass and luminosity \citep{kormendy1995inward,magorrian1998demography}, as well as black hole mass and stellar velocity dispersion \citep{ferrarese2000fundamental,gebhardt2000relationship}. The reliability of these relationships depends on accurate black hole mass measurements. 
	
	M87 is one of the most massive galaxies of the Virgo cluster and sits at the centre of the main sub-cluster \citep[e.g.][fig.~7]{cappellari2011atlas}. It is a prototypical massive slow rotator early-type galaxy, with a large Sersic \citep{1963BAAA....6...41S} index and a core in the nuclear surface brightness profile \citep{kormendy2009structure}, and fits all characteristics for having assembled most of its mass by dry mergers \citep[see review by][]{cappellari2016structure}. Like other galaxies of its type, M87 contains a supermassive black hole at its centre \citep{kormendy2013coevolution} whose sphere of influence has the largest angular size of any known black hole outside of the Milky Way, making it a valuable target for black hole studies. The measurement of the black hole shadow by the Event Horizon Telescope (EHT) \citep{akiyama2019first} makes M87 the first galaxy for which direct imaging of the supermassive black hole has taken place. This has the potential to serve as a powerful test of the general theory of relativity: but only if we can confirm through independent measurements that the black hole mass recovered assuming general relativity is correct. The measurement of the black hole shadow from the EHT \citep{akiyama2019first} assuming general relativity determined the black hole mass to be $(6.5\pm0.7)\times 10^9 \text{M}_\odot$. In the case of M87 there are two other such classes of measurement. Gas dynamical measurements by \cite{harms1994hst}, \cite{macchetto1997supermassive}, and \cite{walsh2013m87} have measured the masses $(2.7\pm0.8)\times 10^9 \text{M}_\odot$, $(3.6\pm1.0)\times 10^9 \text{M}_\odot$, and $(3.3^{+0.8}_{-0.7})\times 10^9 \text{M}_\odot$, respectively. Stellar dynamical measurements \citep{gebhardt2009black,gebhardt2011black,2023arXiv230207884L} using the orbital superposition method of Schwarzschild \citep{schwarzschild1979numerical} have been made which found a black hole masses of $(6.0\pm0.5)\times 10^9 \text{M}_\odot$, $(6.2\pm0.4)\times 10^9 \text{M}_\odot$, and $(5.37^{+0.37}_{-0.25})\times 10^9 \text{M}_\odot$  respectively. There is thus a discrepancy in the recovered black hole masses by a factor of two depending on the method used. \cite{2019ApJ...882...82J,jeter2021reconciling} propose that this may be due to unrealistic assumptions made in the gas modelling. They find that more detailed accounting of the radial motion of the gas as well as allowing for a thick gas disk can alleviate the discrepancy (though it should be noted that their models are not fit to data). Recently, \cite{osorno2023m87} has produced detailed ionized gas maps of M87 using the same MUSE data as in this paper, revealing a complex multi-component gas structure. They suggest that the cause of the discrepancy is likely due to incorrect assumptions about the ionized gas disk inclination, though they comment that it is challenging to make an independent black hole mass measurement with this data. The agreement between stellar dynamical measurements and the measurement of the black hole shadow is reassuring, but it is still important to continue testing and independently verifying stellar dynamical models in order to fully understand possible systematics and to improve the robustness of the measurement. In this paper we derive new independent measurements of the black hole mass using stellar dynamics with two different high-resolution integral-field datasets from two different telescopes and a different dynamical modelling approach than previously used.
	
	
	This paper is laid out as follows: in \autoref{sec:data} we introduce the integral field data and photometric data used in this study. In \autoref{sec:kinext} we describe our spectral fitting and discuss a number of tests we performed and several methods of extracting the kinematics that we use. In \autoref{sec:photom} we use a combination of IFU data with photometry to accurately measure the stellar density profile for M87 down to the region dominated by the AGN. In \autoref{sec:dynmod} and \autoref{sec:discussion} we describe the details of our Jeans modelling and present our black hole mass constraints. We compare these with previous observations and discuss a number of systematic uncertainties. Lastly, in \autoref{sec:conclusion} we summarize our results and comment on the future landscape for studies of M87.
	
	We take the distance to M87 to be 16.8 Mpc \citep{akiyama2019first}. All black hole masses quoted are scaled to this distance. This corresponds to a spatial scale of 81.1 pc per 1 arcsecond.

	\section{Data}\label{sec:data}
	\subsection{Integral Field Spectroscopy}
	We use integral field observations from the Optically Adaptive System for Imaging Spectroscopy (OASIS) spectrograph made on the Canada-France-Hawaii Telescope (CFHT) \citep{mcdermid2006sauron} and the Multi Unit Spectroscopic Explorer (MUSE) \citep{bacon2010muse} in narrow field mode (NFM) on the Very Large Telescope (VLT) with adaptive optics (AO). This gives us two independent views of the central kinematics of M87. To add kinematic information at larger distances, we also use observations from SAURON \cite{bacon2001sauron}.
	
	The OASIS integral field spectrograph (IFS) has a 10\arcsec$\times$8\arcsec \ field of view with a 0\farcs27$\times$0\farcs27 pixel scale. For this measurement the spectrograph was configured to cover the wavelength range of 4760-5558 \AA \ with a resolution of 5.4 \AA \ FWHM (corresponding to an instrumental dispersion of $\sigma_{\rm inst} \approx 134$ km s$^{-1}$) sampled at 1.95 \AA \ per pixel. Three observations were made for 2700 seconds each, which were then combined to form the final image (see \cite{mcdermid2006sauron} for a detailed description of the observations and data reduction).

	The MUSE observations were made as part of program 0103.B-0581 (PI: N. Nagar) in NFM. The NFM covers a field of view of 7.5"x7.5" with a pixel scale of 0.025"/pix and is used together with the the adaptive optics facility GALACSI \citep{arsenault2008galacsi, 2012SPIE.8447E..37S}, providing laser tomographic AO corrections. The observations were carried out on 20 February 2021 and consist of nine dithered exposures of 700s, resulting in a total exposure time of   6300s. The data were reduced with the standard MUSE data reduction pipeline \citep{weilbacher2020data} in the ESO reflex environment \citep{freudling2013eso} using the dedicated offset sky exposures for sky subtraction with the standard pipeline parameters. The parameters for source detection and image alignment were optimized to align the individual exposures based on a combination of point sources and the knots of the jet of M87. We also tested whether the remaining sky residuals could be removed using the Zurich Atmospheric Purge (ZAP, \cite{Soto2016zap}) on sky residual cubes produced by the pipeline, but the improvement was not significant so we proceeded with the original cube. MUSE covers a wavelength range from 4650-9300 \AA \ with a gap between 5780-6050 \AA \ due to a Na notch filter blocking the light from the four laser guide stars facility (4LGSF). The spectrum is sampled at 1.25 \AA \ per pixel with a resolution of about 2.6 \AA \ FWHM, corresponding to an instrumental dispersion $\sigma_{\rm inst}\approx63$ km s$^{-1}$. For our analysis, we restrict the spectral range to cover only 4800-5700 \AA. The range greater than $\sim$7000 \AA \ has significant sky residuals so we choose to exclude this range in our analysis. The remaining range to the right of the notch filter (6050 \AA \ - 7000 \AA) does not have deep stellar features that provide information for stellar kinematics. Furthermore, this region has several gas emission features with multiple kinematic components. One way of dealing with this would be to mask all of the gas. However, this would conceal some kinematic features to the left of the notch filter and would leave only a few disconnected regions of unmasked spectrum to the right of the notch filter. We thus choose to proceed by omiting the region to the right of the notch filter and fit all of the gas to the left of the notch filter. We observe some slight flux calibration issues at the ends of the spectrum which we clip for our analysis.

	The SAURON observation of M87 was first described in \cite{emsellem2004sauron} and later reanalysed as a part of \cite{cappellari2011atlasone}. The field of view of SAURON is $33\arcsec \times 41\arcsec$ with a pixel size of 0\farcs94 $\times$ 0\farcs94. The wavelength range covered is 4800-5380 \AA \ at 4.2 \AA \ spectral resolution (corresponding to an instrumental dispersion of $\sigma _{\rm inst} = 108$ km \ s$^{-1}$) sampled at 1.1 \AA \ per pixel. Four observations were made for 1800 seconds each. The data reduction was performed with \textsc{xsauron} \citep{bacon2001sauron}. Further details are available in \cite{emsellem2004sauron,cappellari2011atlasone}.

	\subsection{Photometry}
	We use HST imaging from \cite{cote2004acs} (HST proposal 9401) to construct our stellar surface brightness model while carefully removing the AGN. This is an F850LP ACS/WFC observation covering a field of view of approximately 211\arcsec$\times$212\arcsec \ with a pixel scale of 0\farcs05. The exposure was made for 90 seconds, guaranteeing that the nucleus does not become saturated. We also use an r-band SDSS mosaic generated with the software \textsc{Montage}\footnote{Available from \url{http://montage.ipac.caltech.edu/}} to constrain the stellar surface brightness at larger radii. The SDSS image covers a spatial scale of approximately 713\arcsec$\times$713\arcsec \ with a pixel scale of 0\farcs396.

	\section{Kinematic Extraction}\label{sec:kinext}
	\subsection{Spectral Fitting}\label{sec:specfit}
	We bin the galaxy spectra spatially using the Voronoi tesselation algorithm and \textsc{VorBin} software package\footnote{Available from \url{https://pypi.org/project/vorbin/}} described in \cite{cappellari2003adaptive}. This algorithm takes the x and y coordinates of a set of data along with the assigned signal and noise and bins neighboring points to a target signal to noise ratio. We define the signal to be the median spectral flux and the noise to be the median error (this is done before logarithmically rebinning the spectra). For the OASIS data, we bin to a target signal to noise ratio of 50 per 1.95 \AA \ spectral pixel. This leaves all of the spaxels in the innermost arcsecond unbinned, allowing for maximum spatial resolution. MUSE has a much higher spatial resolution, requiring some spatial binning in the innermost parts of the galaxy to have sufficient signal for kinematic fitting. We start by masking the innermost 0\farcs1 as these spectra are entirely dominated by the AGN\footnote{Note that we do not mask the jet. This is because the jet contributes a much smaller flux to the total spectrum and can be fit by additive Legendre polynomials}.  We tested binning with a target signal to noise ratio of 50, 40, 30, 20, 15, and 10 per 1.25 \AA \ spectral pixel. The recovered black hole mass is consistent for 50, 40, 30, and 20, but increases sharply for lower signal to noise. This is due to the fact that with this level of noise, a flat fit to the spectra is allowed in the innermost regions, resulting in many central spectra having anomalously large dispersion. For the rest of this work, we use the case with a signal to noise ratio of 50 per 1.25 \AA \ spectral pixel. For both datasets, we remove all data points for which the fraction of the flux due to stars is less than 50\%. This is determined after running the fits by comparing the average flux in the Legendre polynomials to the stars (more on this later).
	
	\begin{figure}
		\centering
		\subfloat{\includegraphics[width = \columnwidth]{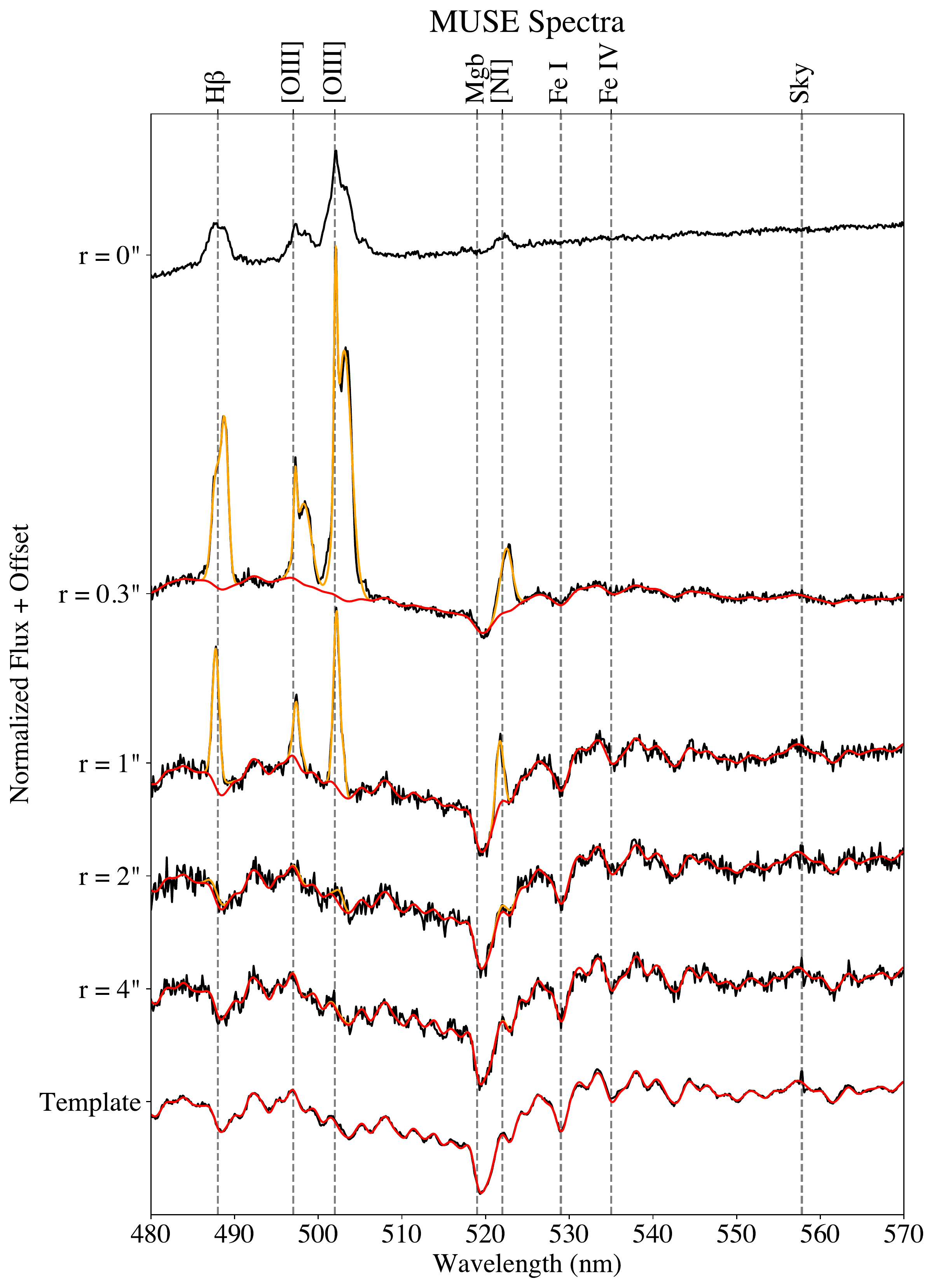}}
		\caption{Plot of the MUSE spectra as a function of radius. The black curve is the data and the red curve is the best \textsc{pPXF} fit to the stars and the orange curve is the best fit to the gas. The $r = 0\arcsec$ spectra is the combined innermost $0\farcs1$ of MUSE. This is not fit as we do not include it in the final analysis. The double peaked structure of the gas appears intermittently at a variety of radii. The bottom spectra shows the single stellar template that we use to fit the stellar kinematics for all of the MUSE spectra, as well as the sum of the gas free spaxels that we use to fit for the single stellar template. The mean luminosity weighted radius of the gas free spectra is 3\farcs3.}
		\label{fig:radial_spectra1}
	\end{figure}
	
	\begin{figure}
		\centering
		\subfloat{\includegraphics[width = \columnwidth]{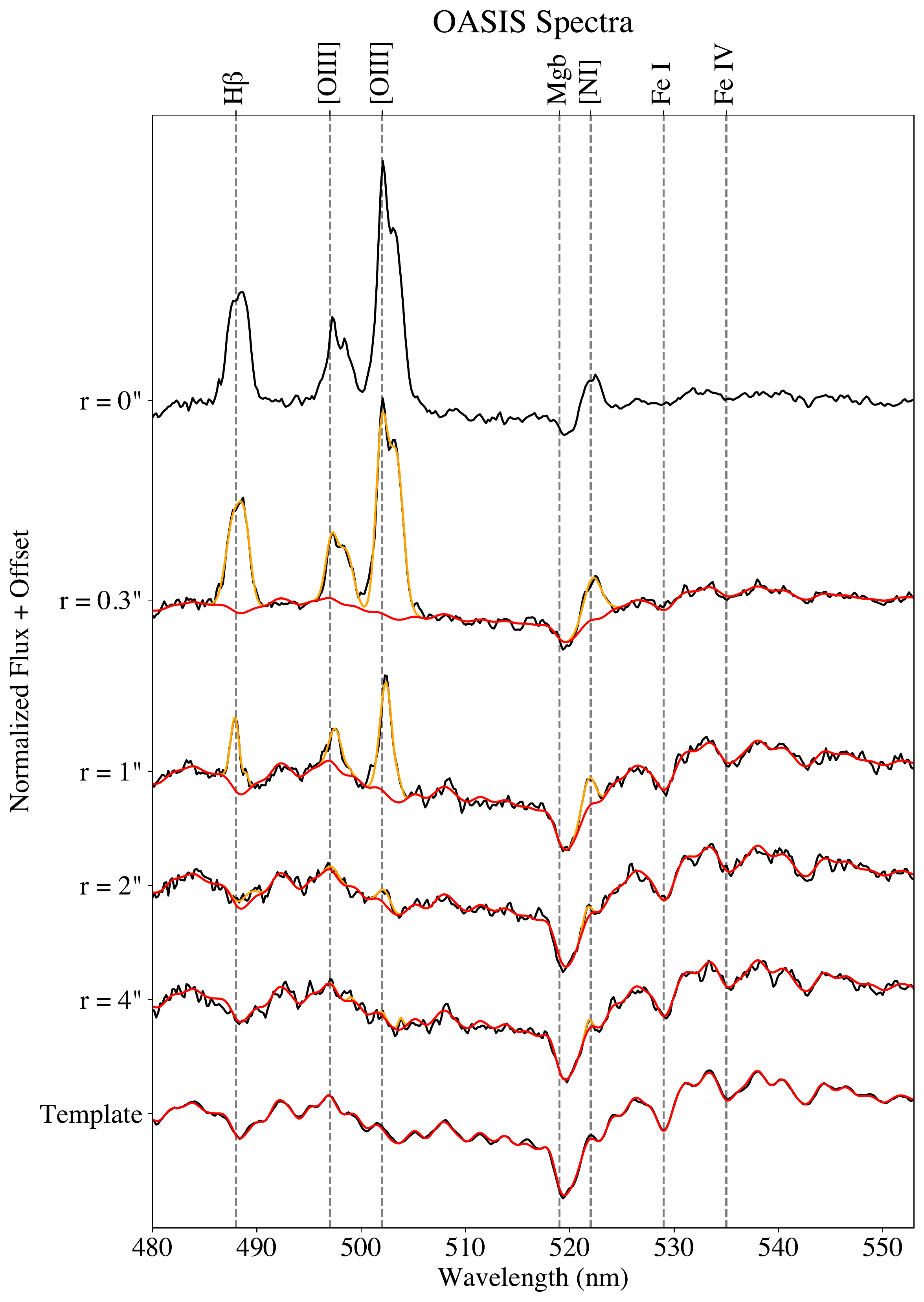}}
		\caption{Plots of the OASIS spectra as a function of radius. The black curve is the data and the red curve is the best \textsc{pPXF} fit to the stars and the orange curve is the best fit to the gas. The $r = 0\arcsec$ spectra is the central spaxel for OASIS. This is not fit as we do not include it in the final analysis. The double peaked structure of the gas appears intermittently at a variety of radii. The bottom spectra shows the single stellar template that we use to fit the stellar kinematics for all of the OASIS spectra, as well as the sum of the gas free spaxels that we use to fit for the single stellar template. The mean luminosity weighted radius of the gas free spectra is 3\farcs7.}
		\label{fig:radial_spectra2}
	\end{figure}
	
	\begin{figure}
		\centering
		\subfloat{\includegraphics[width = 0.5\columnwidth]{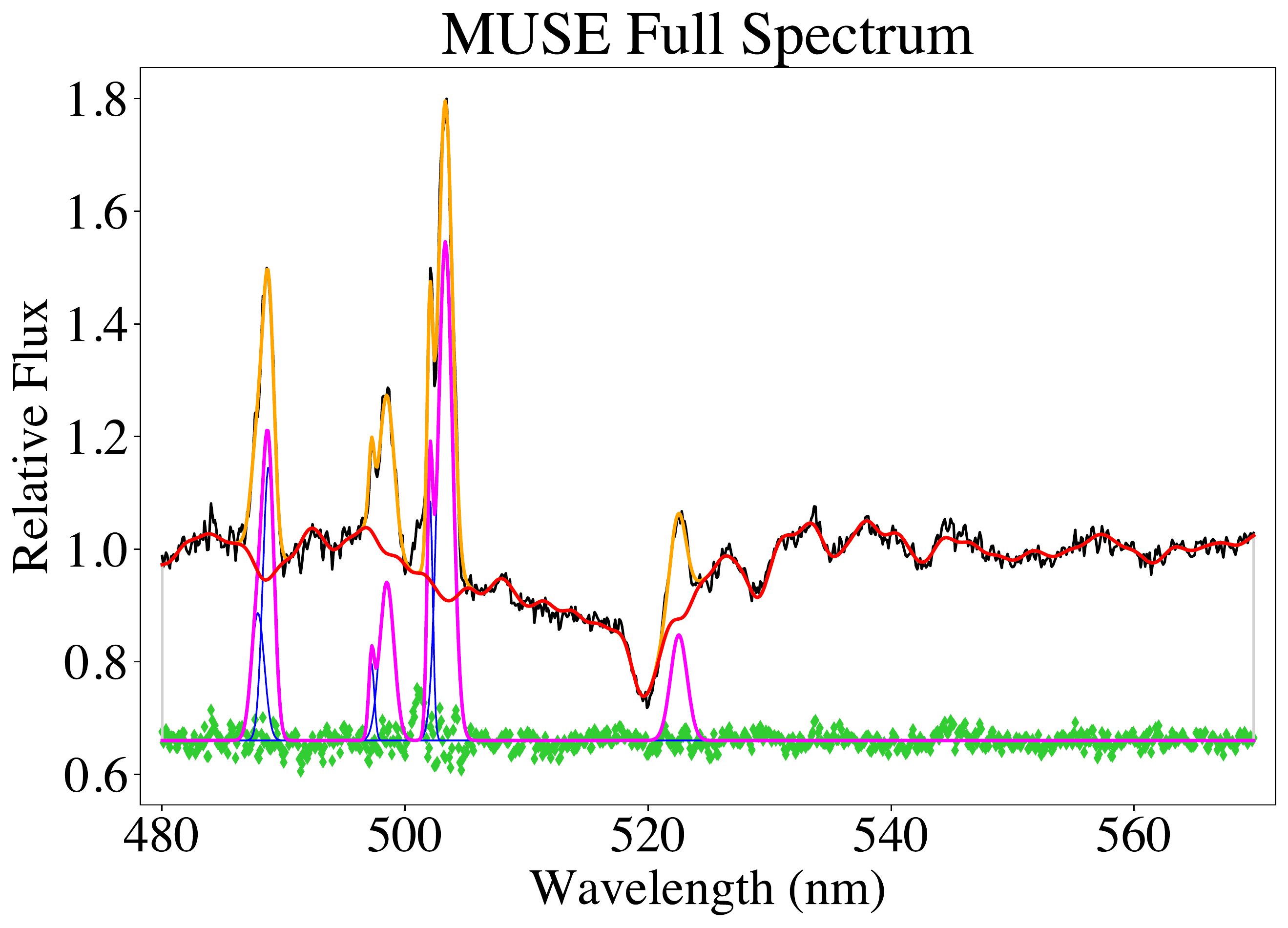}}
		\subfloat{\includegraphics[width = 0.5\columnwidth]{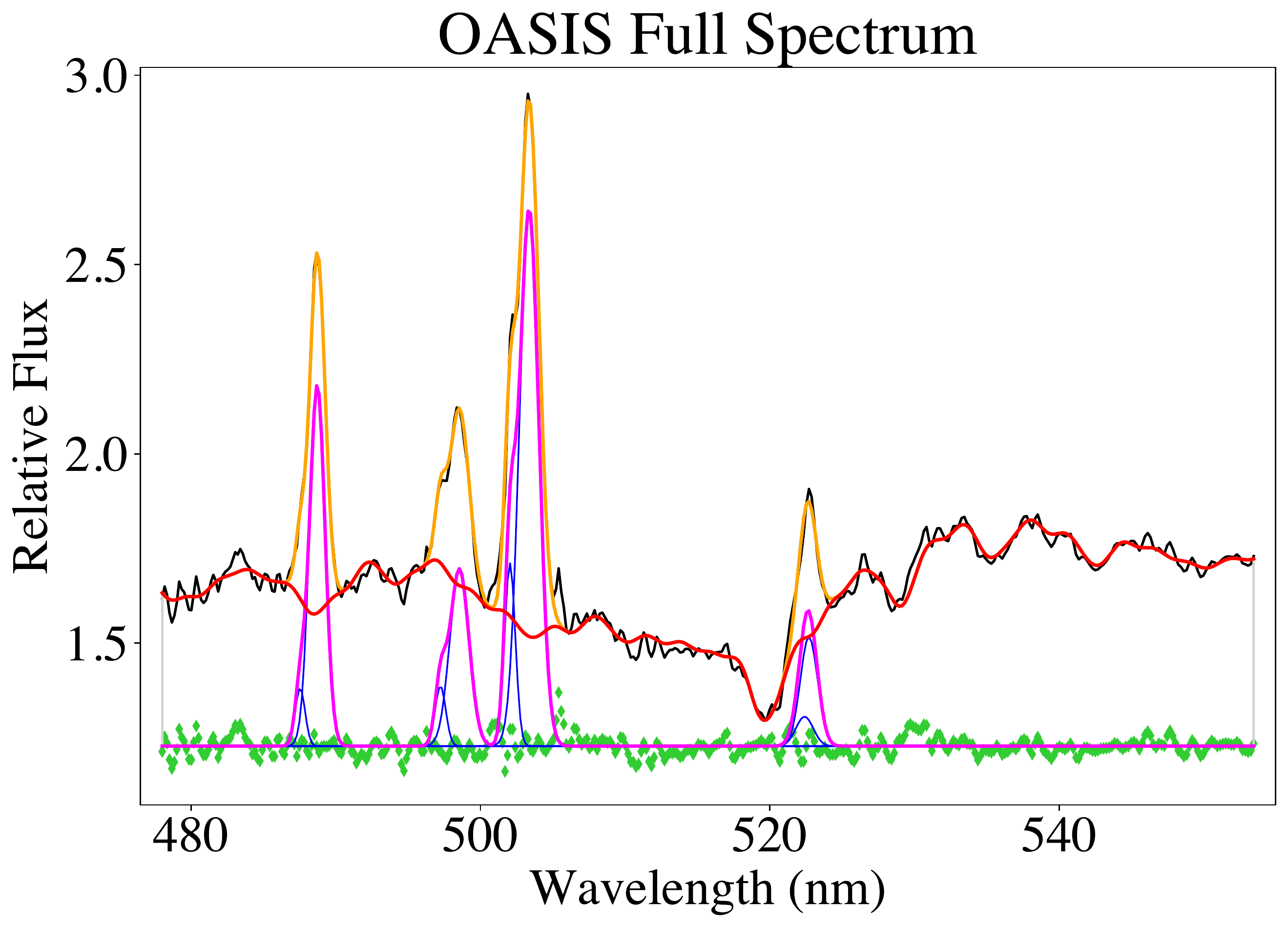}}\\
		\vspace{-10pt}
		\subfloat{\includegraphics[width = 0.5\columnwidth]{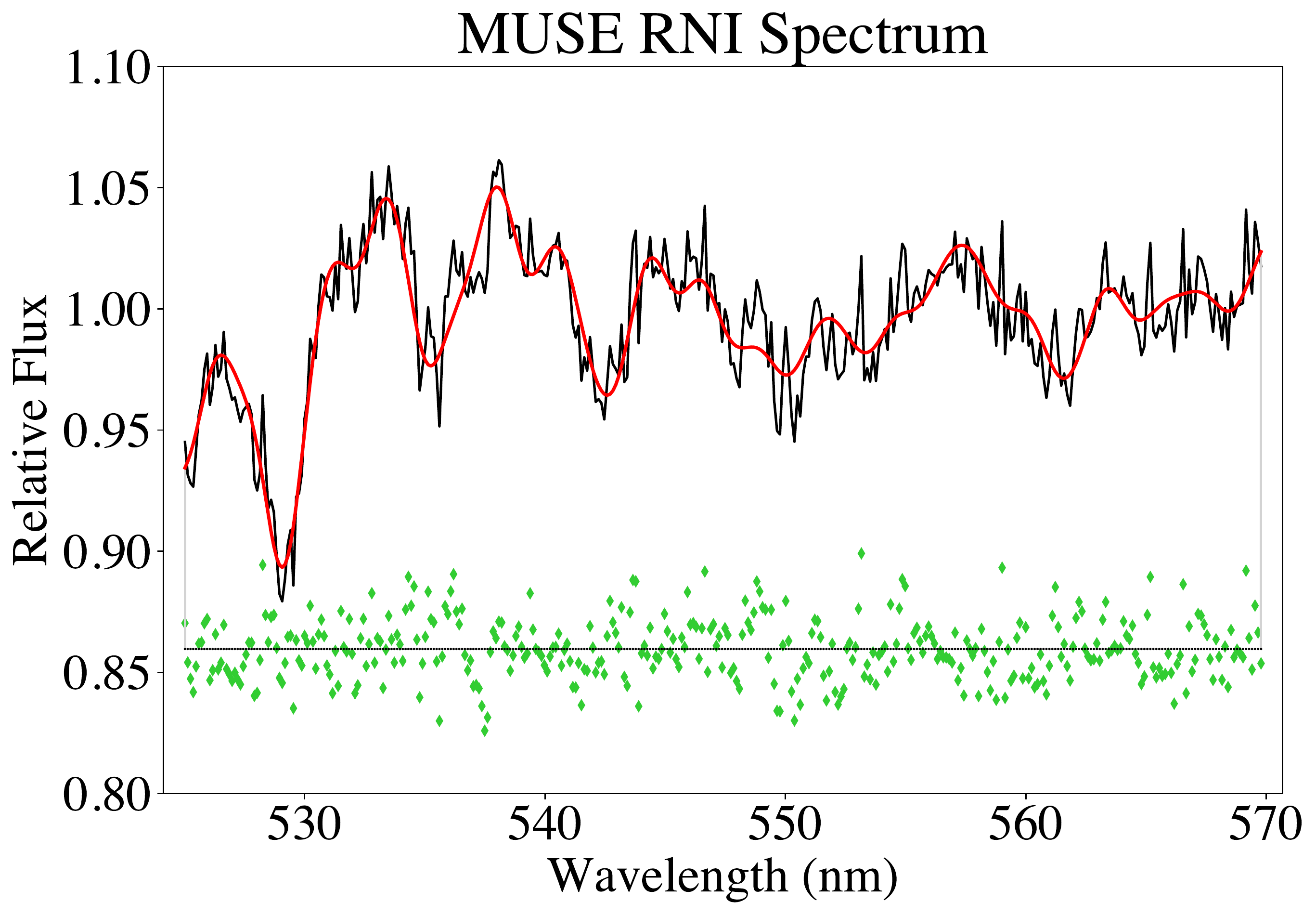}}
		\subfloat{\includegraphics[width = 0.5\columnwidth]{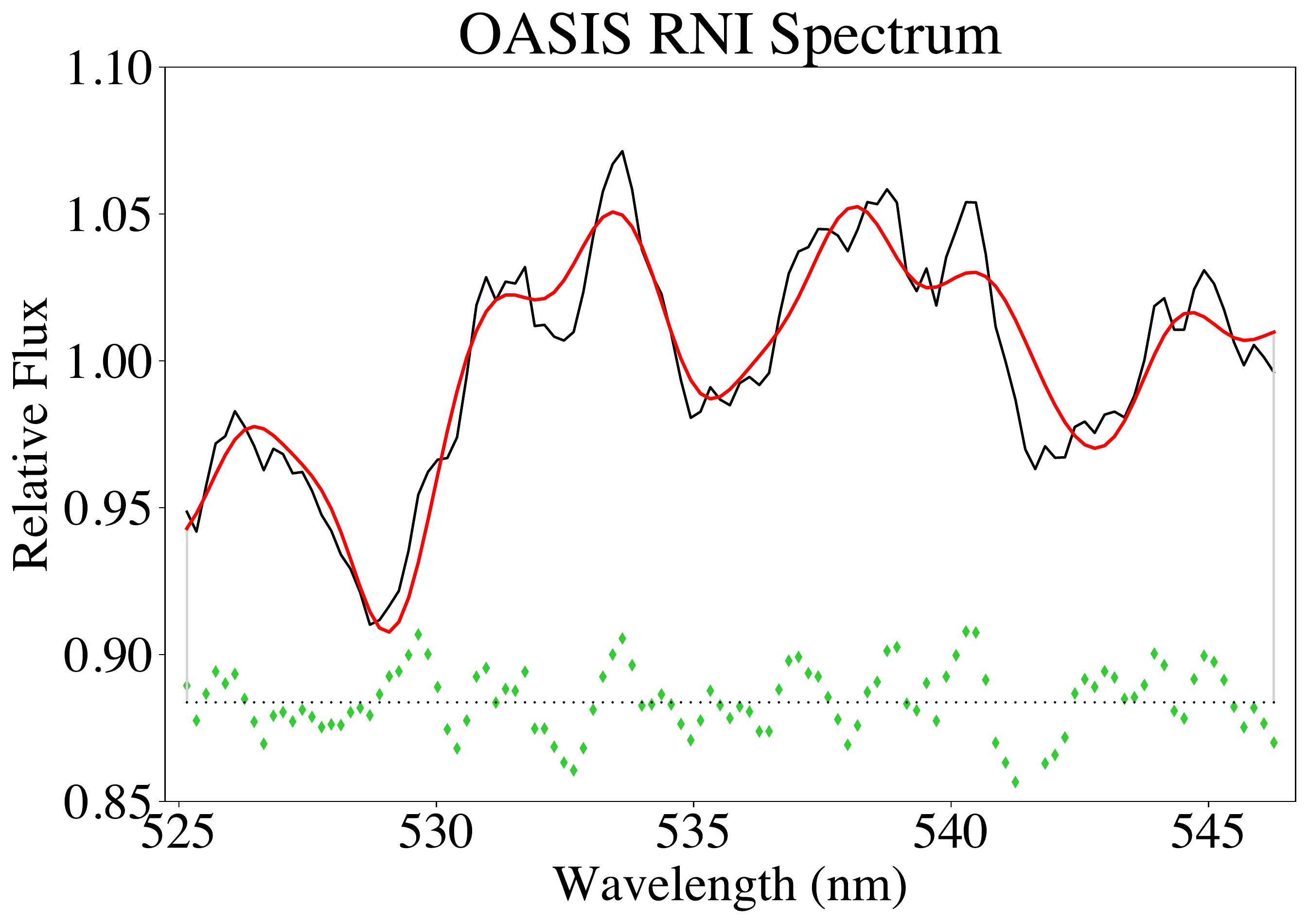}}
		\caption{The top panels show a fit of the MILES stellar spectra to the M87 spectra at large radii without gas emissions. This is done separately for MUSE and OASIS. This fitted spectra is then used to fit the stellar kinematic component for each spaxel. The middle panels show a spectrum from the centre of M87 that has been fitted with the above stellar template, as well as with multiple gas templates. This shows strong gas emissions with a double peaked profile for H$\beta$ and [OIII]. These must be carefully accounted for in order to produce a reliable kinematic fit. The bottom panels show a fit of the stellar template to a spectrum in the central parts of M87 where the spectra has been restricted to start at the right of [NI] (denoted RNI spectra). Here there is no gas, simplifying the fitting, but the noise in the spectrum increases.}
		\label{fig:two_peak}
	\end{figure}

	
	We logarithmically resample the spectra to a velocity scale $\Delta V=c\Delta\ln\lambda$ of 105 and 66 km s$^{-1}$ for OASIS and MUSE respectively and fit the binned spectra using the penalized Pixel Fitting method \textsc{pPXF} software package\footnote{Available from \url{https://pypi.org/project/ppxf/}} of \cite{cappellari2004parametric} and \cite{cappellari2017improving,cappellari2022full}. This method allows for the simultaneous fitting of template stellar spectra, template gas spectra, and continuum contributions/template mismatch with the addition of additive polynomials. We can write an observed galaxy spectrum as\footnote{This ignores sky, attenuation, and multiplicative polynomials. See the full expression in eq. 13 of  \cite{cappellari2022full}} 
	\begin{equation}
		\begin{aligned}
			G = &\sum_i w_i[T_i^s(x) * \mathcal{L}_i^s(cx)]+ \sum_j u_j[T_j^g(x) * \mathcal{L}_j^g(cx)]  \\ & + \sum_k b_k\mathcal{P}_k(x)
		\end{aligned}
	\end{equation}
	Here $T^s$ represents the stellar templates, $T^g$ the gas templates, $\mathcal{L}$ the corresponding line of sight velocity distribution, and $\mathcal{P}$ additive polynomials (in this case taken to be Legendre polynomials).

	The strongest absorption feature in the OASIS and MUSE spectra that contributes to the fits of the stellar kinematics is due to Mgb around 5200 \AA. However, in the innermost arcsecond of the galaxy this feature is contaminated by gas emission from [NI]$\lambda\lambda$5197,5200 (see \autoref{fig:radial_spectra1}, \autoref{fig:radial_spectra2}, and \autoref{fig:two_peak}). When it comes to fitting the absorption feature from Mgb then, there is a degeneracy between the stellar dispersion and the gas kinematics. This is compounded by the fact that it is precisely in the innermost regions that the relative flux in each spectra due to the AGN increases, further increasing the uncertainty in the kinematic extraction. In order to test how the treatment of this affects the extracted kinematics, we consider two separate scenarios. In the first scenario we simply fit the full spectrum. This has the advantage that outside of the innermost arcsecond the kinematic extraction is very reliable with the disadvantage of the innermost arcsecond being less reliable. In the second scenario we restrict the spectra to begin at 5250 \AA \ (these spectra are referred to as spectra to the right of [NI] or RNI spectra) so that we exclude the contaminated region. This has the benefit of avoiding any uncertainty from the fit to Mgb at the cost of further restricting the spectral range. This is done for both OASIS and MUSE. The details of the kinematic extraction for each of these scenarios is significantly different.

	\begin{figure}
		\centering
		\subfloat{\includegraphics[width = 3in]{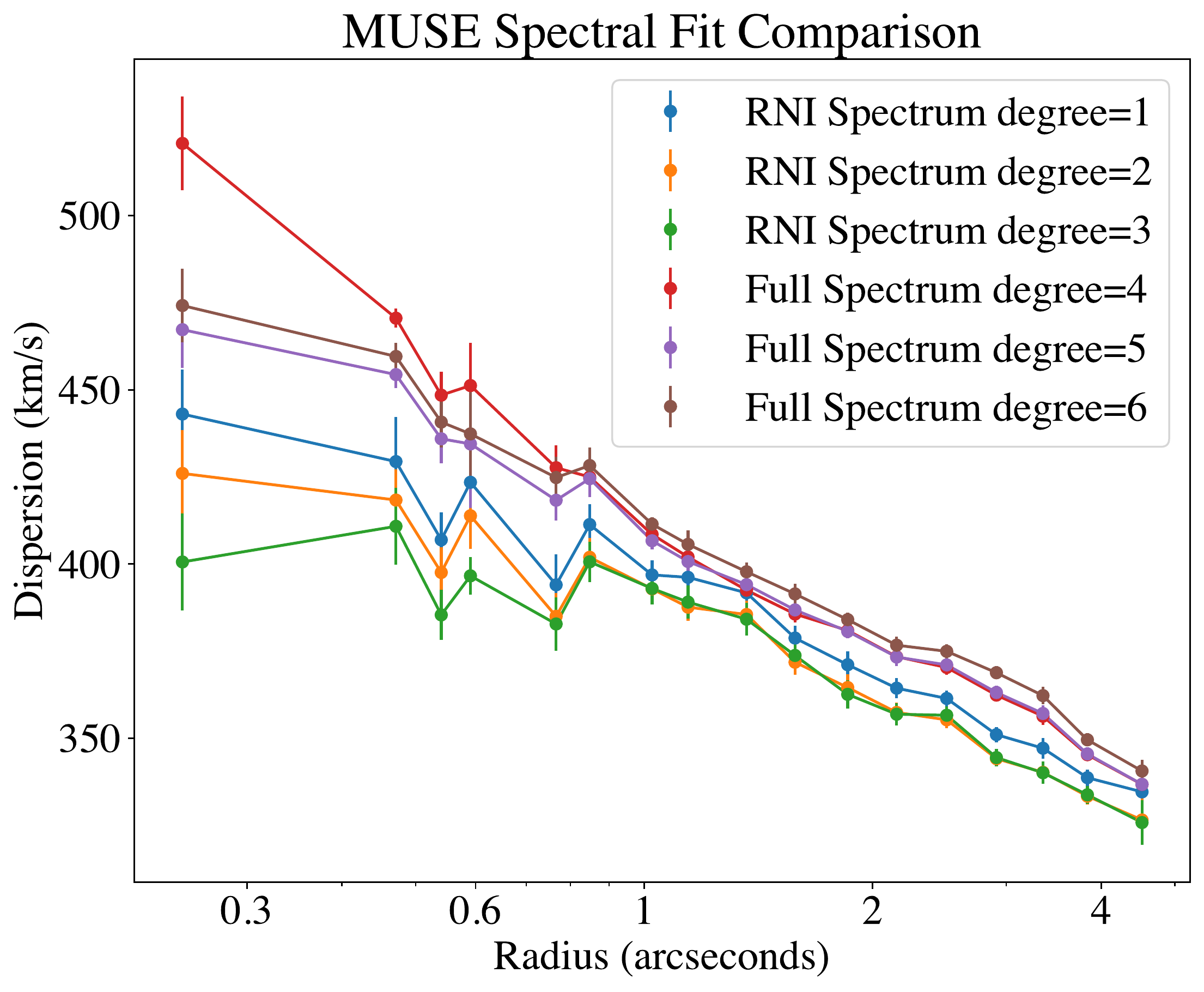}}\\ 
		\vspace{-10pt}
		\subfloat{\includegraphics[width = 3in]{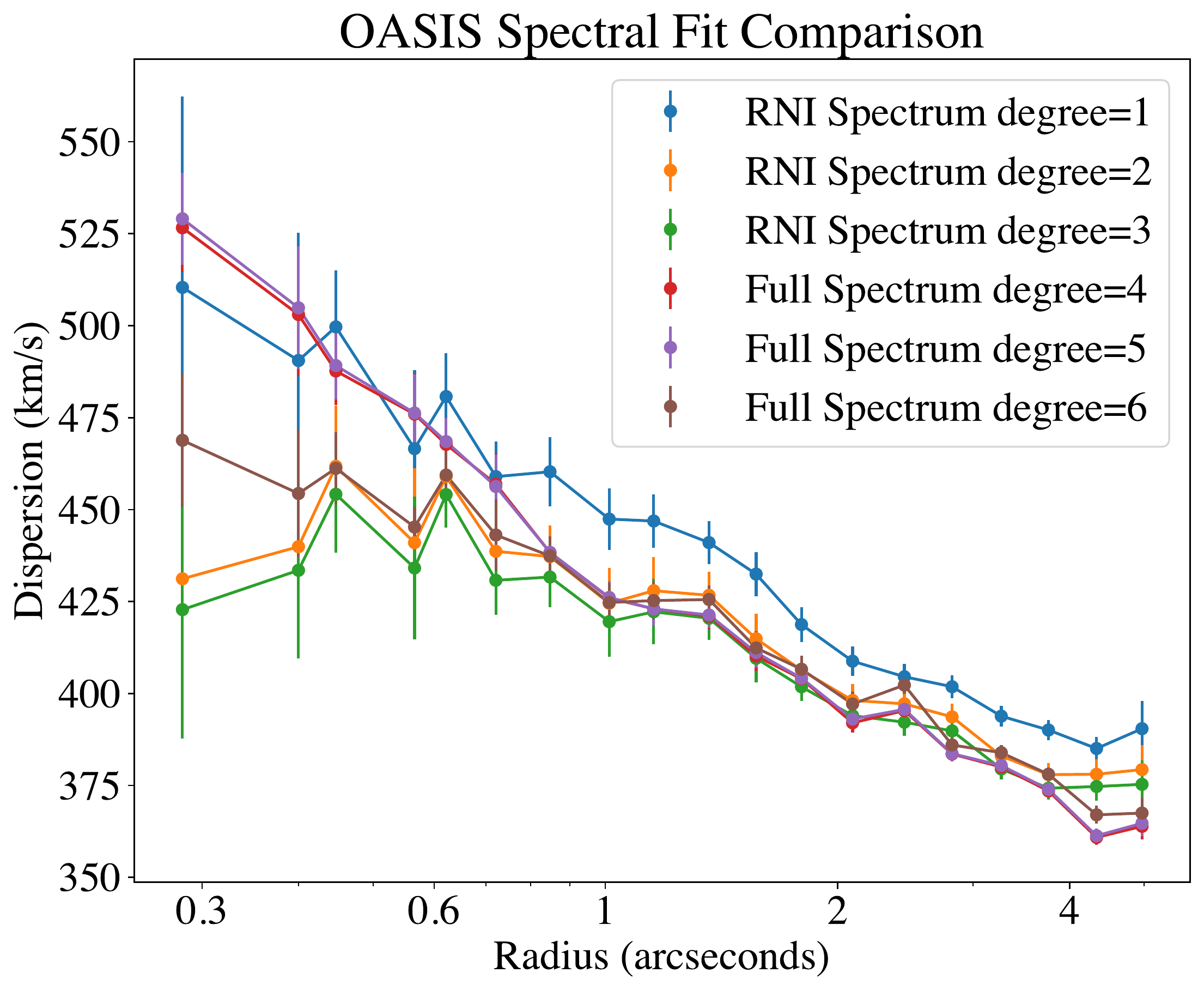}}
		\vspace{-10pt}
		\subfloat{\includegraphics[width = 3in]{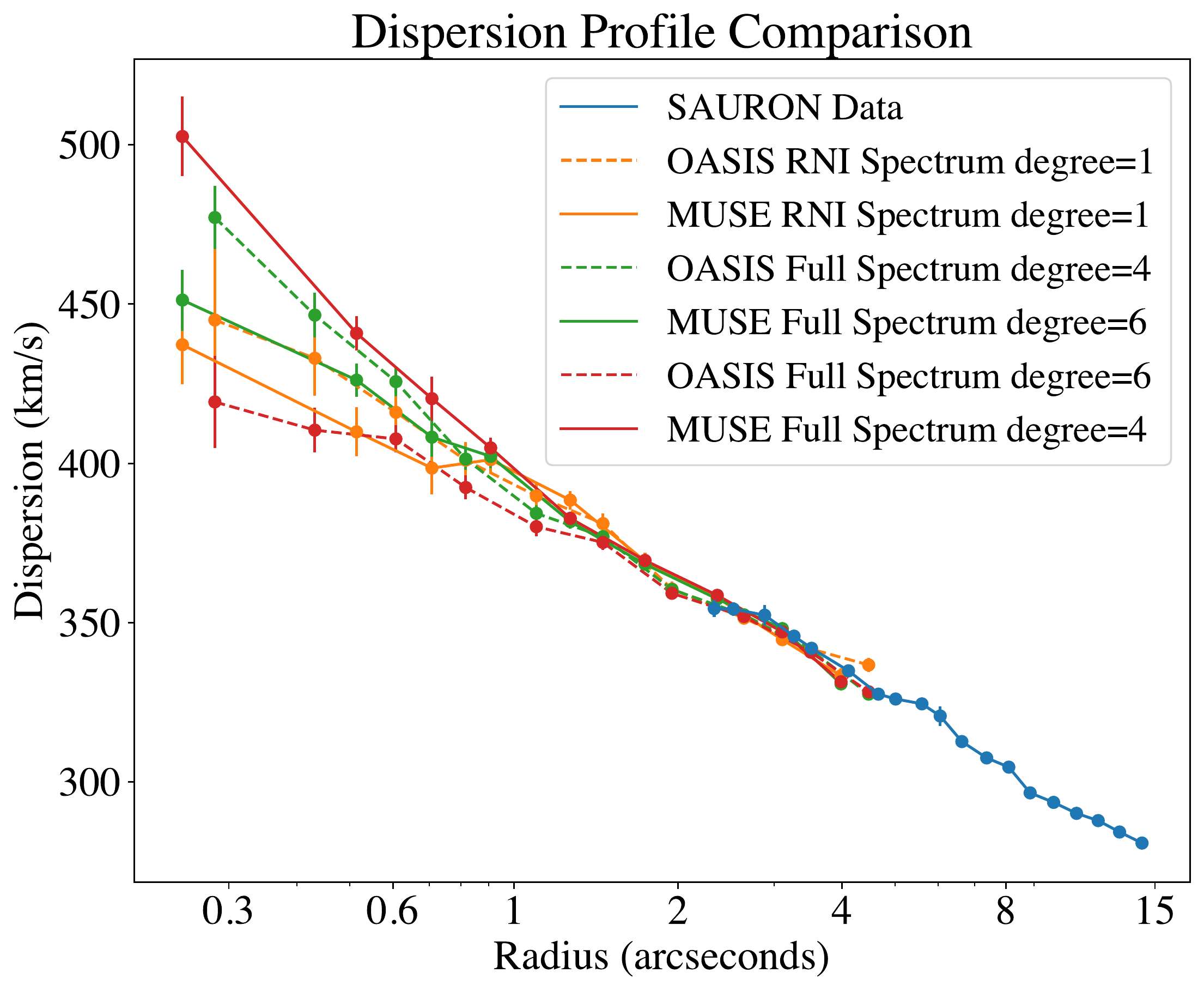}}
		\caption{The top and centre panels show the extracted dispersion for M87 under a number of different assumptions for MUSE and OASIS, respectively. Note that in each case the dispersion profile is consistent up to a constant scaling outside of one arcsecond. Within one arcsecond, however, there is a large deviation in the extracted profiles. This is due to the uncertainty in the kinematic extraction caused by contamination from the AGN and the degeneracy in the fits between the fit to [NI] and the Mgb absorption. The bottom panel shows the dispersion profiles used in the final analysis scaled to match the SAURON profile. Data shown in the same color are combined in the final analysis.}
		\label{fig:all_kin}
	\end{figure}

	Spectral fitting over the full wavelength range is challenging to perform due to the presence of strong gas emission lines from H$\beta$, [OIII]$\lambda\lambda$4959,5007, and [NI]$\lambda\lambda$5197,5200. Furthermore, we observed multiple gas kinematic components for H$\beta$ and [OIII] in the inner most part of the galaxy (see \autoref{fig:two_peak}). Additionally, the presence of the central AGN dilutes the stellar features of the central spectra making a reliable extraction of the kinematics without an assumption of the spectra shape difficult. Furthermore, since we fit the gas, there is a degeneracy between the stellar flux and the flux from the gas. In order to increase the robustness of the extracted kinematics, we allow for only one single stellar component in our fits (i.e. we fit one linear combination of template spectra to the data, as opposed to many variably weighted template spectra). We determine this single stellar template by co-adding the spectra with no gas emission lines and fitting this with stars from the MILES stellar library\footnote{Available from \url{http://miles.iac.es/}} \citep{sanchez2006medium,falcon2011updated}. Note that one single stellar template does not mean one constant stellar population across the field of view since the Legendre polynomials allow for significant variations in the line strength of the spectral features of the stellar population (note, we know from \cite{sarzi2018muse} that there are no sharp changes to the stellar population in the innermost parts of M87). We use the MILES stellar library consisting of nearly 1000 individual stellar spectra. We also tried this using SSPs from the MILES stellar library \citep{vazdekis2010evolutionary} including synthetic spectra with alpha enhancement \citep{vazdekis2015evolutionary} allowing for maximum possible variation in the parameters, as well as the MUSE stellar library \citep{ivanov2019muse}, consisting of 35 individual stellar spectra, but we ultimately find the best fit using the MILES stars. 
	
	The choice to fit only a single stellar template as opposed to the full spectral library is very important for galaxies with an AGN. \cite{silge2005cena} measured the supermassive black hole mass of Centaurus A using Gemini NIFS data, allowing for a varying stellar template and found a steeply increasing dispersion profile in the center of the galaxy and measured a black hole mass between (1.5-2.4)$\times 10^8 \ M_\odot$. This was a significant outlier of the $M-\sigma$ relation and did not agree with later measurements of the black hole mass using gas \citep{neumayer2007cenagas}. Later, \cite{cappellari2009mass} performed the same measurement with SINFONI data using a fixed single stellar template and found a black hole mass equal to $5.5\times10^7 \ M_\odot$. This was no longer a significant outlier of the $M-\sigma$ relationship and is in excellent agreement with the determination of the black hole mass from gas \citep{neumayer2007cenagas}. The case where there is no AGN was tested in \cite{westfall2019mangapipeline}. There, the authors show in figure 13 that the difference in the extracted kinematics for high mass red galaxies between those using a single or varying template is at most 3 per cent. This is in contrast to the case of blue galaxies for which there is a much stronger deviation, likely due to the fact that these galaxies are undergoing star formation and thus have stronger population gradients. We also note that, while we independently determine the single stellar template for both MUSE and OASIS, the two agree quite closely (less than one percent RMS deviation over the relevant wavelength range). Using the OASIS template to fit the MUSE data and vice versa returns effectively the same kinematics as just using the OASIS template for OASIS and the MUSE template for MUSE.
	
	The OASIS spectral resolution is much larger than that of the stars in the MILES stellar library which have an instrumental resolution of 2.51 \AA \ FWHM corresponding to $\sigma_{\rm instr}\approx61$ km s$^{-1}$ at 520nm. We account for this by degrading the resolution of the template spectra (before logarithmically rebinning) with a gaussian to a constant resolution per angstrom so that the spectral resolution for the two are the same. The spectral resolution for MUSE over the wavelength range we use is nearly the same as that of the MILES stars, so we do not apply a correction. We determine the single stellar template separately for both OASIS and MUSE. We then allow gas templates in the \textsc{pPXF} fits for $H\beta$, [OIII] and [NI], but allow H$\beta$ and [OIII] to have two distinct kinematic components each. Given the number of templates (1 stellar spectrum + 2 H$\beta$ + 2 [OIII] + 2 [NI] spectra) and the fact that the gas could be challenging to fit due to the possibility of there being multiple local best fits, we experimented with a number of constraints, such as treating the gas components of H$\beta$ and [OIII] as a part of the same kinematic component by fixing their velocity and dispersion to be equal. Ultimately, we found that the most reliable fit to the stellar spectra is comes from allowing maximum freedom in the gas fit. That is, treating each gas template as having its own velocity and dispersion. This is because even slight offsets in the gas fits for central spectra result in large residuals that end up driving the stellar fit. This means having a total of six kinematic components: one for the stars, one for each set of H$\beta$ and [OIII], and one for [NI].

\begin{figure}
		\centering
		\subfloat{\includegraphics[width = \columnwidth]{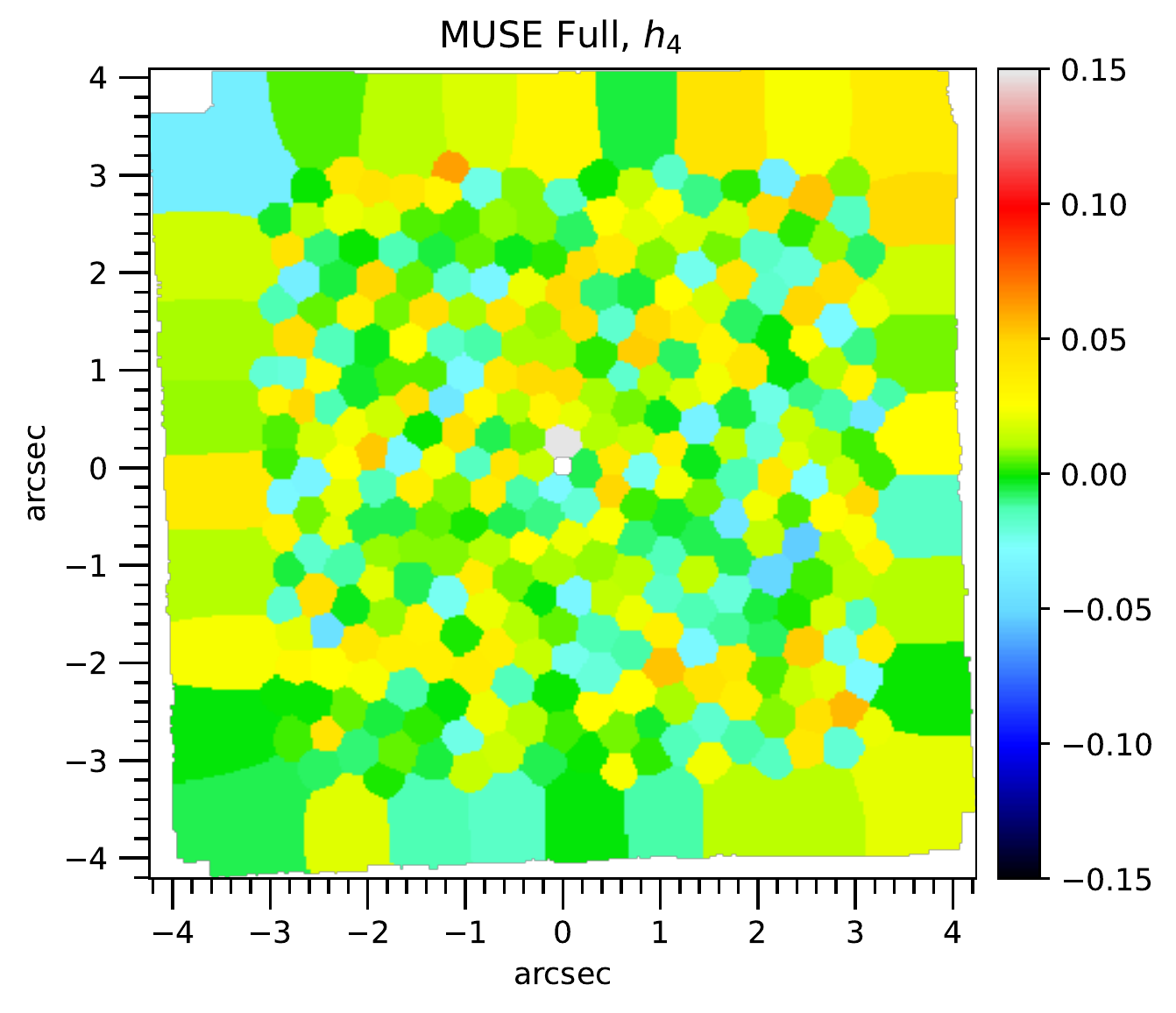}}
		\caption{Map of $h_4$ measured by running \textsc{pPXF} on the Full MUSE spectra with Legendre polynomial degree of 4. The values are centered on zero with some scatter. The results are unchanged under different Legendre Polynomial assumptions.}
		\label{fig:h4}
	\end{figure}

	We assume that the line of sight distribution can be treated as a Gaussian without the inclusion of higher order Gauss-Hermite moments. We made this assumption because (i) \citet[sec.~2]{cappellari2007sauron} found using synthetic galaxy models that the sigma obtained from a Gaussian fit (moments=2) with \textsc{pPXF} provides a better approximation to the second velocity moments than computing of the second moment by integrating the LOSVD from a fit which includes higher Gauss-Hermite moments (e.g. moments=4); (ii) making this assumption one is able to accurately predict with JAM the observed $V_{\rm rms}$ of hundreds of real galaxies (e.g. fig.~1 of \cite[e.g.][fig.~1]{cappellari2015small}; and (iii) in the specific case of the stellar kinematics of M87, previous studies have found that, within the range of our data, the lowest order Gauss-Hermite moments are either consistent with zero or small \cite[fig. D2]{2023arXiv230207884L}, implying that the LOSVD is essentially Gaussian. To test this, we run \textsc{pPXF} for our MUSE data where we include Gauss-Hermite moments up to $h_4$. We find evidence for some offset in $h_3$ (median between 0.015 and 0.018 depending on the choice of Legendre Polynomial) which suggests that there is some small template mismatch. We find that $h_4$ is consistent with zero (see \autoref{fig:h4}). This is consistent with the result of \cite{2023arXiv230207884L}. This holds across the choice of Legendre Polynomials. Thus we feel confident not including the Gauss-Hermite moments in our analysis.
	
	One issue that we faced was some of the fits returning large stellar velocities in a couple of the central spaxels of the galaxy. M87 is well known to be a slow rotator \citep{emsellem2007sauron,emsellem2014kinematically}, so any large deviations in the velocity suggest an error in the fitting. We handle this by fixing the velocity across the field for both datasets to equal the recession velocity\footnote{Note that \cite{emsellem2014kinematically} finds evidence for a kinematically decoupled core with rotation velocity $\pm$5 \ km s$^{-1}$. This is within the errors of this analysis.}. This serves as a realistic prior that helps decrease the noise in our kinematic extraction. In order to confirm that this does not impact the extraction of the dispersion, we compared the dispersion measured before and after fixing the velocity. We found that, for both MUSE and OASIS, this did not meaningfully impact the dispersion over the field of view (less than one percent RMS deviation between the two) and the difference for the central spaxels is less than $\sim$2 per cent. It is worth noting that the SAURON data we use does not fix the velocity at any point over the field of view. Given that we exclude SAURON data from the central regions and we know that fixing the velocity in the center does not significantly impact the dispersion at large radii, we believe that it is consistent to do this.
	
	Previous work has not run into this issue due to differences in the spectral range considered. \cite{gebhardt2011black} uses data covering the CO band head, which features several deep absorption structures ideal for determining stellar kinematics. \cite{emsellem2014kinematically} uses MUSE data in wide field mode without AO which does not feature the notch filter designed to block light from laser guide stars used in AO and thus can see the absorption due to the sodium doublet around 5900 \AA. Additionally, \cite{2023arXiv230207884L} uses much bluer data with spectral features such as CA II H, CA II K in addition to Mgb that add valuable kinematic information.
	
	We calculate errors in the dispersion using wild bootstrapping \citep{davidson2008wild} of the spectra residuals and repeating the pPXF fits 100 times on the bootstrapped spectra. Lastly, we perform this analysis three separate times using Legendre polynomials of degrees 4, 5, and 6. We find that the extracted dispersion in the centre most region depends on the choice of Legendre polynomial (see \autoref{fig:all_kin}). This is likely due to the degeneracy between [NI] and Mgb absorption. Different degrees of Legendre polynomial, especially those with very high degree given the total wavelength range, can go beyond accounting for template mismatch and can start reproducing parts of the stellar spectrum. 
	
\begin{figure}
		\centering
		\subfloat{\includegraphics[width = 3in]{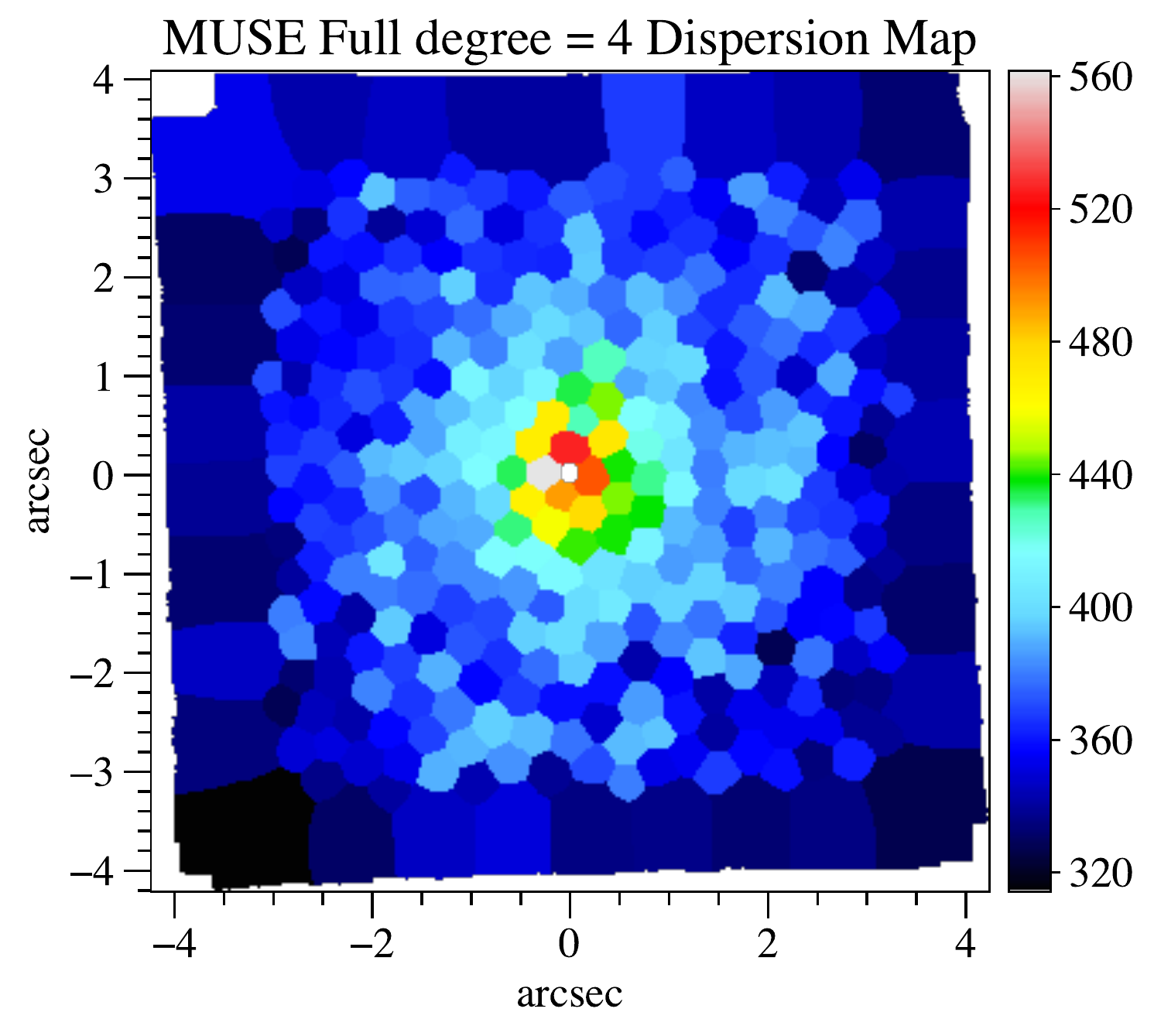}}\\
		\vspace{-15pt}
		\subfloat{\includegraphics[width = 3in]{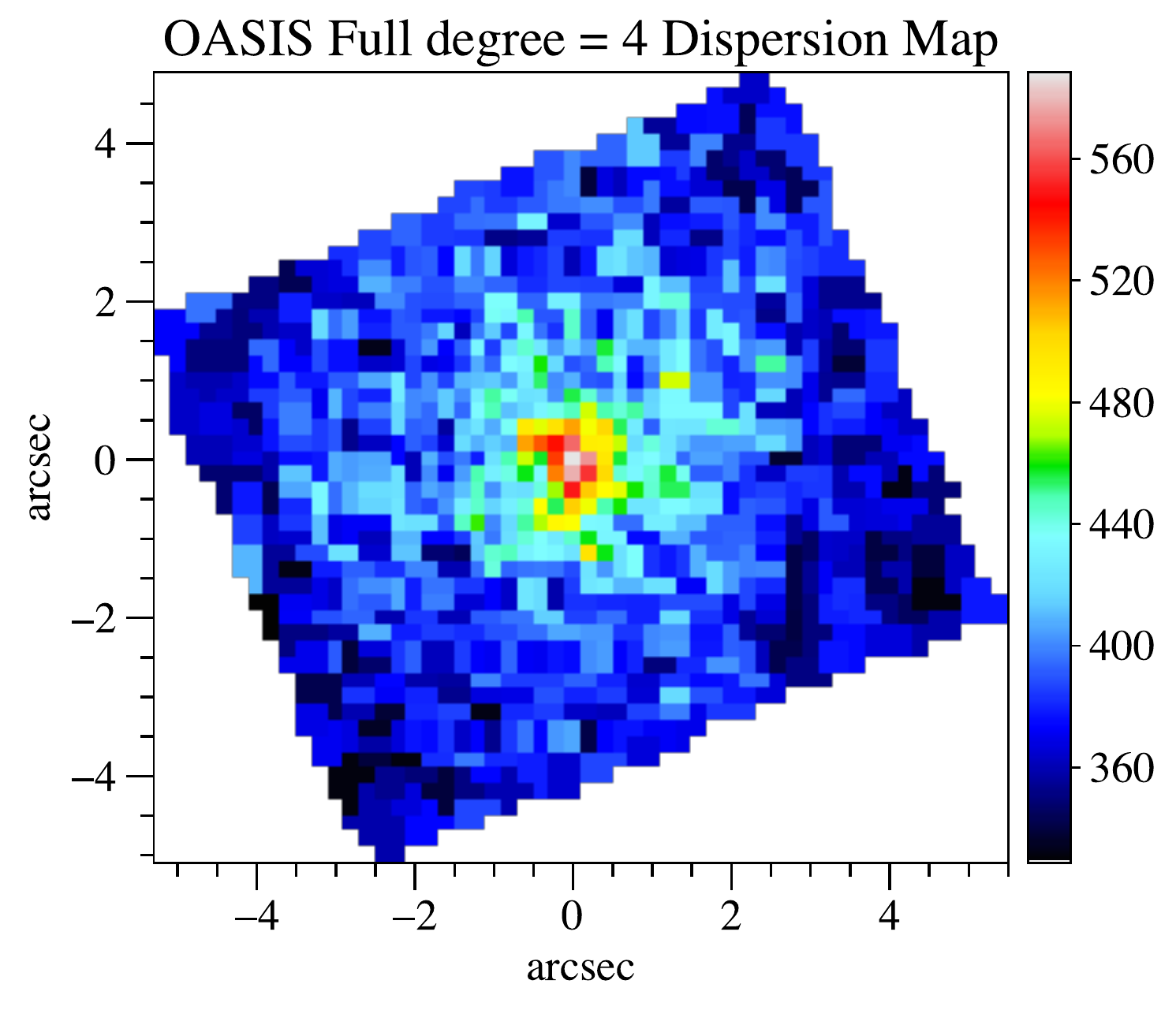}}
		\caption{Top panel: MUSE kinematics extracted over the full spectrum with additive Legendre polynomial degree of 4. The centremost 0.1\arcsec \ is masked (this is the circle in white at the origin). Bottom panel: OASIS kinematics extracted over the full spectrum with additive Legendre polynomial degree of 4. Both are oriented so that north is facing up and east is facing to the left. }
		\label{fig:Disp_map}
	\end{figure}

	Kinematic extraction redwards of [NI] is done using the same single stellar template as in the case for the full spectra, no gas templates, and additive Legendre polynomials of degree 1, 2, and 3. As in the previous case, we also fix the velocity of each spectra to the recession velocity, which we determined as the median over the field of a free fit. A dispersion map of the extracted kinematics in the case where we fit the full spectrum with additive Legendre polynomials of degree 4 is shown in \autoref{fig:Disp_map}. The map appears symmetric, as we expect since the core of M87 is highly spherical. As a result, we plot the dispersion profile for the remaining scenarios as a function of radius in \autoref{fig:all_kin}.
	
	The supermassive black hole in M87 has the largest angular size from Earth of any known black hole outside of the Milky Way. We define the sphere of influence $r_{\rm BH}$ as  
	\begin{equation}
	M_{\rm BH} = M^*(<r_{\rm BH})
	\end{equation}
	i.e. the radius such that the black hole mass equals the mass in stars. Assuming the range of black hole masses and $M/L$ values determined in this work, we determine the $r_{\rm BH}$ is between $\sim$5 and 6\arcsec and possibly even larger. Thus the black hole sphere of influence dominates much of the field of view for both OASIS and MUSE. This makes measuring parameters such as the stellar mass to light ratio challenging as, within this field of view, there is a large degeneracy between the black hole mass and the stellar mass. We can break this degeneracy by adding larger field kinematic data that is more sensitive to the mass of the stars. We do this by including SAURON data \citep{emsellem2004sauron} for M87 as reanalysed for the ATLAS$^{\rm 3D}$ project\footnote{Available from \url{https://purl.org/atlas3d}} by \cite{cappellari2011atlasone}. Here the spaxels were binned to a target signal to noise ratio of 40 (as opposed to 60 in the original SAURON reduction). The gas was masked during the fits and the degree of additive Legendre polynomials was set to 4. Similar to this work, the stellar kinematics were fit using the MILES stellar library with a single optimal linear combination of templates.
	
	When considering large field data, it is important to be conscious of the fact that any subsequent parameter studies will be heavily influenced by information provided at larger radii because there is a greater volume of data points at large radius than at small radius. To account for this we only consider SAURON data out to a radius of 15\arcsec. To avoid accounting for uncertainties in the SAURON PSF, we do not include any SAURON data within the innermost 2\arcsec of the galaxy. We observe an offset between our extracted dispersions and the SAURON dispersion. This is expected due to systematic uncertainties in the kinematic extraction, especially due to template mismatch. To account for this, we apply a multiplicative scale to the MUSE/OASIS dispersions so that they match the SAURON data between 2\arcsec \  and 4\arcsec. Scaling the velocity axis by a factor $\Upsilon$ is equivalent to changing the overall mass normalization of the model by a factor $\sqrt{\Upsilon}$. For MUSE, the factor required to scale the dispersion to match the SAURON data varies between 0.95 and 1 (given the choice of Legendre polynomial). For OASIS, the range is between 0.87 and 0.9. If one wanted to scale each data set to either the OASIS or MUSE data, this would correspond to an increase in the black hole mass of between $\sim 5-6$ per cent for OASIS and up to $\sim 2-3$ per cent for MUSE.
	
	Given the large number of dispersion profiles generated, we have to make a choice of which ones to study. We rule out using the degree 2 and 3 RNI spectra as we expect that using such a high degree of Legendre polynomial over such a small wavelength range will lead to an unreliable kinematic extraction (indeed, for both MUSE and OASIS, we see the dispersion profile either completely flattening out or decreasing in the centre). For the degree 4,5,6 polynomials note that the degree 5 polynomial for MUSE closely resembles the degree 6, and for OASIS the degree 5 dispersion profile closely resembles the degree 4 dispersion profile. As such, we choose to throw out the degree 5 profile from each data set and are left with 6 dispersion profiles as shown in the bottom panel of \autoref{fig:all_kin}. In the final analysis, we combine different data sets in order to draw a reliable result. We choose the combinations:
	\begin{enumerate}
		\item  SAURON + OASIS degree 6 + MUSE degree 4
		\item SAURON + OASIS degree 4 + MUSE degree 6
		\item SAURON + OASIS RNI degree 1 + MUSE RNI degree 1
	\end{enumerate}
	In the latter case we refer to the kinematics as the RNI spectra.

	\subsection{PSF from AGN Spectrum}
	The PSF from IFS data has often been determined by comparing convolved HST photometry to the flux from the IFS cube (e.g. \citealp{mcdermid2006sauron,krajnovic2018quartet}, fig.B1). This method is most reliable when the centre of the galaxy has a cusp since the observed profile will be more strongly affected by the PSF. Since the nuclear region of M87 has a core, we expect this method to produce a less reliable PSF. However, we can circumvent this by noting that M87 has a bright AGN that can be assumed to be unresolved and can be used as a point source to infer the PSF directly. The profile of the PSF can thus be measured if we can extract out the component of each spectra due to the AGN. This can be done using the additive polynomials in our spectral fitting. The AGN likely has a flat non-thermal continuum that can be well approximated by polynomials while the underlying galaxy does not.

\begin{figure}
		\centering
		\subfloat{\includegraphics[width = 2.8in]{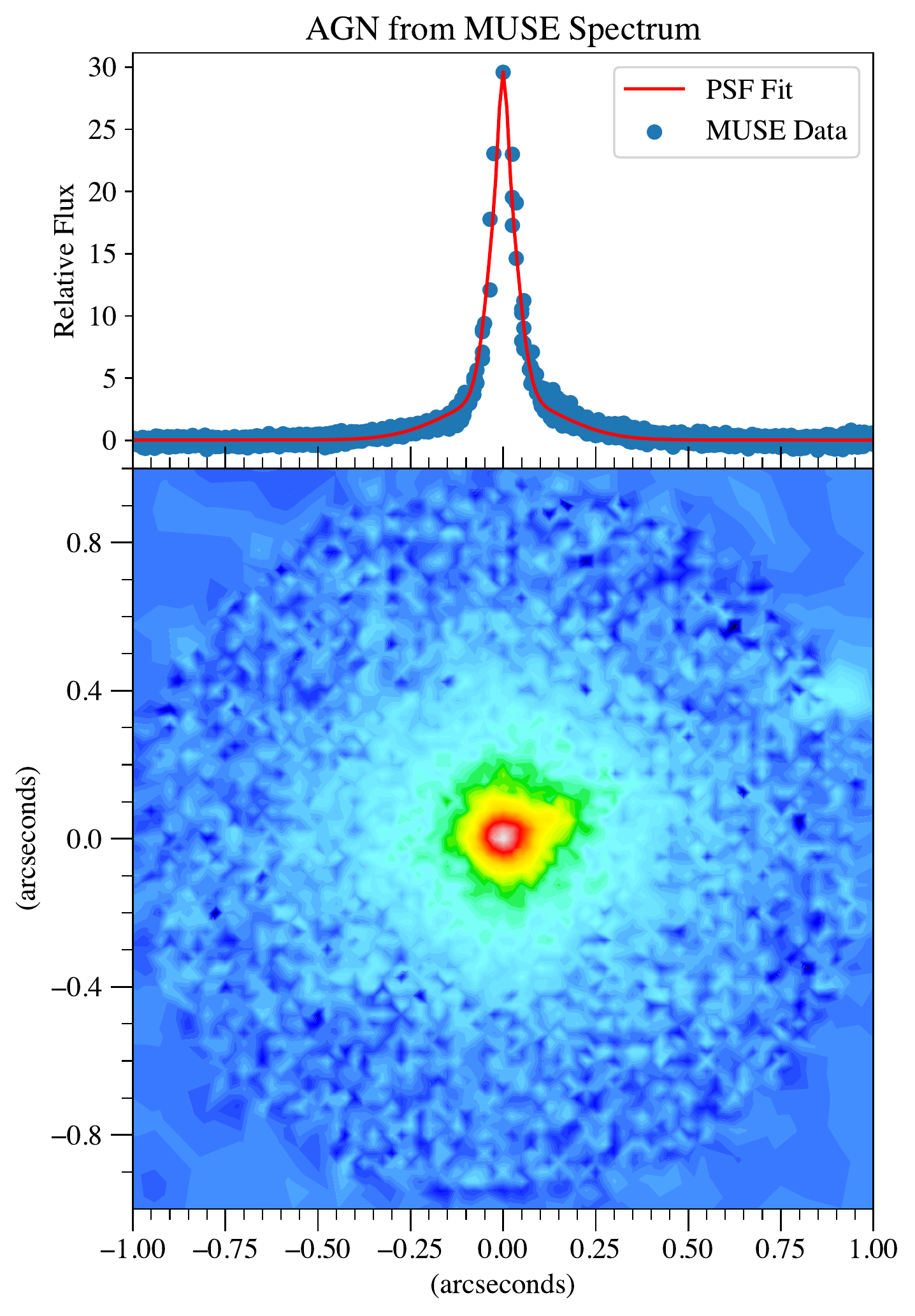}}\\
		\vspace{-15pt}
		\subfloat{\includegraphics[width = 2.8in]{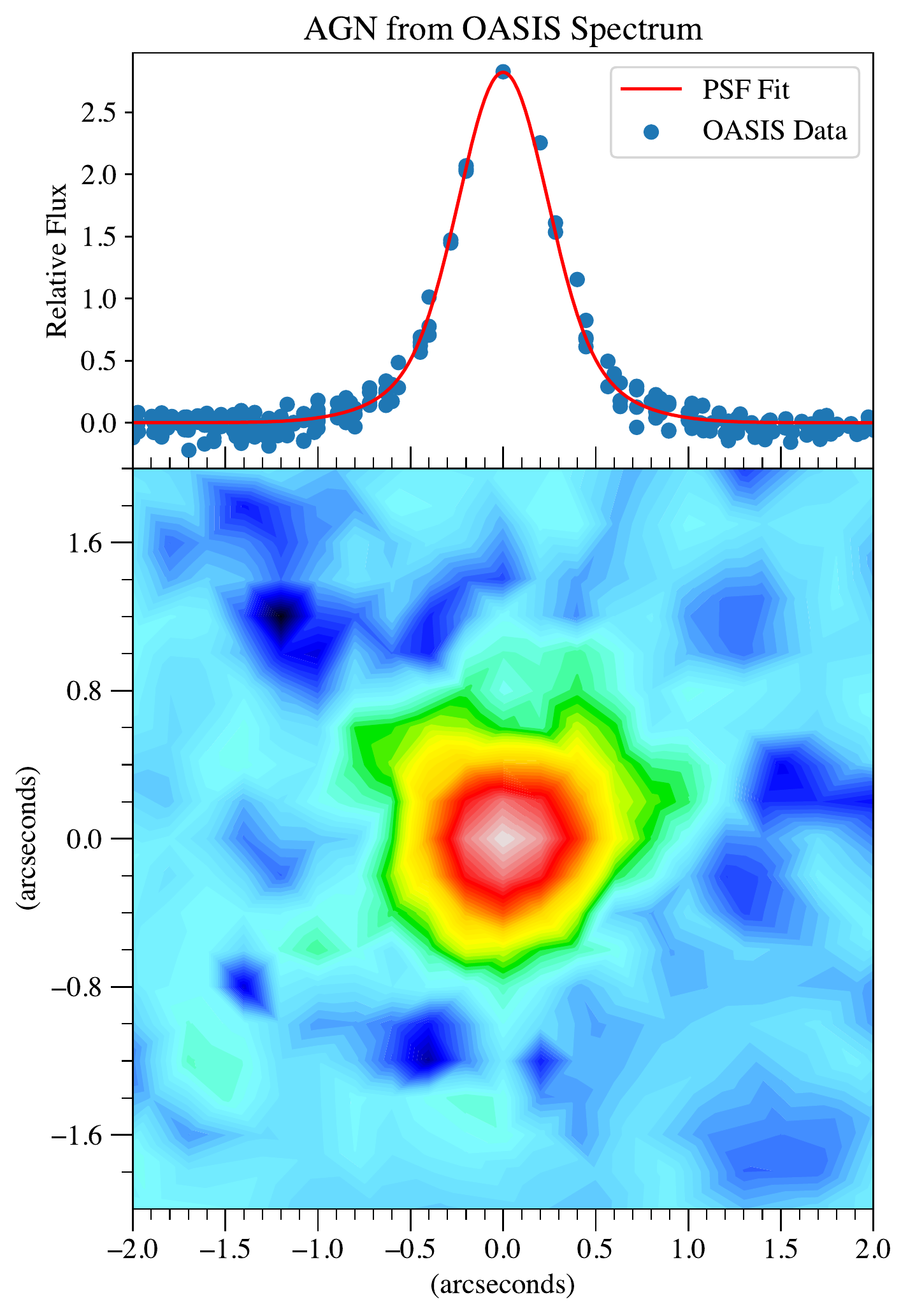}}
		\caption{The top panel shows a plot of the spectroscopic data of the AGN with the best fit multi-gaussian expansion of the PSF after being integrated over the MUSE/OASIS lenslet. This is presented as a one dimensional function where the x-axis is the radius of each data point, with those lying to the left of the origin having $x<0$ and those lying to the right having $x>0$. The colormap shows the 2D image of the spectrally extracted AGN on a log scale. The bottom panel shows this for OASIS. These results show the reliability of using the AGN contribution to the spectra as measurement of the PSF. }
		\label{fig:PSF_slice}
	\end{figure}

	We extract the shape of the PSF by performing a \textsc{pPXF} fit to the unbinned central spaxels in each dataset over the largest possible wavelength range (4800-5700\AA \ for MUSE and for 4760-5558\AA \ OASIS) using the fixed stellar template determined before from the gas-free and AGN-free spectra with multiple gas components and additive Legendre polynomials. We also tried restricting the wavelength range to that of the RNI data but found that we are unable to obtain reliable results due to the spectral range lacking distinct features in the center to anchor the total weight of the stars compared to the AGN. Additionally, we experimented with fixing the kinematics though we found that this did not impact the results. We allow the degree of the additive Legendre polynomials to vary from 1 to 6. The FWHM of the PSF varies between 0\farcs042 to 0\farcs061 for MUSE. For OASIS this range is 0\farcs527 to 0\farcs586. In this case we do not mask any of the central spaxels. We also test the effects of masking the jet but find only small diferences to the extracted FWHM. We adopt the PSF in the degree 4 case for both the MUSE and OASIS data. It is worth noting that, in the case of MUSE, we do not expect the uncertainty in the PSF to impact the our final results since we mask an inner region that is about the same size as the PSF. In \autoref{fig:PSF_slice}, we show the results for MUSE and OASIS using degree 4 Legendre polynomials, as well as a plot of the one dimensional profile as a function of radius to the left or right of the origin. We parametrize the PSF using a multi-gaussian expansion in the form of 
	\begin{equation}
		\rm{PSF}(R) = \sum_{i=1}^Q\frac{G_i}{2\pi \sigma_i^2}\exp\left(\frac{-R^2}{2\sigma_i^2}\right)
	\end{equation}
	with $R$ the radius, $\sigma_i$ the standard deviation of gaussian $i$, and $G_i$ the normalization of each gaussian satisfying $\sum_{i=1}^Q G_i = 1$. We model the observed AGN profile by integrating the PSF over the lenslet size of OASIS/MUSE. We do this by first noting that a PSF convolved observable $S_{\rm{obs}}(x,y)$ can be written as \cite[e.g.][appendix. D]{qian1995axisymmetric}
	\begin{equation}
		\label{eq:obs}
		S_{\rm{obs}}(x,y) = \int_{-\infty}^{\infty} \int_{-\infty}^{\infty}S(x,y)K(x-x^\prime,y-y^\prime)dxdy
	\end{equation}
	with
	\begin{equation}
		\begin{split}
			K(x,y) = \sum_{i=1}^Q \frac{G_i}{4} \left[ \rm{erf}\left( \frac{L_x/2 - x}{\sqrt{2}\sigma_i} \right) +  \rm{erf}\left( \frac{L_x/2 + x}{\sqrt{2}\sigma_i} \right) \right] \\
			\times \left[ \rm{erf}\left( \frac{L_y/2 - y}{\sqrt{2}\sigma_i} \right) +  \rm{erf}\left( \frac{L_y/2 + y}{\sqrt{2}\sigma_i} \right) \right]
		\end{split}
	\end{equation}
	Note that $K$ is the analytic expression of a gaussian integrated over a lenslet of size $L_x$ by $L_y$ centred on the point $(x,y)$. As our observable is unresolved, we can treat it as a delta function. Substituting that into \autoref{eq:obs} gives that the model of the PSF is simply $K(x,y)$, with the lenslet size substituted for that of MUSE or OASIS. 
	
	We then fit the PSF parameters by matching this model to the observed spectrally determined AGN by employing the \textsc{optimize.least\_squares} function of \textsc{SciPy} \citep{2020SciPy-NMeth}. We assume the errors are constant except in the centre, where we set them to be small so as to force a good fit at both large and small radii. We use the PSF extracted with degree 4 Legendre polynomials for the rest of the analysis. The measured parameters for each PSF is given in \autoref{tab:PSF}. The FWHM for OASIS is 0\farcs561, which is consistent with the seeing on the night of the observation \citep{mcdermid2006sauron}. The FWHM for MUSE is 0\farcs049, which is close to diffraction limited.
	
	
\begin{table}
	\centering
	\caption{Table of fitted MGE parameters for OASIS, MUSE, and the HST F850LP \textsc{TinyTim} PSF, respectively. The FWHM for OASIS is 0\farcs561, for MUSE is 0\farcs049, and for HST is 0\farcs063. The HST PSF is larger than the MUSE PSF because the wavelength observed in the HST observation is longer than that in the MUSE observation.}
	\begin{tabular}{ |c|c|c| } 
		\hline
		Number & $G_i$ & $\sigma_i$ (arcsec) \\ 
		\hline
		1 & 0.515 & 0.215 \\ 
		2 & 0.485 & 0.423 \\ 
		\hline 
		1 & 0.00769 & 0.00793 \\ 
		2 & 0.20493 & 0.03595 \\ 
		3 & 0.59847 & 0.14573 \\ 
		4 & 0.18891 & 0.79289 \\ 
		\hline
		1 &  0.34266 & 0.02454 \\
		2 & 0.36807 & 0.08022 \\
		3 & 0.13923 & 0.09772 \\
		4 & 0.04820 & 0.21524 \\
		5 & 0.10184 & 0.47409 \\
		\hline
	\end{tabular}
	\label{tab:PSF}
\end{table}

\section{Photometry and Mass Modeling}\label{sec:photom}
\subsection{Stellar Tracer Distribution}\label{sec:stellartracer}
Modelling the stellar tracer distribution of M87 is challenging due to non-stellar contributions in the photometric data, namely the AGN and jet. It is sufficient to mask the jet, but masking the AGN would mean masking the location where the supermassive black hole is. This is the most important region to model for black hole studies, implying that we must take another approach. The way that we circumvent this is by using the information from our spectral fits. Spectral fitting determines the contribution of stars, gas, and the AGN (through additive polynomials) such that one can extract out the pure stellar contribution. 

To do this, we measure the flux due to the stars in the \textsc{pPXF} fit to the MUSE data from section 3.2. We do this by subtracting from each spectra the best fit gas lines and additive Legendre polynomials (which approximate the AGN spectrum). We test this using Legendre polynomial degrees between 1 and 6. The main difficulty is in the central spaxels where some of the fits of the stellar spectrum become so diluted that they are completely degenerate with the Legendre polynomials. In order to test this, we ran one set of fits for each polynomial degree and with the kinematics fixed to the largest value in the bottom panel of \autoref{fig:all_kin}, as well as the smallest. We show the results of this in \autoref{fig:MGE_radial}. Here you can see that, while some of the choices of polynomial degree lead to decreasing stellar densities in the center of the galaxy, they approach an upper bound which we take to be the true stellar density. We ignore those profiles with decreasing inner stellar density as this behavior is not observed in other core galaxies either observationally or in simulations. We fit the profile over the region of the MUSE data using a double power law and match it to the radial surface brightness from Hubble outside of the innermost 0.5\arcsec. From this we create a modified Hubble image which has the innermost arcsecond replaced with our fitted AGN-free profile. 

We parametrize the galaxy surface brightness using the Multi-Gaussian Expansion method \citep{emsellem1994multi,cappellari2002efficient}. In order to model the full extent of the galaxy, we match an SDSS r-band mosiac of M87 to the Hubble image and fit them simultaneously using the robust MGE fitting algorithm and the \textsc{MgeFit}\footnote{Available from \url{https://pypi.org/project/mgefit/}} software package of \cite{cappellari2002efficient}. This algorithm fits the projected surface brightness using a multi-gaussian expansion of the form 
\begin{equation}
	\Sigma(x,y) = \sum_{j=1}^N I_j\exp\left[ -\frac{1}{2\sigma_j^2} \left( x_j^2 + \frac{y^2}{q_j^2} \right) \right]
\end{equation}
We mask the jet and gap between the detectors in the HST image, and we mask a prominent star in the SDSS image. We also provide \textsc{MgeFit} with the Hubble ACS/WFC PSF in order to obtain the PSF-deconvolved stellar distribution as opposed to the observed distribution. We generate this Hubble PSF using the tool \textsc{TinyTim} \citep{krist2010tiny}. We record our MGE expansion for this in \autoref{tab:PSF}. The MGE/galaxy contours are shown in \autoref{fig:MGE_contour} and the parameters we fit are shown in \autoref{tab:MGE_m87}. We express the MGE in the AB photometric system and the F850LP band of HST where we have used the absolute magnitude of the sun $M_{\odot,{\rm F850LP}}$=4.50 mag from \cite{willmer2018absolute} and the galactic extinction $A=0.029$ mag from \cite{schlafly2011measuring}. The zero point at the time of the observation was $ZP = 24.873$.

Our spectral-decomposition approach allows for a much more reliable measurement of the stellar surface brightness near the black hole than was possible before. The most recent stellar dynamical determinations of the black hole mass in M87 \citep{gebhardt2009black,gebhardt2011black,2023arXiv230207884L} have used the surface density profile provided in \cite{kormendy2009structure}. This work determines the surface density profile of M87 by combining observations across a number of different distance scales and photometric bands. The data determining the central profile of M87 comes from \cite{lauer1992m87stellar}, in which the authors fit a power law starlight model, central nonthermal point source, and optical counterparts for the jet knots in order to determine the shape of the stellar distribution. \cite{kormendy2009structure} notes that there are some gradients and discontinuities in the ellipticity and principal axis inside the core of M87 which may be due to issues with the AGN treatment in \cite{lauer1992m87stellar}. We find a much flatter core in the innermost region which, holding everything else constant, will result in a larger black hole mass. This is discussed in more detail later.

\subsection{Mass Modelling with M/L Gradients}
In order to perform dynamical modelling, one has to parametrize the gravitational potential. \cite{sarzi2018muse} used stellar population models to measure gradients in the stellar $M^*/L$ in M87 by allowing for both ages, metallicity and stellar initial mass function (IMF) variations. Although these measurements are assumption dependent and quite uncertain, they provide an estimate for a possible stellar $M^*/L$ variation within the innermost 30-40\arcsec \ of M87, implying that it is not sufficient to assume that mass follows light without testing the alternative. This is further corroborated by \cite{oldham2018m87imf} which also finds evidence for a radially decreasing $M^*/L$. We allow for variations away from mass follows light in this region by allowing for a radially dependent $M^*/L$ profile given by
\begin{equation}\label{eq:masstolight}
	\left(\frac{M^*}{L}\right)(r) =	\begin{cases}		\left(M/L\right)_1 & r<1''\\		\left[\left(\frac{M}{L}\right)_1\!\! - \left(\frac{M}{L}\right)_2\right]\frac{\lg(r/30)}{\lg(1/30)} + \left(\frac{M}{L}\right)_2  & 1'' \le r \le 30''\\		\left(M/L\right)_2 & r>30''	\end{cases}
\end{equation}
This functional form is motivated by the fact that, within the error bars, figure 11 of \cite{sarzi2018muse} is well fit by this function. The observed nearly linear variation in $M^*/L$ with $\lg(r)$ cannot represent the true $M^*/L$ variation since it is unbounded at 0 and infinity, so we set it to be constant outside of the regions constrained by the data. We implement this within the MGE formalism in the following way: we evaluate $(M^*/L)(r)$ at each $\sigma_j$ (where $\sigma_j$ is the standard deviation of each gaussian in the MGE) and multiply the surface brightness $I_{F850LP,j}$ by this value. In principle, $(M^*/L)_1$ and $(M^*/L)_2$ are free parameters, but we can determine possible upper and lower bounds by studying the effect of varying the IMF. \cite{sarzi2018muse} finds that the largest possible M/L ratio assuming a Kroupa IMF is close to 5.0 in the r-band. From figure 2 of \cite{cappellari2012imf}, we see that, empirically in the galaxy population, the largest  $M^*/L$ increase one can expect due to the IMF normalization is a factor ~2.6 heavier than the value corresponding to a Kroupa IMF. Taking the r-band $M^*/L$ with Kroupa IMF as reference, the heaviest $M^*/L$ one can realistically expect for M87, allowing for extreme IMF gradients, correspond to $M^*/L$=13 in the r-band. Lastly, we can convert this to the SDSS z-band (which we take to be approximately the same as the ACS/WFC F850LP band) using the conversion formula
\begin{equation}
	M/L_z = M/L_r \times 10^{0.4([z-r]_{\rm M87} - [z-r]_{\odot})}
\end{equation}
Using M87 colors $z=9.92$ and $r=10.70$ from SDSS DR7, and solar magnitudes $z_\odot=4.50$ and $r_\odot=4.65$ from \cite{willmer2018absolute}, we find that the upper bound is $(M^*/L)_z\la7.23$. This becomes important later in the analysis where we introduce this cut off when our modelling would otherwise prefer an unphysically large $M^*/L$ ratio.

This choice of parametrization restricts the steepest possible $M^*/L$ variation. One could further increase the freedom of this parametrization by allowing the innermost radius at which the $M^*/L$ ratio becomes constant to vary. We discuss the impact of this in \autoref{sec:disc_ml}.


\subsection{Dark Matter}
Several studies of the dark matter halo of M87 have been previously made using globular clusters \cite{murphy2011galaxy,agnello2014m87,oldham2016dm,li2020m87}. \cite{oldham2016dm} finds a preference for a somewhat cored halo, whereas \cite{murphy2011galaxy,li2020m87} does not find a preference for either a cored profile or a cuspy halo (though \cite{murphy2011galaxy} has a slight preference for a cored halo). \cite{agnello2014m87} finds a preference for a very steep cusp, but this may be due to choices in their modelling (See \cite{oldham2016dm} for an extensive discussion). Our data only covers up to 15\arcsec, which is about a fifth of a half light radius, so we cannot make any reliable inference about the dark matter contribution. We include dark matter and vary $M/L$ in our study to account for the possible range of shapes of the total density profiles, which leads to a more conservative estimate of the black hole mass. As such, we proceed with a NFW dark matter halo \citep{navarro1997universal}
\begin{equation}
	\rho(r) = \frac{\rho_0}{(r/r_s)(1+r/r_s)^2}
\end{equation}
The range of our data is much less than $r_s$ so we set it arbitrarily to 20kpc because its precise value is irrelevant for the modelling results. We parametrize the overall magnitude as the fraction of matter within a sphere of one half light radius consisting of dark matter, $f_{\rm dm}$. We calculate the enclosed masses from the MGE analytically using the routine \textsc{mge\_radial\_mass} from \textsc{Jampy}\footnote{Available from \url{https://pypi.org/project/jampy/}}. We calculate the half light radius from the MGE of \autoref{tab:MGE_m87} using the routine \textsc{mge\_half\_light\_radius} in \textsc{Jampy} and find a value of 70\arcsec.

\begin{figure}
	\includegraphics[width=\columnwidth]{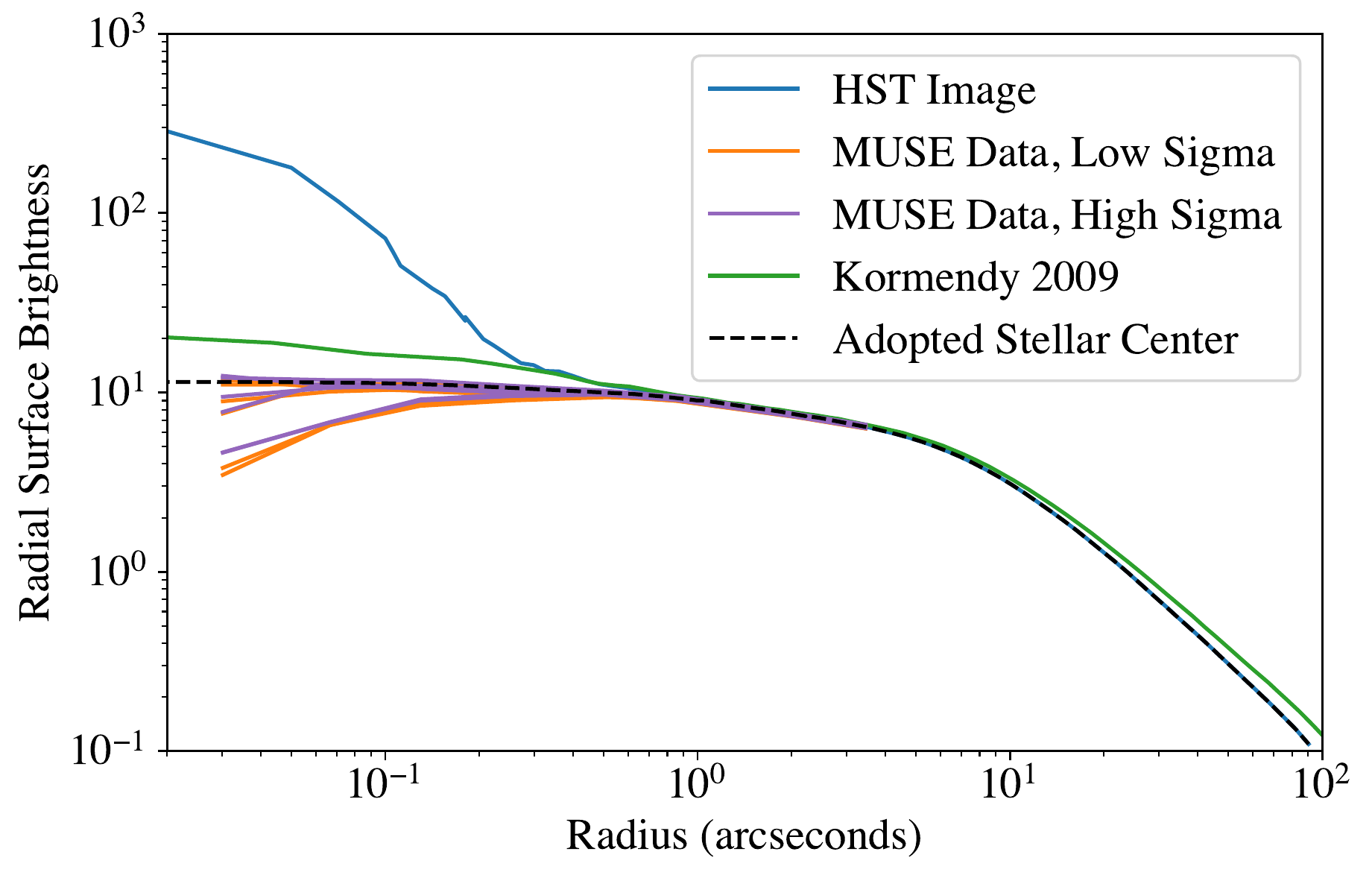}
	\caption{Comparison of a scaled HST radial profile with a profile where the centre region is set by the stellar profile measured spectroscopically from MUSE.  The purple and orange curves show the spectroscopically extracted stellar profile from the MUSE data under different assumptions. The orange curve fixes the stellar kinematics to the lowest values seen in the bottom panel of \autoref{fig:all_kin}, and similarly for the purple curve with the largest values. The different lines of the same color show different choices of Legendre polynomial degree. The black dashed curve shows the profile we adopt. We also compare with the profile used in the previous stellar dynamical determinations of \protect\cite{gebhardt2009black,gebhardt2011black,2023arXiv230207884L}. Their profile is a factor of 2 larger than ours in the centre.}
	\label{fig:MGE_radial}
\end{figure}

\begin{figure}
	\includegraphics[width=\columnwidth]{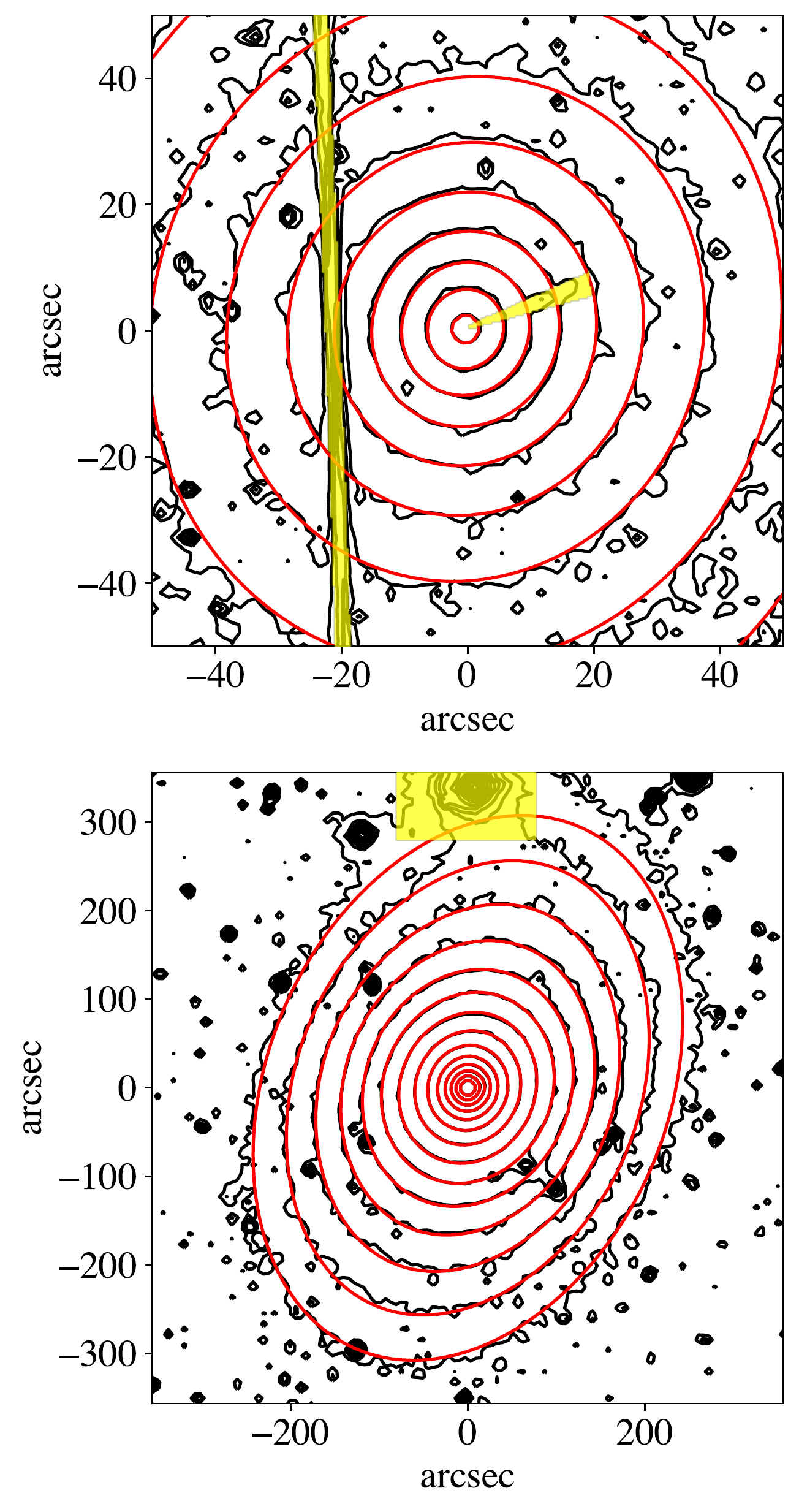}
	\caption{Top panel: best fit MGE contours in the innermost 100 arcseconds of the HST image. The regions in yellow are the masked jet and gap between the detectors. Bottom panel: best fit MGE contours over the full SDSS field. We mask a prominent star at the top of the field. You can see that the MGE fit covers the full shape of the galaxy and provides an excellent fit in the innermost region where our data is the most sensitive. The galaxy is oriented so that north is at the top and east is to the left.}
	\label{fig:MGE_contour}
\end{figure}

\begin{table}
	\centering
	\caption{MGE parameters for the deconvolved ACS/WFC F850LP surface brightness. This corresponds to the orange profile in \autoref{fig:MGE_radial} and the red contours in \autoref{fig:MGE_contour}. The principal photometric axis measured is 156.7 degrees east of north.}
	\begin{tabular}{ |c|c|c|c| } 
		\hline
		$j$ & $\lg(I_{j})$ & $\lg(\sigma_j)$ & $q_j$ \\ 
		& ($L_{\odot} \ $pc$^{-2}$) & (arcsec) & \\
		\hline
		1 & 3.070 & -0.834 & 0.957 \\ 
		2 & 3.372 & -0.147 & 0.957 \\ 
		3 & 3.354 & 0.246 & 0.957 \\ 
		4 & 3.366 & 0.624 & 0.957 \\ 
		5 & 3.435 & 0.820 & 0.957 \\ 
		6 & 3.310 & 1.056 & 0.957 \\ 
		7 & 3.001 & 1.314 & 0.957 \\ 
		8 & 2.389 & 1.583 & 0.847 \\ 
		9 & 2.380 & 1.745 & 0.936 \\ 
		10 & 1.894 & 2.018 & 0.743 \\ 
		11 & 1.488 & 2.317 & 0.743 \\
		\hline
	\end{tabular}
	\label{tab:MGE_m87}
\end{table}

\section{Dynamical Modeling}\label{sec:dynmod}
\subsection{Jeans Modelling}
The Jeans Anisotropic Modelling (JAM) method \citep{cappellari2008measuring,cappellari2020efficient} has been used to model the stellar dynamics of galaxies and study their stellar mass-to-light ratios and dark matter content in large integral-field spectroscopic surveys such as ATLAS$^{\rm 3D}$ \citep{cappellari2013atlas3d}, SAMI \citep{scott2015sami}, MaNGA \citep{li2018sdss} as well as surveys at high redshift e.g. for the LEGA-C survey \citep{van2021stellar}. It was applied to the study of galaxies' total density profiles out to large radii \citep{cappellari2015small} and in several studies of smaller galaxy samples. More recently, JAM was employed to accurately predict Gaia kinematics of the Milky Way using all six-dimensional components of the stellar phase space \citep{nitschai2020first, nitschai2021dynamical}.

Tests of JAM using high-resolution N-body simulations \citep{lablanche2012atlas3d} and lower-resolution but more extensive cosmological hydrodynamical simulations \citep{li2016assessing} have shown that, with high-S/N data, JAM recovers accurate total density profiles with negligible bias. More recently, JAM was compared in detail against the \cite{schwarzschild1979numerical} method using samples of both observed galaxies, with circular velocities from interferometric observations of the CO gas, and numerical simulations respectively. Both studies consistently found that the JAM method produces even more accurate (smaller scatter vs the true values) density profiles \citep[fig.~8]{leung2018edge} and enclosed masses \citep[fig.~4]{jin2019evaluating} than the more general Schwarzschild models. More quantitatively, between 0.8-1.6 effective radii, where the gas is well-resolved and the $V_{\rm c}$ is better determined, \cite{leung2018edge} reports a mean 1 $\sigma$ error 1.7$\times$ smaller for JAM over the equivalent Schwarzschild model. Similarly, when considering all 45 model fits to the N-body simulations by \cite{jin2019evaluating}, the 68th percentile deviation (1 $\sigma$ error) is 1.6$\times$ smaller for JAM than the equivalent Schwarzschild model. The increased accuracy of JAM in extracting density distributions may be due to JAM assumptions acting as an empirically-motivated prior and reducing the degeneracies of the dynamical inversion. 

For supermassive black hole studies, JAM was found to accurately recover the “known” mass of the two most accurate benchmark black holes in NGC4258 \citep{drehmer2015benchmark} and the Milky Way \citep[sec. 4.1.2]{feldmeier2017triaxial}. Moreover, extensive tests of a few tens of galaxies have found that JAM and Schwarzschild methods recover black hole masses that are generally consistent with one another \citep{cappellari2010testing,seth2014supermassive,thater2017low,thater2022cross,krajnovic2018quartet}. However, not all BH measurements from different methods agree within the uncertainties and further comparisons between different approaches are still needed.

There are two implementations of JAM: one where the velocity ellipsoid is assumed to be aligned with the cylindrical-polar coordinate system \citep{cappellari2008measuring}, and one where the velocity ellipsoid is assumed to be aligned with the spherical-polar coordinate system \citep{cappellari2020efficient}. The choice of which implementation to use depends on the galaxy's intrinsic shape. M87 is a slow rotator early type galaxy \citep{emsellem2011census} and slow rotators as a class are weakly triaxial, or nearly spherical, inside the half-light radius, becoming more triaxial at larger radii (see review by \cite{cappellari2016structure}). M87 has a specific angular momentum $\lambda_{\rm R_{\rm e}}\approx0$ within the uncertainties \citep{emsellem2011census}. It showed some barely detectable misaligned stellar rotation from SAURON data \citep{emsellem2004sauron}, which became more clearly visible from high-S/N MUSE data \citep{emsellem2014kinematically}. The ellipticity of M87 increases at large radii (see \autoref{fig:MGE_contour}), where misaligned stellar rotation and triaxiality become evident \citep{2023arXiv230207884L}. Within the slow-rotators class, M87 was classified as a non-rotator \citep{krajnovic2011atlas3d}. As a class, these are massive early type galaxies generally found at the centre of clusters, which tend to be rounder than $\epsilon\la0.15$ in projection \cite[fig.6]{emsellem2011census}, indicating that, although triaxial, they must be intrinsically close to spherical with a ratio between the minor to major axes of the ellipsoidal density $c/a\ga0.85$. This excludes the possibility that M87, which is nearly round in projection, may appear as such due to a special viewing angle. Instead, M87 must be intrinsically close to spherical in the region where it shows circular isophotes.

JAM models, unlike the Schwarzschild models, do not require large-radii kinematics to constrain the models because they are nearly insensitive to the mass distribution at radii outside the region where kinematics are available. For this reason, one can expect axisymmetric JAM models with the velocity ellipsoid aligned with spherical-polar coordinates, to provide an accurate description of the inner dynamics of M87, even though the galaxy, like all slow rotators, becomes more strongly triaxial at much larger radii than those we model.


\subsection{Priors on the Anisotropy}\label{sec:param}
JAM takes as parameters the galaxy inclination, anisotropy profile, stellar tracer distribution, and mass distribution, including the black hole mass. The kinematic axis is poorly constrained due to M87 being dominated by unordered motion in the central region, \citep{emsellem2014kinematically,sarzi2018muse} so we fix the kinematic axis to align with the galaxy photometric axis as given in \cite{krajnovic2011atlas3d}. The value reported there is 151.3 degrees east or north, which is consistent with our value of 156.7 degrees east of north. The stellar tracer and mass distributions are described in \autoref{sec:stellartracer}. As the inner regions of M87 are nearly spherically symmetric with little ordered motion, we fix the inclination to 90 degrees. Changing this has a negligible effect on the measured black hole masses, because a nearly spherical model appears spherical from any inclination. We can exclude M87 being an intrinsically flat but nearly face-on disk, because of the general shape distribution of slow rotators \citep{cappellari2016structure,2018ApJ...863L..19L}.

The last thing to specify is the anisotropy profile. It is well known that for a spherical system, there is a degeneracy between the anisotropy and the density profile. This so-called mass-anisotropy degeneracy implies that for a range of assumed density profiles, one can adopt a corresponding anisotropy profile in such a way that the model reproduces the same profile of second velocity moments \citep{binney1982m,gerhard1993line}. The degeneracy, however, is not complete, and the range of allowed profiles depends on the specific situation because the anisotropy is limited by the two extreme cases where the orbital distribution is fully radial or fully tangential respectively. For this reason, without further assumptions, one would generally expect large uncertainties in BH masses from spherical models based on the Jeans equations. 

The situation, however, has improved dramatically from the days when the mass-anisotropy degeneracy was first discovered. Since then, many studies have modelled the inner dynamics of galaxies using general models that allow one to account for the full shape of the line-of-sight velocity distribution, rather than the moments alone. We think we now even have a good understanding of the underlying physics of the orbital distributions we have measured. In particular, we have found that massive slow-rotator galaxies with a core in their surface brightness profile, like M87, are consistently characterized by a nearly isotropic, or just slightly radially anisotropic orbital distribution outside the break radius, while orbits start becoming tangentially biased inside that radius, reaching the peak tangential anisotropy well inside the BH sphere of influence (\citealp{gebhardt2003axisymmetric} fig.10; \citealp{cappellari2008supermassive} fig.2; \citealp{thomas2014dynamical}). 
The observations are quantitatively well reproduced by models in which both the cores in the surface brightness and the tangentially biased orbits are due to gas-free mergers of galaxies with supermassive black holes in their centres. The black holes sink towards the centre of the gravitational potential via dynamical friction, while ejecting stars on radial orbits (e.g. \citealp{milosavljevic2001formation,milosavljevic2002galaxy,rantala2019simultaneous,frigo2021two}). 

In the case of M87, due to its very flat inner core, one can place constraints on its orbital anisotropy even from theoretical arguments alone. The cusp-slope vs central anisotropy theorem by \cite{an2006cusp} states that for a spherical power-law tracer population $\rho\propto r^{-\gamma}$ in a Keplerian potential, there is a relation between the anisotropy $\beta=1-\sigma_t^2/\sigma_r^2$ with $\sigma_t$ and $\sigma_r$ the tangential and radial dispersion, respectively, and the logarithmic slope $\gamma$ of the tracer, such that $\beta < \gamma - 1/2$. The inner slope of M87 varies from nearly flat in the centre ($\gamma\approx 0$ see \autoref{fig:MGE_radial}) to $\gamma\approx0.27$ between 1-5\arcsec \citep{lauer2007centers}. We can thus conservatively conclude that the inner anisotropy has an upper limit of $\sigma_r/\sigma_t \la 0.9$ and possibly even $\sigma_r/\sigma_t \la 0.8$ for $\gamma = 0$. The assumptions of the theorem are satisfied well inside the sphere of influence of the BH of M87 and for this reason the theorem provides additional support for the expected significant tangential anisotropy near the BH of M87.

For all these theoretical and empirical reasons, nowadays, it does not make sense to assume complete freedom in the orbital anisotropy of Jeans models as done in the past. Instead, the knowledge we accumulated on the galaxies anisotropy can be used as a Bayesian prior, which is easy to enforce to our JAM models. The ability to place priors is an important feature of stellar dynamical codes. All Schwarzschild codes use regularization to enforce smoothness in the orbital distribution (e.g. \cite{richstone1988maximum,vandermarel1998axisymmetric,gebhardt2000axisymmetric,cappellari2002counterrotating,valluri2004difficulties,thomas2005regularized}). A recent study suggested that the fine tuning of regularization is essential for accurate results \citep{neureiter2022accuracy}.This regularisation is mathematically equivalent to a prior.

In the next section we describe a new way of specifying the anisotropy variations in JAM models, which is ideally suited to enforce anisotropy priors.	

\begin{figure*}
	\includegraphics[width=\textwidth]{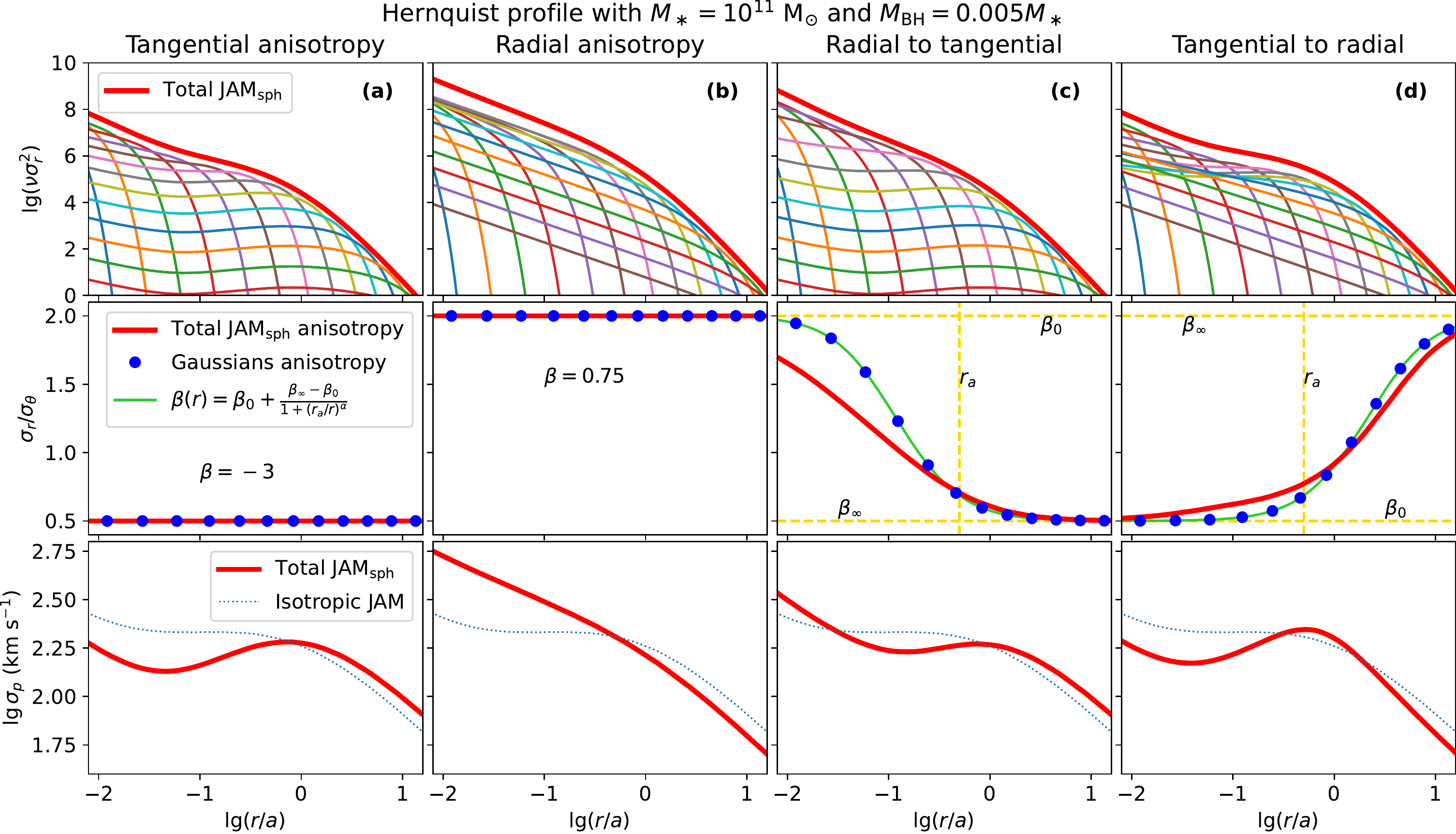}
	\caption{Top row: the contributions to the total luminosity weighted second moment from different MGE components. Middle row: intrinsic anisotropy ratio compared with the anisotropy assigned to each gaussian. Bottom row: projected velocity dispersion. The columns correspond to different choices of the total anisotropy, ranging from constant to radially varying.}
	\label{fig:jam_approximate_anisotropy_profiles}
\end{figure*}

\begin{figure}
	\centering
	\includegraphics[width=0.8\columnwidth]{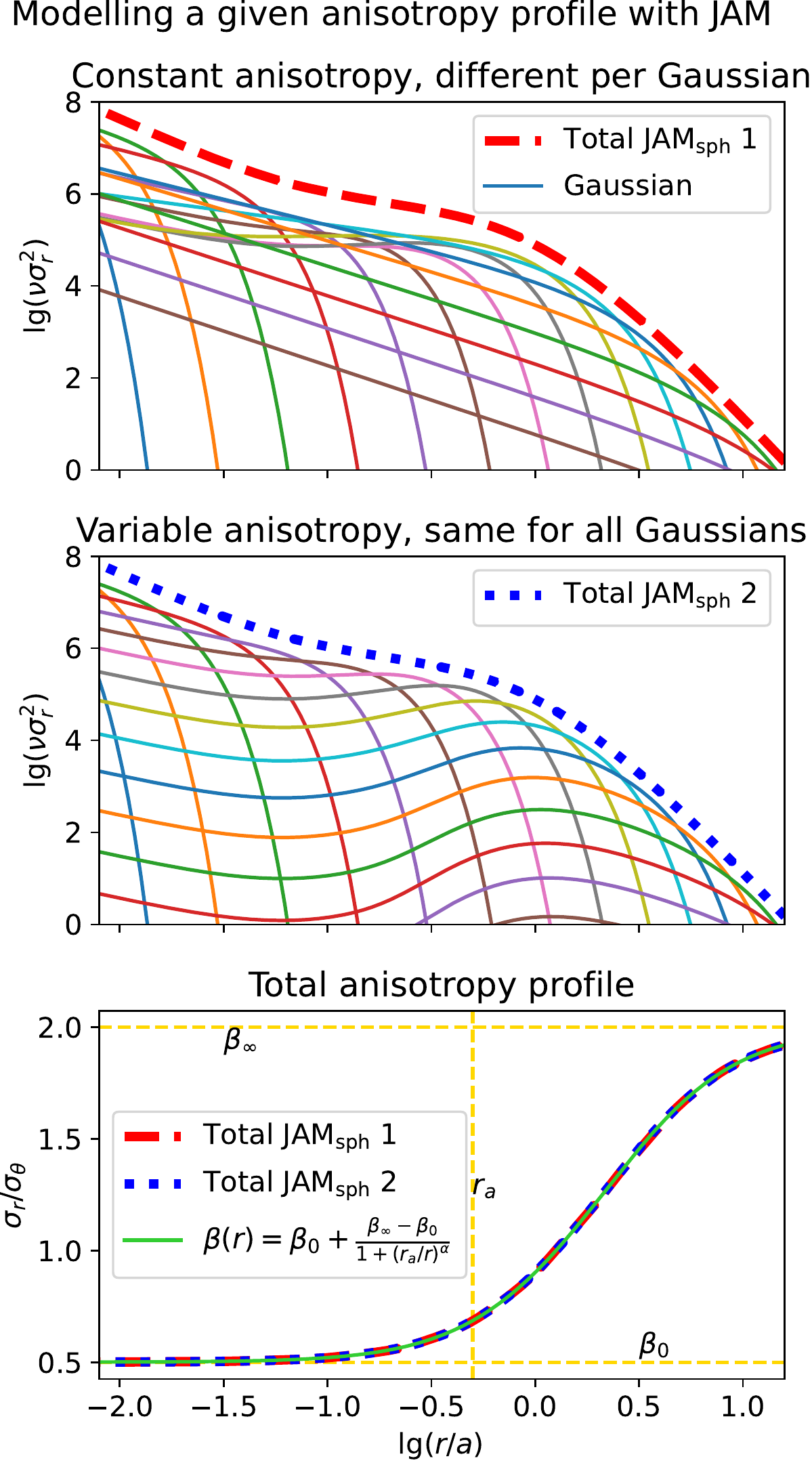}
	\caption{The top panel shows the contributions to the radial velocity dispersion where the desired anisotropy profile is reached by fitting the anisotropy for each gaussian in the MGE. The center panel shows the same thing using the analytic implementation described in equations 12-14. The bottom panel shows the intrinsic anisotropy ratio and confirms that the result using the two methods is the same, the key difference being that the top one requires fitting the anisotropy for each gaussian while the center one requires a small modification to the analytic solution.}
	\label{fig:jam_precise_anisotropy_profiles}
\end{figure}

\subsection{Fitting a given anisotropy profile with JAM}\label{sec:intro}

Let's consider for simplicity a spherical non-rotating JAM model. The intrinsic stellar dispersion of the model is given by the luminosity-weighted sum of the dispersion of the individual Gaussians making up the MGE
\begin{equation}
	\nu\sigma_r^2 = \sum_k [\nu\sigma_r^2]_k.
\end{equation}
Where $\nu$ is the deprojected MGE oblate axisymmetric luminous density (see equation 13 of \cite{cappellari2008measuring}). \autoref{fig:jam_approximate_anisotropy_profiles} shows the contribution of the individual $[\nu\sigma_r^2]_k$ for different anisotropies, for a \cite{hernquist1990analytical} model with mass $M_\ast=10^{11}$ M$_\odot$ containing a typical nuclear supermassive black hole of mass 0.5\% that of the stellar mass \citep[eq.~11]{kormendy2013coevolution}. When the Gaussians in a JAM model are isotropic or have tangential anisotropy, each Gaussian essentially contributes to the total $\nu\sigma_r^2$ only near a radius close to its dispersion $r\approx\sigma_k$ (\autoref{fig:jam_approximate_anisotropy_profiles}a). In these cases, one can construct a desired anisotropy profile $\beta(r)$ by simply assigning the anisotropy $\beta_k=\beta(\sigma_k)$ to the Gaussians with dispersion $\sigma_k$, as pointed out in \citet[sec.~3.2.2]{cappellari2008measuring}. \autoref{fig:jam_approximate_anisotropy_profiles} shows that this approximation works quite well in general. However, when the Gaussians are significantly tangential anisotropic, or near the central supermassive black hole, the total $\nu\sigma_r^2$ rises steeply at small radii and a single Gaussian does not contribute to the total JAM model only around $r\approx\sigma_k$ (\autoref{fig:jam_approximate_anisotropy_profiles}c,d). In those situations, there is no simple precise relation between the anisotropy of a given Gaussian and the total anisotropy of the JAM model at $r\approx\sigma_k$. This is generally not a problem, if one is not interested in the fitted anisotropy. However, if one wants to quantitatively reproduce a specific total anisotropy profile one has to numerically fit for the anisotropies $\beta_k$ of the different Gaussians. 

In this paper we parametrize the anisotropy using a rather flexible logistic function of logarithmic radius
\begin{equation}\label{eq:logistic}
	\beta(r)=\beta_0+\frac{\Delta\beta}{1+(r_a/r)^\alpha},
\end{equation}
with $\Delta\beta=\beta_\infty-\beta_0$. This anisotropy function was also used by \citet[eq.~30]{baes2007dynamical}. For $\beta_0=0$, $\beta_\infty=1$ and $\alpha=2$ it reduces to the Osipkov-Merritt special form \citep{osipkov1979spherical,merritt1985distribution}. For $\alpha=1$ it specializes to the homographic anisotropy function used by \cite{bacon1985anisotropic}.

In the top panel of \autoref{fig:jam_precise_anisotropy_profiles} we show how one can reproduce our logistic anisotropy profile with JAM. In the figure, we adopted a rather extreme anisotropy variation, with inner tangential anisotropy $\sigma_r/\sigma_\theta=1/2$ ($\beta_0=-3$), outer radial anisotropy $\sigma_r/\sigma_\theta=2$ ($\beta_\infty=0.75$), anisotropy radius $r_a=a/2$, where $a$ is the break radius of the \citet{hernquist1990analytical} profile, and the sharpness of the transition is $\alpha=1.5$. One can see that by fine-tuning the anisotropy of the different Gaussians, one can reproduce quite general anisotropy variations. The problem with this approach is that the anisotropy $\beta_k$ of the individual Gaussians has to be fitted non-linearly in a least-square sense, while repeatedly computing the intrinsic velocity moments of the model, to obtain the total anisotropy for a given choice of $\beta_k$ parameters.

An alternative is to solve the original Jeans equations for an axisymmetric model with spherically-aligned velocity ellipsoid, in \citet[eq.~8]{cappellari2020efficient} by relaxing the assumption of a constant anisotropy per Gaussian. This is only possible analytically for special choices of the anisotropy function $\beta(r,\theta)$. We found that, adopting the parametrization of \autoref{eq:logistic} for the anisotropy, the solution of \citet[eq.~10]{cappellari2020efficient}, using the same notations, generalizes to
\begin{subequations}\label{eq:sig_r}
	\begin{align}
		&\nu\overline{v_r^2}(r,\theta) = \int_r^\infty \left(\frac{r'}{r}\right)^{2\beta_0}
		\left[\frac{1 + (r'/r_a)^\alpha}{1 + (r/r_a)^\alpha}\right]^\frac{2\Delta\beta}{\alpha} \Psi(r',\theta')\; \dd r'\\
		&\theta'=\arcsin\left\{\left(\frac{r'}{r}\right)^{\beta_0-1}\left[\frac{1 + (r'/r_a)^\alpha}{1 + (r/r_a)^\alpha}\right]^\frac{\Delta\beta}{\alpha}\!\!\sin\theta\right\}.
	\end{align}
\end{subequations}
In the special case $\alpha=1$, this solution reduces the one using homographic functions in \citet[eq.~3]{bacon1985anisotropic}\footnote{The published expression has a typo, with an extra density $\nu(\rho,\alpha)$.}. 

The application of a generic varying anisotropy function to the axisymmetric cylindrically-aligned Jeans solution of \citet[eqs.~8, 9]{cappellari2008measuring}, is straightforward if one make the cylindrical anisotropy a function $\beta_z(|z|)$ of the modulus of the cylindrical coordinate $z$. One simply has to replace the constant axial anisotropy $\beta_z$, with the corresponding varying expressions $\beta_z(z)$ parametrized by the function of \autoref{eq:logistic}. In the case of the tangential anisotropy $\gamma$, the situation is identical regardless of the alignment of the velocity ellipsoid, and one can just replace the constant with an arbitrary function of the coordinates $\gamma(R,z)$, without having to change anything else.

When applying the axisymmetric solution to a model with both the tracer distribution and the total density described by an MGE \citep{emsellem1994multi,cappellari2002efficient}, following the steps outlined in \citet[sec.~5.1]{cappellari2020efficient}, the resulting solution requires only minimal changes and the numerical algorithm can be left unchanged. One only needs to replace the following two expressions with the corresponding one indicated by the arrows, in all the expressions of \citet[eqs.~46--53]{cappellari2020efficient}
\begin{align}
	&\beta_k\rightarrow\beta_0+\frac{\Delta\beta}{1+(r_a/r)^\alpha}\\
	&\left(\frac{r'}{r}\right)^{2\beta_k}\rightarrow
	\left(\frac{r'}{r}\right)^{2\beta_0}
	\left[\frac{1 + (r'/r_a)^\alpha}{1 + (r/r_a)^\alpha}\right]^\frac{2\Delta\beta}{\alpha}.
\end{align}

In the spherical limit, the expression for the intrinsic radial velocity dispersion in \citet[eq.~B2]{cappellari2020efficient} generalizes to
\begin{equation}
	\nu\overline{v_r^2}(r) = \int_r^\infty \left(\frac{r'}{r}\right)^{2\beta_0}
	\left[\frac{1 + (r'/r_a)^\alpha}{1 + (r/r_a)^\alpha}\right]^\frac{2\Delta\beta}{\alpha} 
	\frac{\nu(r')M(r')}{r'^2}\; \dd r'\\.
\end{equation}
Unlike the constant-anisotropy case, when projecting this model along the line-of-sight one cannot remove one of the two resulting integrals, except for some special cases \citep{mamon2005dark}, which are not very useful for practical applications.

We implemented these changes in v7.0 of the Python \textsc{JamPy} package\footnote{\url{https://pypi.org/project/jampy}}, which now allows one to compute axisymmetric or spherical models with the logistic radial anisotropy variation, for both the spherically-aligned and cylindrically-aligned solutions. An application in the spherical limit is shown in the middle panel of \autoref{fig:jam_precise_anisotropy_profiles}. As expected the JAM model with the same variable-anisotropy for all Gaussians produces the same dispersion profile as the model with constant anisotropy for each individual Gaussian, as they both follow by design the same given anisotropy profile.

\subsection{MCMC Analysis}
For each of the models and dispersion profiles we perform an MCMC analysis to carefully assess the influence of the modelling and systematic effects of the kinematic extraction on the recovered BH mass. We define the $\chi^2$ to be
\begin{equation}
	\chi^2 = \sum_{\ell = 1}^{N} \left(\frac{\sigma^\ell_m - \sigma^\ell_d }{\Delta \sigma^\ell}\right)^2
\end{equation}
where $\sigma_d^\ell$ is the extracted dispersion in binned spaxel $\ell$, $\sigma_m^\ell$ is the model dispersion, and $\Delta \sigma^\ell$ is the bootstrapped uncertainty in the extracted dispersion. We thus express the $\chi^2$ for the combined data sets as
\begin{equation}
	\chi^2 = \chi^2_{\rm MUSE} + \chi^2_{\rm OASIS} + \chi^2_{\rm SAURON}
\end{equation}

From here, we perform an MCMC analysis using the code \textsc{emcee} of \cite{foreman2013emcee}. For each of the three combinations of different kinematic extractions described in \autoref{sec:specfit}, we test four models, for a total of 12 different combinations of data and models. In each model we adopt as free parameters the four anisotropy parameters $(\beta_0, \beta_\infty, r_a, \alpha)$ described in \autoref{sec:param}, the black hole mass $M_{\rm BH}$, and make the following assumptions:
\begin{itemize}
	\item Constant stellar $M/L$ without NFW dark matter. This adds an extra parameter $(M/L)_{\rm tot}$ for a total of 6 model parameters.
	\item Constant stellar $M/L$ with NFW dark matter. This add two extra free parameters, the stellar $M*/L$ and the normalization of the halo, quantified by the dark matter fraction within one effective radius $f_{\rm dm}(<R_{\rm e})$, for a total of 7 model parameters.
	\item Varying $M/L$ without NFW dark matter. This adds two extra free parameters: $(M/L)_1$ and $(M/L)_2$ which parametrize the mass-to-light variation (see \autoref{eq:masstolight}), for a total of 7 parameters. 
	\item Varying M/L with NFW dark matter. This combines both a NFW dark matter halo, parametrized by the dark matter fraction $f_{\rm dm}$, along with the parameters $(M/L)_1$ and $(M/L)_2$ which parametrize the mass-to-light variation, for a total of 8 parameters.
\end{itemize}
These varying assumptions allow us to make contact with previous work which have made various assumptions on the parametrization of the gravitational potential, and also allow us to test the impact of each model assumption on the final result.

\begin{table}
	\centering
	\caption{Table of free parameters and their permitted upper and lower bounds. The limits on $(\sigma_r/\sigma_t)_0$ and  $(\sigma_r/\sigma_t)_\infty$ come from the large literature of observations and simulations of core galaxies. The limits on $M/L$ come from considering the heaviest possible IMF combined with the results from \protect\cite{sarzi2018muse}}
	\begin{tabular}{ |c|c|c| } 
		\hline
		Parameter & Lower Bound & Upper Bound  \\ \hline
		$(\sigma_r/\sigma_t)_0$ & 0.5 & 1  \\ 
		$(\sigma_r/\sigma_t)_\infty$ & 1 & 1.3 \\ 
		$r_a$ & 0 & $\infty$  \\ 
		$\lg \alpha$ & -0.3 & 0.6  \\ 
		$M_{\rm bh}$ & 0 & $\infty$  \\ 
		$(M/L)_1$ & 0 & 7.23  \\ 
		$(M/L)_2$ & 0 & 7.23  \\
		$f_{\rm dm}$ & 0 & 1 \\
		\hline
	\end{tabular}
	\setlength{\tabcolsep}{.3em}
	\label{tab:mcmc}
\end{table}

In order to best sample the space of possible parameters, we re-express the anisotropy parameters in terms that result in a more efficient and uniform sampling of the model posterior. Namely, we sample the anisotropy parameters defined such that
\begin{align}
	(\sigma_r/\sigma_t)_0 &= \frac{1}{\sqrt{1-\beta_0}}\\
	(\sigma_r/\sigma_t)_\infty  &= \frac{1}{\sqrt{1-\beta_\infty}}\\
	a &= \lg \alpha
\end{align}
Given what we know about the anisotropy profile in M87, we restrict these parameters to the ranges $0.5\leq (\sigma_r/\sigma_r)_0 \leq 1$ and $1 \leq (\sigma_r/\sigma_r)_\infty \leq 1.3$. The bound at 1 comes from the enforced condition that the anisotropy becomes tangential in the center and radial at large radii. The bounds at 0.5 and 1.3 represent the largest range of anisotropy values reliably observed in early type galaxies \cite{gebhardt2003axisymmetric,cappellari2008supermassive,cappellari2009mass,mcconnell2012dynamical,thomas2014dynamical,krajnovic2018quartet}, as well as in simulations of slow rotators \cite{rantala2019simultaneous,frigo2021two}. This range also encompasses the range of anisotropy profiles previously determined for M87 \citep{cappellari2005nuclear,gebhardt2011black}. In \autoref{tab:mcmc}, we list all of the parameters along with the corresponding bounds on their values. We also restrict the parameter $\lg\alpha$ to be between -0.3 and 0.6. This is due to the fact that small values of $\alpha$ correspond to no anisotropy transition, which we want to exclude, and large values of $\alpha$ give rise to an infinitely large parameter space where the spatial transition of the anisotropy takes place nearly instantaneously. In practice, the preferred range of parameter space almost always lies between -0.3 and 0.6 so this does not impact our final results.

\begin{table*}
	\centering
	\setlength{\tabcolsep}{.3em}
	\caption{Table of best fit parameters for each combination of instrument, spectral range, and legendre polynomial degree. The values presented are the median of the posteriors, with the upper and lower bounds corresponding to the 1$\sigma$ interval. The black hole masses are clustered between 8 and 10 billion solar masses. In the data column M stands for MUSE, O for OASIS, and S for SAURON.}
	\begin{tabular}{|c|c|c|c|c|c|c|c|c|c| } 
		\hline
		Data  & Model & $(\sigma_r/\sigma_t)_0$ & $(\sigma_r/\sigma_t)_\infty$ & $r_a$ & $\lg \alpha$ & $M_{\rm bh}$ & $(M/L)_1$ & $(M/L)_2$ & $f_{\rm dm}$ \\ \hline
		
		M RNI + O RNI + S & Constant $M/L$ & $0.55^{+0.08}_{-0.04}$&$1.24^{+0.03}_{-0.05}$&$1.4^{+0.5}_{-0.3}$&$0.08^{+0.04}_{-0.04}$& ($9.7^{+0.7}_{-0.9}$) $ \times 10^9 \ M_\odot $ &$3.4^{+0.1}_{-0.1}$& N/A & N/A \\ [1ex] 
		
		M RNI + O RNI + S & NFW DM & $0.54^{+0.07}_{-0.03}$&$1.26^{+0.03}_{-0.05}$&$1.6^{+0.5}_{-0.4}$&$0.09^{+0.04}_{-0.04}$& ($10.2^{+0.8}_{-1.0}$) $ \times 10^9 \ M_\odot $ &$3.2^{+0.1}_{-0.2}$& N/A &$0.06^{+0.07}_{-0.04}$ \\ [1ex] 
		
		M RNI + O RNI + S & Varying $M/L$ & $0.58^{+0.14}_{-0.06}$&$1.21^{+0.06}_{-0.06}$&$1.2^{+0.6}_{-0.5}$&$0.04^{+0.07}_{-0.08}$& ($8.2^{+1.7}_{-1.7}$) $ \times 10^9 \ M_\odot $ &$4.7^{+0.8}_{-0.9}$&$3.1^{+0.2}_{-0.2}$& N/A \\ [1ex] 
		
		M RNI + O RNI + S & Varying $M/L$ + NFW DM & $0.56^{+0.09}_{-0.05}$&$1.19^{+0.07}_{-0.07}$&$1.1^{+0.5}_{-0.5}$&$0.02^{+0.07}_{-0.08}$& ($8.7^{+1.2}_{-1.2}$) $ \times 10^9 \ M_\odot $ &$6.5^{+0.5}_{-0.8}$&$1.5^{+0.6}_{-0.5}$&$0.36^{+0.14}_{-0.14}$ \\ [1ex] 
		
		M d=4 + O d=6 + S & Constant $M/L$ & $0.53^{+0.08}_{-0.02}$&$1.24^{+0.04}_{-0.04}$&$1.1^{+0.4}_{-0.3}$&$0.05^{+0.07}_{-0.04}$& ($9.2^{+0.7}_{-1.0}$) $ \times 10^9 \ M_\odot $ &$3.4^{+0.1}_{-0.1}$& N/A & N/A \\ [1ex] 
		
		M d=4 + O d=6 + S & NFW DM & $0.53^{+0.05}_{-0.02}$&$1.26^{+0.03}_{-0.05}$&$1.2^{+0.3}_{-0.2}$&$0.05^{+0.03}_{-0.04}$& ($9.6^{+0.6}_{-0.7}$) $ \times 10^9 \ M_\odot $ &$3.3^{+0.1}_{-0.1}$& N/A &$0.03^{+0.05}_{-0.02}$ \\ [1ex] 
		
		M d=4 + O d=6 + S & Varying $M/L$ & $0.59^{+0.15}_{-0.07}$&$1.19^{+0.07}_{-0.07}$&$0.9^{+0.6}_{-0.4}$&$0.02^{+0.1}_{-0.08}$& ($7.7^{+1.2}_{-1.6}$) $ \times 10^9 \ M_\odot $ &$5.0^{+0.9}_{-0.8}$&$3.1^{+0.2}_{-0.2}$& N/A \\ [1ex] 
		
		M d=4 + O d=6 + S & Varying $M/L$ + NFW DM & $0.57^{+0.11}_{-0.05}$&$1.14^{+0.07}_{-0.06}$&$0.9^{+0.4}_{-0.4}$&$0.03^{+0.09}_{-0.09}$& ($8.3^{+1.3}_{-1.6}$) $ \times 10^9 \ M_\odot $ &$6.5^{+0.5}_{-0.9}$&$2.0^{+0.6}_{-0.6}$&$0.24^{+0.14}_{-0.12}$ \\ [1ex] 
		
		M d=6 + O d=4 + S & Constant $M/L$ & $0.67^{+0.17}_{-0.09}$&$1.22^{+0.06}_{-0.06}$&$2.2^{+1.5}_{-0.7}$&$0.12^{+0.07}_{-0.06}$& ($8.9^{+1.4}_{-1.4}$) $ \times 10^9 \ M_\odot $ &$3.5^{+0.1}_{-0.1}$& N/A & N/A \\ [1ex] 
		
		M d=6 + O d=4 + S & NFW DM & $0.54^{+0.04}_{-0.03}$&$1.26^{+0.02}_{-0.04}$&$1.2^{+0.3}_{-0.2}$&$0.05^{+0.03}_{-0.04}$& ($9.6^{+0.6}_{-0.8}$) $ \times 10^9 \ M_\odot $ &$3.3^{+0.1}_{-0.1}$& N/A &$0.03^{+0.05}_{-0.02}$ \\ [1ex] 
		
		M d=6 + O d=4 + S & Varying $M/L$ & $0.68^{+0.19}_{-0.12}$&$1.17^{+0.07}_{-0.07}$&$2.0^{+1.2}_{-0.7}$&$0.14^{+0.08}_{-0.09}$& ($8.7^{+1.6}_{-1.7}$) $ \times 10^9 \ M_\odot $ &$4.5^{+0.9}_{-0.9}$&$3.2^{+0.2}_{-0.2}$& N/A \\ [1ex] 
		
		M d=6 + O d=4 + S & Varying $M/L$ + NFW DM & $0.66^{+0.13}_{-0.11}$&$1.16^{+0.09}_{-0.07}$&$1.7^{+1.1}_{-0.5}$&$0.08^{+0.09}_{-0.1}$& ($8.9^{+1.4}_{-1.6}$) $ \times 10^9 \ M_\odot $ &$6.2^{+0.8}_{-1.0}$&$1.7^{+0.8}_{-0.7}$&$0.33^{+0.16}_{-0.17}$ \\ [1ex] 
		
		M RNI & Varying $M/L$ + NFW DM & $0.58^{+0.14}_{-0.06}$&$1.24^{+0.04}_{-0.09}$&$1.7^{+1.23}_{-0.58}$&$0.00^{+0.08}_{-0.05}$& ($10.0^{+1.3}_{-1.7}$) $ \times 10^9 \ M_\odot $ &$6.9^{+0.3}_{-0.5}$&$1.8^{+0.2}_{-0.3}$&$0.07^{+0.08}_{-0.05}$ \\ [1ex] 
		
		O RNI & Varying $M/L$ + NFW DM & $0.56^{+0.1}_{-0.04}$&$1.23^{+0.05}_{-0.07}$&$2.1^{+1.12}_{-0.54}$&$0.06^{+0.07}_{-0.06}$& ($11.1^{+0.8}_{-1.2}$) $ \times 10^9 \ M_\odot $ &$6.9^{+0.3}_{-0.5}$&$1.6^{+0.3}_{-0.3}$&$0.06^{+0.08}_{-0.04}$ \\ [1ex] 
		\hline
	\end{tabular}\\
	{\raggedright {The values in this table are computed after the kinematics have been scaled to match the SAURON data (see \autoref{sec:specfit}). In order to convert to the MUSE scaling, multiply the black hole mass, $(M/L)_1$, and $(M/L)_2$ by 1.018, 1.019, 1.025, and 1.007 for the the Full spectra degree 4, 5, 6 and RNI spectrum, respectively. To convert to the OASIS scaling, multiply by 1.051, 1.051, 1.057, and 1.053 for the Full spectra degree 4, 5, 6 and RNI spectrum, respectively. 	}	\par}
	\label{tab:MCMC_vals}
\end{table*}

Running JAM for the required number of steps necessary to generate reliable posteriors for all of these combinations of models and data is very computationally expensive. The final contours exhibit strong covariances that require a long burn in time for the walkers to sample and populate the posterior. The likelihood also has multiple local minima which further increases the run time if the chain is started further away from the global minimum. In order to speed up this process, we start by running \textsc{emcee} for 300000 steps with 100 walkers for the JAM model of a spherical galaxy using the routine \textsc{jam\_sph\_proj} in \textsc{Jampy}. \footnote{Note that this is not the same as JAM with a spherically aligned velocity ellipsoid. The final routine used is \textsc{jam\_axi\_proj} which solves the axisymmetric Jeans equation assuming the velocity ellipsoid is spherically aligned. Here we assume that the system truly is spherically symmetric, thus simplifying the computation required to solve the Jeans equation.} This is a good approximation to the true Jeans solution as M87 is highly spherical, especially in the range of our data. Once this step is complete, we have a good approximation of the posterior. From there, we run \textsc{emcee} for 50000 steps with 100 walkers for the JAM model assuming axisymmetry and a spherically aligned velocity ellipsoid using the routine \textsc{jam\_axi\_proj}. These final 50000 steps are what is used for the final analysis. 

\begin{figure*}
	\centering
	\subfloat{\includegraphics[width = 2.2in]{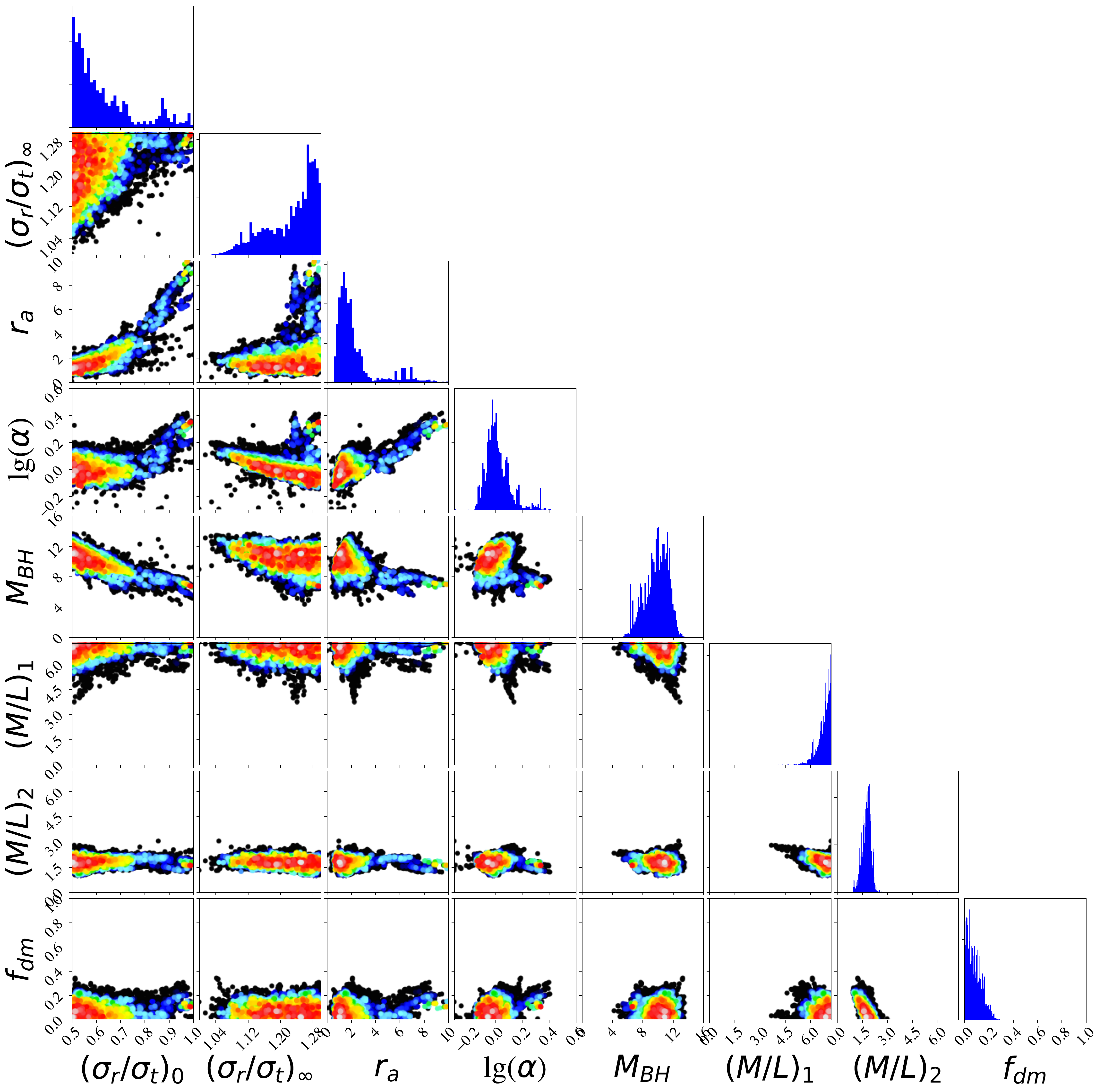}}
	\subfloat{\includegraphics[width = 2.2in]{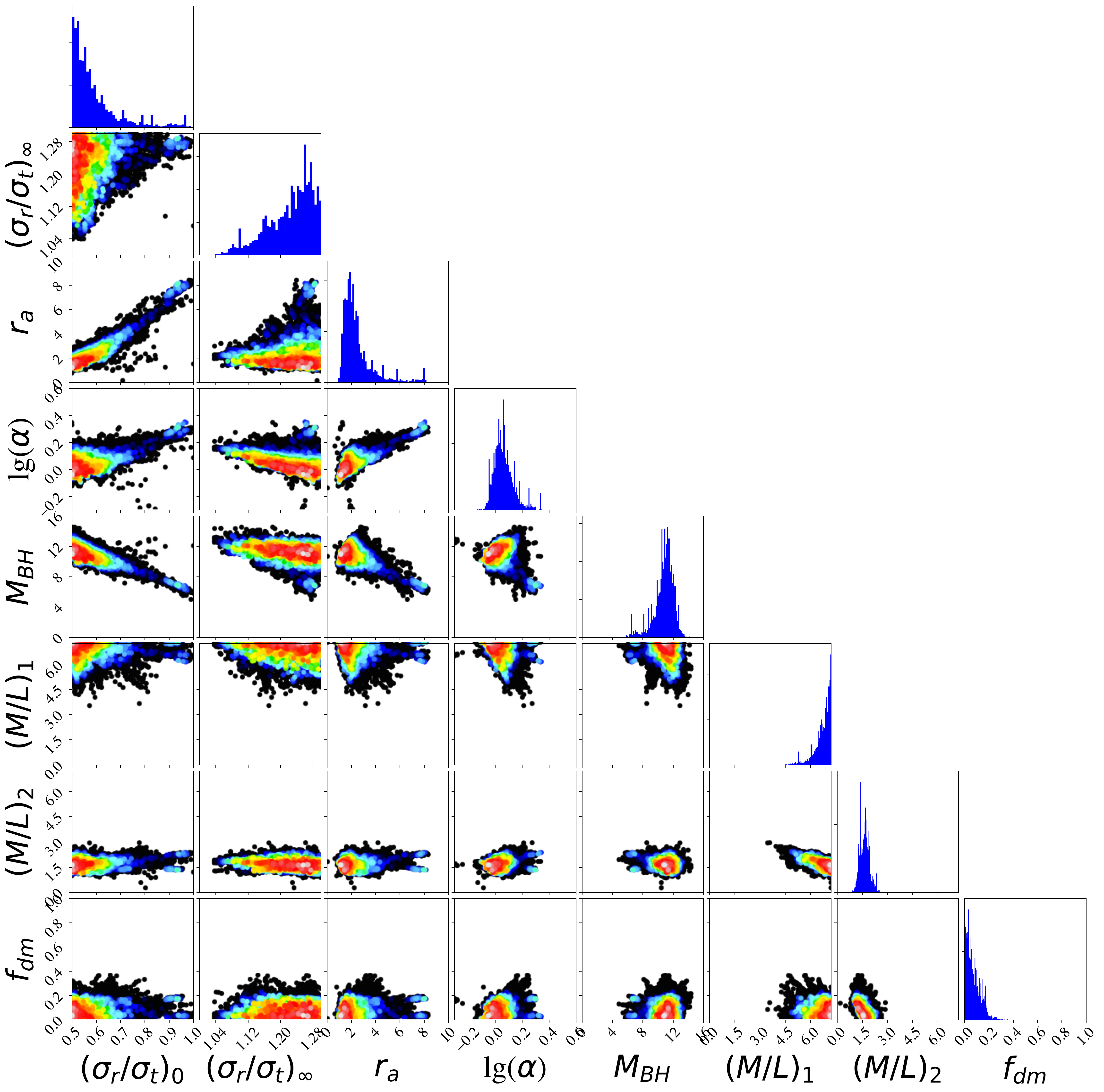}}
	\subfloat{\includegraphics[width = 2.2in]{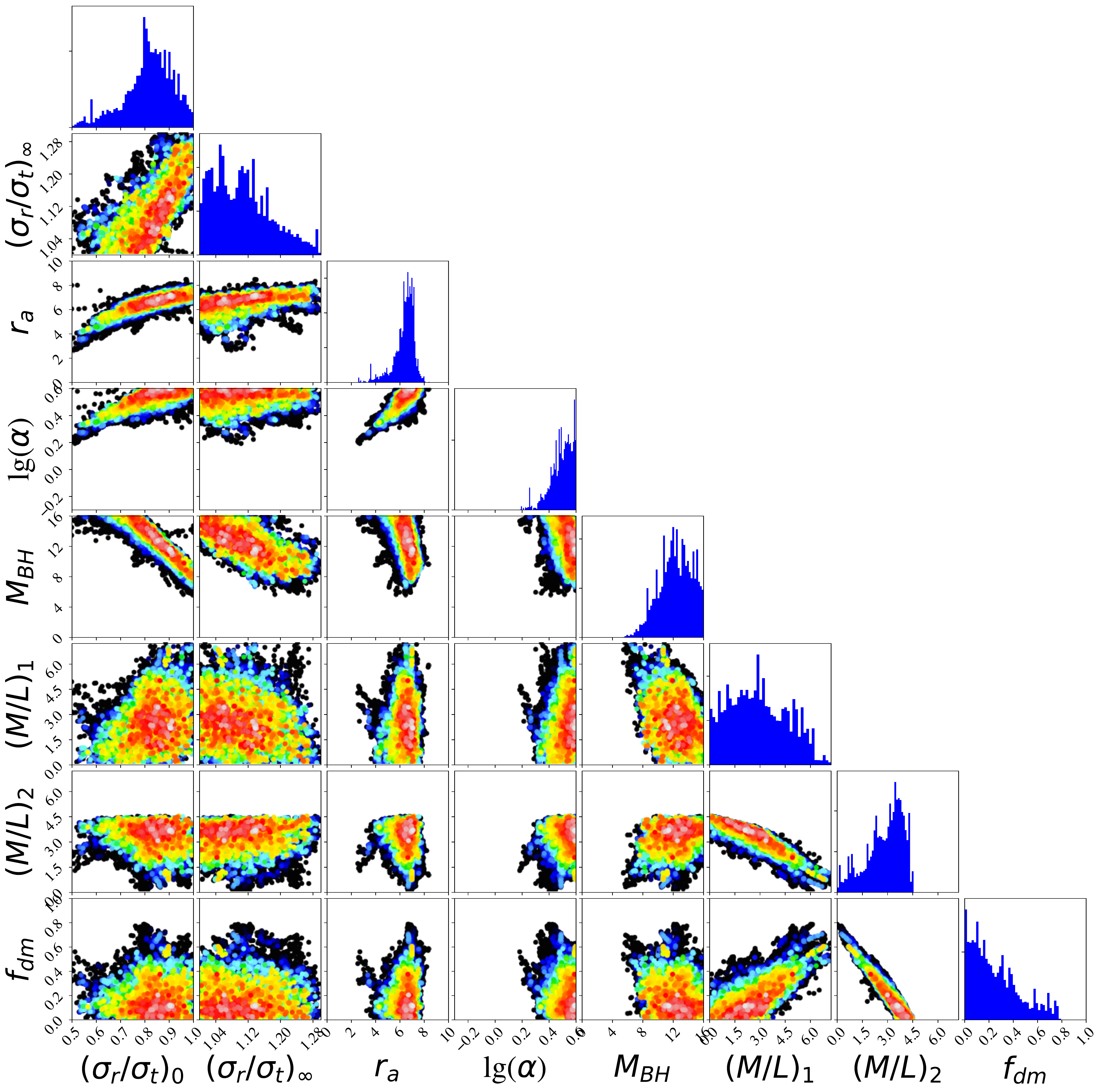}}
	\caption{Corner plots for the individual MUSE, OASIS, and SAURON data sets, respectively, using the RNI spectra. The MUSE and OASIS corner plots feature a secondary minimum corresponding to a near isotropic anisotropy profile in the centre of the galaxy with a lower black hole mass. Bestfit parameters and errors for MUSE and OASIS are shown in the last two rows of \autoref{tab:MCMC_vals}.}
	\label{fig:MCMC_dmml}
\end{figure*}

We present the posteriors for this in \autoref{fig:MCMC_4mod}. We also show the posteriors assuming the most general model for the individual MUSE RNI, OASIS RNI, and SAURON data sets in \autoref{fig:MCMC_dmml}. Given our choice of physically motivated priors, some of the posteriors are not symmetric and even run into the boundary. As such, we report the posterior median and left and right 1$\sigma$ confidence interval for all of the combinations of models and combined data in \autoref{tab:MCMC_vals}. As we expect, the results for M87 change very little between using spherical JAM and axisymmetric JAM. We also show the model fits to the observed dispersion in \autoref{fig:dispersion_vary}.



\begin{figure}
	\centering
	\subfloat{\includegraphics[width = 3.1in]{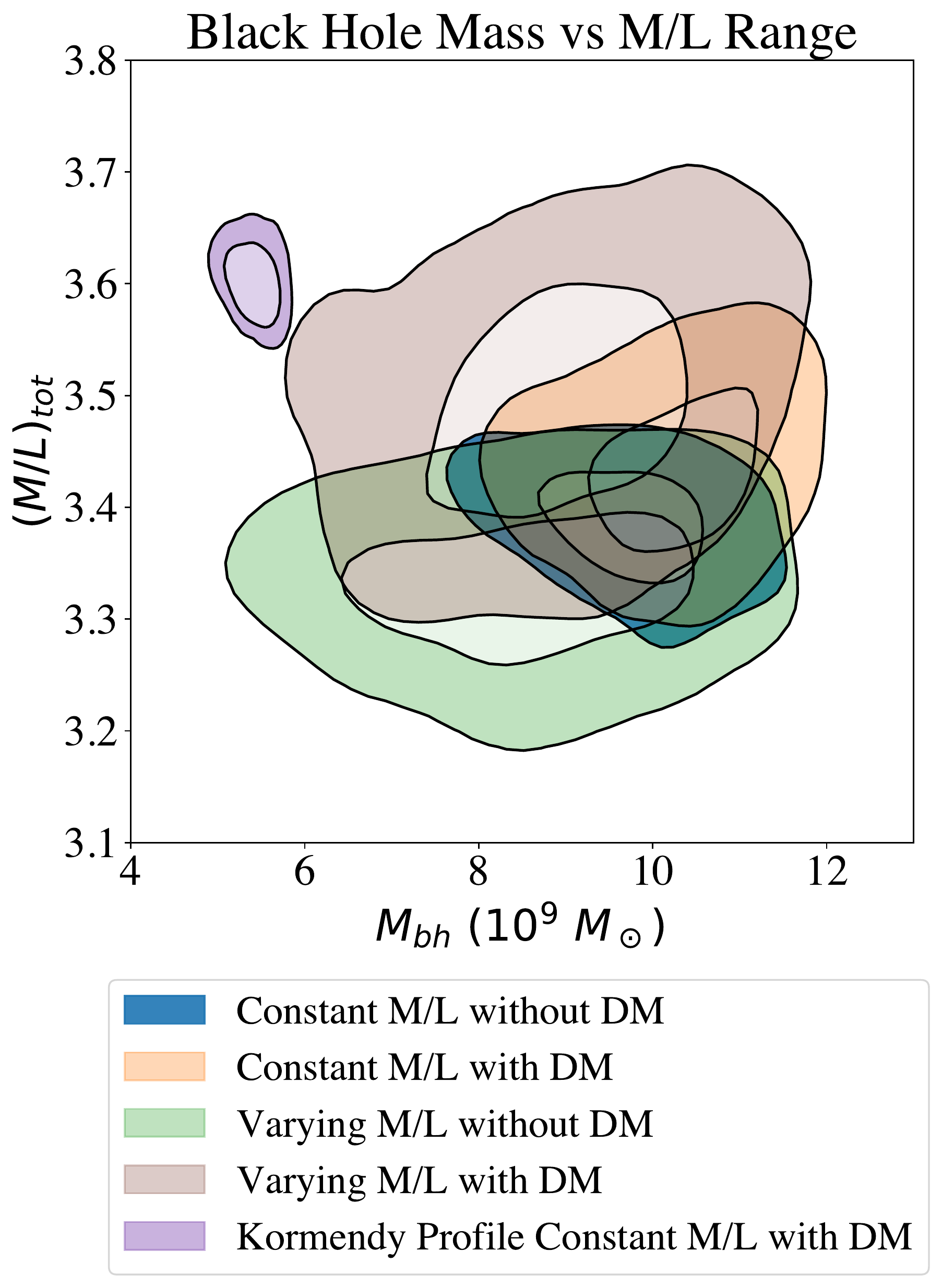}}
	\caption{Plot of the 1 and 3$\sigma$ confidence intervals for black hole mass and mass to light ratio within a sphere of one half light radius for each of the scenarios shown in \autoref{fig:MCMC_4mod}.  }
	\label{fig:bhml_contour}
\end{figure}

\section{Discussion}\label{sec:discussion}
\subsection{Black Hole Results}
In \autoref{fig:MCMC_4mod} we show the full posteriors using the RNI spectra for each of the four models. In \autoref{fig:bhml_contour} we show $1\sigma$ and $3\sigma$ contours of the black hole mass and total $M/L$ within one half light radius for the four different models using the RNI spectra. We choose to focus on the RNI spectra as this is the case where the OASIS and MUSE data have the closest agreement, suggesting that this choice of kinematics best reflects the true kinematics. The impact of the different choices of kinematics on the following results is some scatter, as opposed to single systemic shifts in the measured values. From \autoref{fig:MCMC_dmml} we can deduce the individual contributions of each data set to the final black hole masses. There we see that the OASIS and MUSE data are remarkably similar, with the SAURON data providing a unique contribution. 
	
We find a large range of permissible black hole mass values, from nearly $6\times 10^9 M_\odot$ to greater than $11\times10^{9} M_\odot$ at one sigma uncertainty across different choices of kinematics and models. The values obtained using the most general model (the median of the posterior) varies between $8.3-8.9\times 10^9 \ M_\odot $. The errors are asymmetric and range in magnitude between 1.2 to 1.6. Using  the kinematics derived from RNI spectra gives $M_{\rm BH} = (8.7\pm1.2\pm 1.3) \times 10^9 \ M_\odot $ where we calculate the second error to be half the difference between the largest and smallest black hole mass across all kinematics and models (i.e. the largest and smallest values in \autoref{tab:MCMC_vals}). Note that restricting this just to the most general model across each choice of kinematics decreases the systematic error to 0.3.

The final allowed range of black hole masses depend strongly on the model assumptions. The most simple model with only constant $M^*/L$ has the smallest range of permitted black hole masses. Including dark matter slightly shifts this range to the right. This is because the SAURON data strongly constrains the kinematics at large radii, so there is a tight correlation between $f_{\rm dm}$ and $M^*/L$ which allows one to interchange stars and dark matter. Decreasing stars at large radii in favor of dark matter, however, must be compensated at small radii with an increased black hole mass. Throughout this the anisotropy parameters remain fairly constant. This changes significantly with the introduction of varying $M^*/L$. In the model with a $M^*/L$ variation and no dark matter, this results in a much larger range of viable black hole masses at the lower mass end. This is due to a correlation between $(M/L)_1$ and the black hole mass which effectively results in the mass of the black hole being exchanged for mass in stars. The results for the black hole mass are similar for the most general model featuring both varying $M/L$ and DM. 

The most interesting difference compared with previous studies is that the black hole mass we measure is more massive than that found in previous stellar dynamical studies. One key difference between our analysis and previous work is that we directly measure the stellar distribution within the influence of the AGN and find the stellar profile to be flatter than that used in all previous black hole studies of M87 using stellar dynamics. In order to determine if this could explain the increase in the black hole mass, we performed an MCMC run using the stellar profile of \cite{kormendy2009structure} as this is used for the tracer distribution in previous studies \citep{gebhardt2009black,gebhardt2011black,2023arXiv230207884L}. We do our modelling with the RNI spectra and in the model with DM and constant M/L, as this most closely approximates the models used in \cite{gebhardt2009black} and \cite{gebhardt2011black}. The 1 and 3$\sigma$ posteriors for this are shown in \autoref{fig:bhml_contour}. We find a preferred black hole mass of $M_{\rm BH} = (5.5^{+0.5}_{-0.3})$ $ \times 10^9 \ M_\odot $, which agrees much more closely with previous measurements which range between $5.4\times 10^9 \ M_\odot$ \citep{2023arXiv230207884L} to $6.2\times 10^9 \ M_\odot$ \citep{gebhardt2011black}. The equivalent model using our stellar distribution gives a black hole mass of $M_{\rm BH} = (10.2^{+0.8}_{-1.0})\times 10^9 \ M_\odot$. Naively one might think that the difference in the black hole mass is purely due to the difference in the gravitational potential between the two models. On its own, our stellar profile should increase the measured black hole mass over previous determinations as it is exchanging the mass of the stars in the central regions of the galaxy with the mass of the black hole. However, one can calculate the decrease in stellar mass in the centre of the galaxy between our model and the model of \cite{kormendy2009structure} and we find that, assuming a constant stellar $M/L$ ratio of 3.4 without DM, that within 5\arcsec (approximately the sphere of influence) the decrease in stellar mass is $5.6\times 10^7 \ M_\odot$, or around 1 per cent of the total stellar mass within 5\arcsec. This implies that the modification to the gravitational potential due to our model cannot be the sole cause of the difference between the two results. 

\begin{figure}
	\includegraphics[width=\columnwidth]{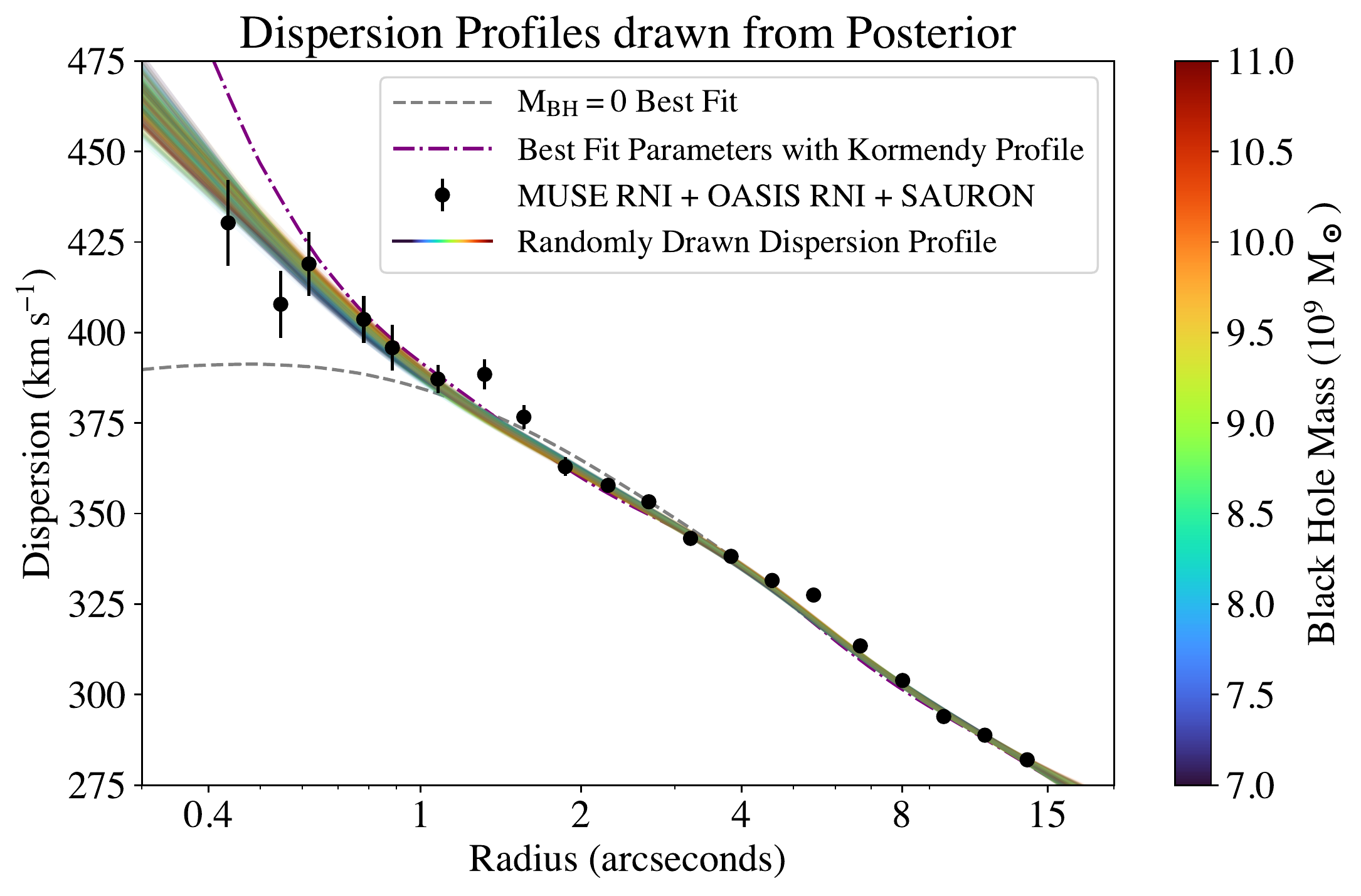}
	\caption{Plot of 1000 dispersion profiles randomly sampled from the MCMC chain with MUSE RNI + OASIS RNI + SAURON and colored according to their supermassive black hole mass. This is compared with a set of binned data points combining the MUSE RNI, OASIS RNI, and SAURON data. We also show a model with no black hole mass. It is technically possible to fit the data with no black hole mass if there is a highly radial anisotropy in the centre. To ensure this does not happen, we fix $r_a = 1$ and find the best fit model enforcing $M_{\rm BH} = 0$. The dotted and dashed purple line shows the dispersion after exchanging our stellar distribution with the one from \protect\cite{kormendy2009structure} but still using the parameters from the best fit model assuming our stellar distribution. You can see that the profile becomes much steeper in the center. The way to remedy this is by decreasing the black hole mass to the value determined in previous studies and adjusting the anisotropy correspondingly.}
	\label{fig:dispersion_vary}
\end{figure}

\begin{figure}
	\includegraphics[width=\columnwidth]{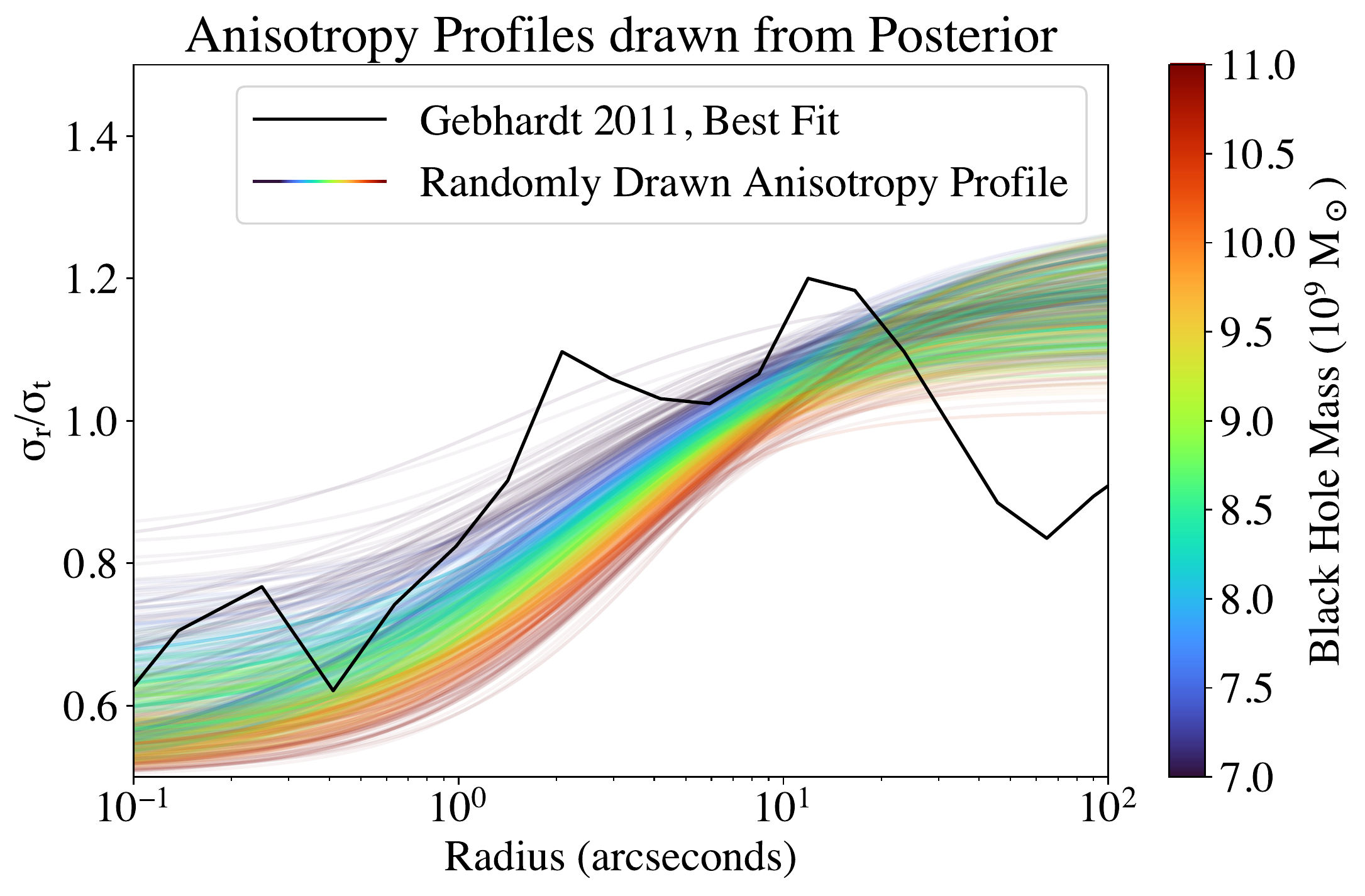}
	\caption{Plot of 1000 anisotropy profiles randomly sampled from the MCMC chain with MUSE RNI + OASIS RNI + SAURON and colored according to their supermassive black hole mass. The best fit anisotropy profile from \protect\cite{gebhardt2011black} is shown in black. We find strong evidence for a radially increasing anisotropy ratio while varying strongly due to the mass anisotropy degeneracy. }
	\label{fig:aniso}
\end{figure}

One possible alternative is that this is a result of the very flat core. As pointed out by \citet[sec.~3.1]{kormendy2013coevolution}, measuring BHs with stellar dynamics in galaxies with flat cores is intrinsically less accurate than in galaxies with steep inner profiles, because of the increase importance of orbital anisotropy. Additionally, in centrally cuspy galaxies, the line of sight velocity distribution along the photometric centre of the galaxy receives its largest contribution from the three dimensional origin of the galaxy. In the case of a very flat core, the line of sight velocity distribution along the photometric centre receives an even contribution from a larger range of radii. This further increases the uncertainty in the kinematic modelling (note the much larger uncertainties using our profile over the \cite{kormendy2009structure} profile) and has the potential to impact the extracted kinematics. Furthermore, a steeper profile implies more stars near the black hole, where the intrinsic sigma is higher. This leads to a higher observed sigma after projecting along the line-of-sight. In order to better demonstrate this effect, we show a plot of the model dispersion assuming our best fit parameters but using the stellar distribution from \cite{kormendy2009structure} as opposed to our stellar distribution in \autoref{fig:dispersion_vary}. We find that, given the same set of parameters, using the old stellar distribution leads to a sharp rise in the dispersion within 1\arcsec. This suggests that in order to fit the data, previous studies required a smaller black hole mass with a more radially biased anisotropy closer to the black hole. This is exactly what we see in \autoref{fig:aniso} where we plot our anisotropy and that of \cite{gebhardt2011black}. Further study of this will be required in future work.

\subsection{Mass to Light Ratio Constraints}\label{sec:disc_ml}
\begin{figure}
	\includegraphics[width=\columnwidth]{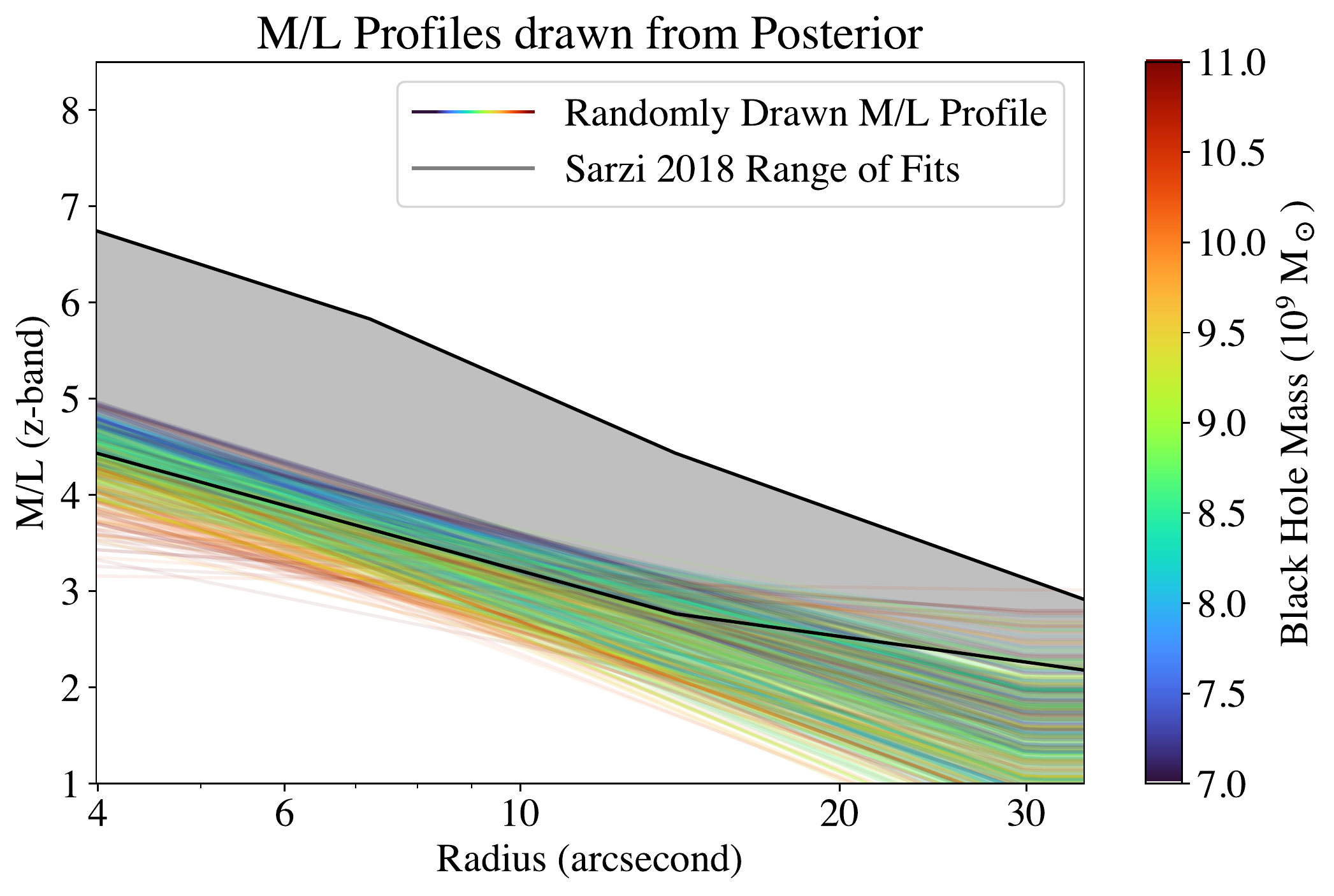}
	\caption{Plot of 1000 M/L profiles randomly sampled from the MCMC chain with MUSE RNI + OASIS RNI + SAURON and colored according to their supermassive black hole mass. This is compared with range of M/L variations assuming Kroupa IMF from \protect\cite{sarzi2018muse} which is shown in gray and outlined with black lines. Our data strongly prefers an increasing M/L variation towards the centre of the galaxy consistent with that in \protect\cite{sarzi2018muse}.}
	\label{fig:ML_vary}
\end{figure}

One important feature in this work is the inclusion of stellar $M/L$ variations. In  \autoref{fig:ML_vary} we show 1000 $M/L$ profiles randomly sampled from the posterior of the most general model using the RNI data. We find a strong preference for an increasing $M/L$ ratio in the central regions of the galaxy. Our best fit $M/L$ variation agrees with the bottom end of figure 11 of \cite{sarzi2018muse} assuming a Kroupa IMF. Previous studies have either failed to take $M/L$ variations into account \citep{gebhardt2009black,gebhardt2011black}, or directly used the profile measured in \cite{sarzi2018muse} \citep{2023arXiv230207884L}. Our result suggests that it is important to treat the profile as a free parameter.

One important limitation of these results is the fact that the $M/L$ variation parametrization we use does not permit particularly steep variations due to it fixing the $M/L$ ratio at 1\arcsec and 30\arcsec. To test the effect of this, we ran a model where the inner radius at which the $M/L$ ratio becomes fixed is free to vary. We find a preference for a range of values between 1\arcsec and 4\arcsec but do not find significant changes in the distribution of $M/L$ profiles in the range of the \cite{sarzi2018muse} data. Our work thus establishes a preference for a $M/L$ gradient at the lower end of what was reported in \cite{sarzi2018muse}. However, we caution that the recovered $M_*/L$ profile is degenerate with the slope of the dark matter profile and the one we derive relies on a fixed NFW halo. Obviously no dynamical model can distinguish between a variation in the total density due to the stellar $M_*/L$ or the dark matter without assumptions.

The total $M/L$ ratio (defined as $M/L$ within one $R_{\rm e}$) that we measure exhibits some model dependence. In \autoref{fig:bhml_contour}, we see that the total $M/L$ ranges between 3.2 up to 3.7 depending on the model assumptions. This range of values is primarily due to the uncertainty in the anisotropy profile at large radius. The model with the largest total $M/L$ is the model with varying $M/L$ and DM. We see in \autoref{fig:MCMC_4mod} that the preferred value of $(\sigma_r/\sigma_t)_\infty$ is smaller than in the other models. This decrease in anisotropy at large radius must be made up by an increase in the mass. This further highlights the importance of the mass-anisotropy degeneracy when studying M87.

The total $M/L$ ratio of M87 in the I-band has been previously measured in \cite{cappellari2006sauron} and in the r-band from \cite{cappellari2013atlas3d} (using mass follows light models in both cases). Converting these to the SDSS-z band (as a proxy for the ACS/WFC F850LP band) gives $M/L$ ratios of 4.2/4.8\footnote{$(M/L)_{\rm tot,z}=4.2$ is determined from isotropic Jeans modelling whereas $(M/L)_{\rm tot,z}=4.8$ is determined from Schwarzschild modelling}, and 4.0. This is slightly larger than our range of $M/L$. In both previous determinations, the models were fit to the data out to R$\approx$35\arcsec, while we only fit the data out to R=15\arcsec. The slightly larger total $M/L$ of previous determinations may be explained as due to both their smaller adopted BH and an increase in the dark matter fraction between these radii. \cite{gebhardt2009black} and \cite{murphy2011galaxy} also present measurements of the V-band stellar $M/L$ ratio of 6.3 and 8.2. Converting these to SDSS-z band gives 2.8 and 3.7, respectively. These cannot be directly compared to the results in \autoref{fig:bhml_contour} as we present the total $M/L$ rather than just the stellar $M/L$. However, we can still conclude that a stellar $M/L$ of 3.7, after including dark matter, will lead to a total $M/L$ that is slightly above the range of what we have determined. Likewise, a stellar $M/L$ of 2.8, combined with a dark matter fraction of $\sim15$ will result in a total $M/L$ of $\sim3.2$, which is at the lower end of what we measure.

In this work we measure the stellar distribution within the influence of the AGN and find the stellar profile to be flatter than in previous work. The difference to the enclosed mass within 5\arcsec is close to 1\%, implying that this does not significantly modify the gravitational potential. However, M87 is a large galaxy where the AGN covers only a small fraction of the stellar distribution relative to the size of the black hole. For other galaxies, such as NGC 4151, one of the key uncertainties in the black hole mass determination is uncertainty on the cuspiness of the inner stellar distribution \citep{roberts2021black}. Applying this technique to that case or similar cases could significantly reduce the uncertainties in the final black hole mass measurement.

\subsection{Dark Matter}

In this work we assume a NFW dark matter halo with break radius equal to 20kpc and find, in our most general models, a preference for a dark matter fraction within one effective radius of around $f_{\rm dm}(<R_{\rm e})\approx0.2$. This closely agrees with a result from \cite{murphy2011galaxy} which determined the dark matter fraction within one effective radius to be approximately 17\%. Our results, however, are strongly model and data dependent. In the less general models we consistently find a preference for no or very little dark matter (see \autoref{fig:MCMC_4mod}). Additionally, on their own, the MUSE, OASIS, and SAURON data do not have a preference for a dark matter halo (\autoref{fig:MCMC_dmml}). This data only goes out to ~1.2 kpc, so we do not expect to very strongly constrain the dark halo. This highlights the importance of including large scale kinematic data for constraining information on the dark matter halo.

\subsection{Anisotropy Profile Constraints}
In \autoref{fig:aniso} we show 1000 anisotropy profiles randomly chosen from the posterior of the NFW DM + Varying $M/L$ model using the RNI spectra. We find a remarkable agreement with previous work. The profiles all display the radially increasing behavior expected of slow rotators. We also visually see that the profiles tend to transition from constant on the right hand side to lower values near 10\arcsec. This is close to the size of the core of M87 (5\farcs66 according to \cite{lauer2007centers}), and agrees with the results of previous studies demonstrating that the size of the core in slow rotators is close to the radius at which the velocity anisotropy ratio becomes tangential \citep{thomas2014dynamical}. Another observation we should make is the strong correlation between the black hole mass and the anisotropy profile in \autoref{fig:aniso}. This clearly demonstrates the strong role that the mass-anisotropy degeneracy plays in this analysis.

\begin{figure*}
	\centering
	\subfloat{\includegraphics[width = 3.0in]{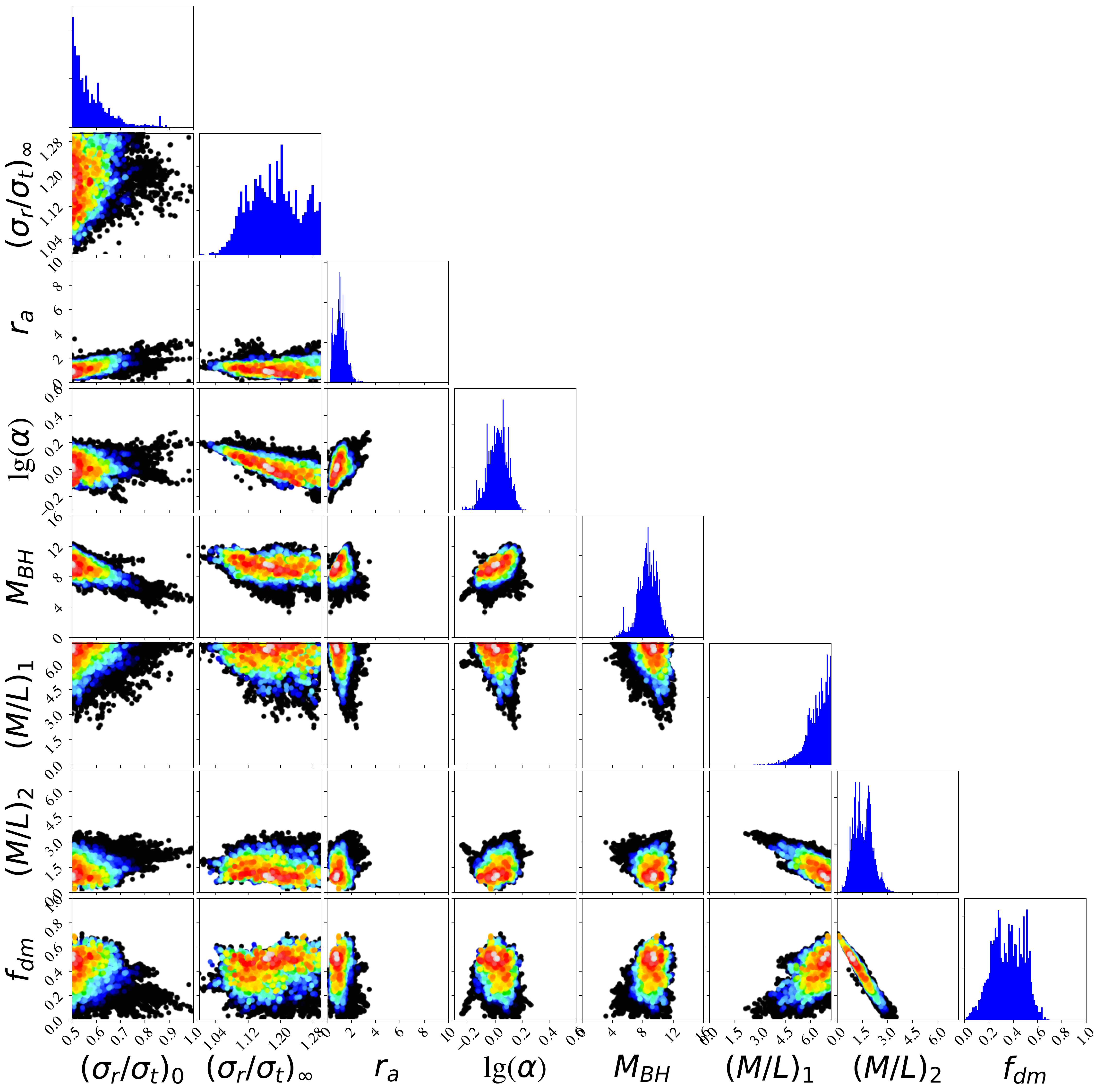}} \hspace{.2cm}
	\subfloat{\includegraphics[width = 3.0in]{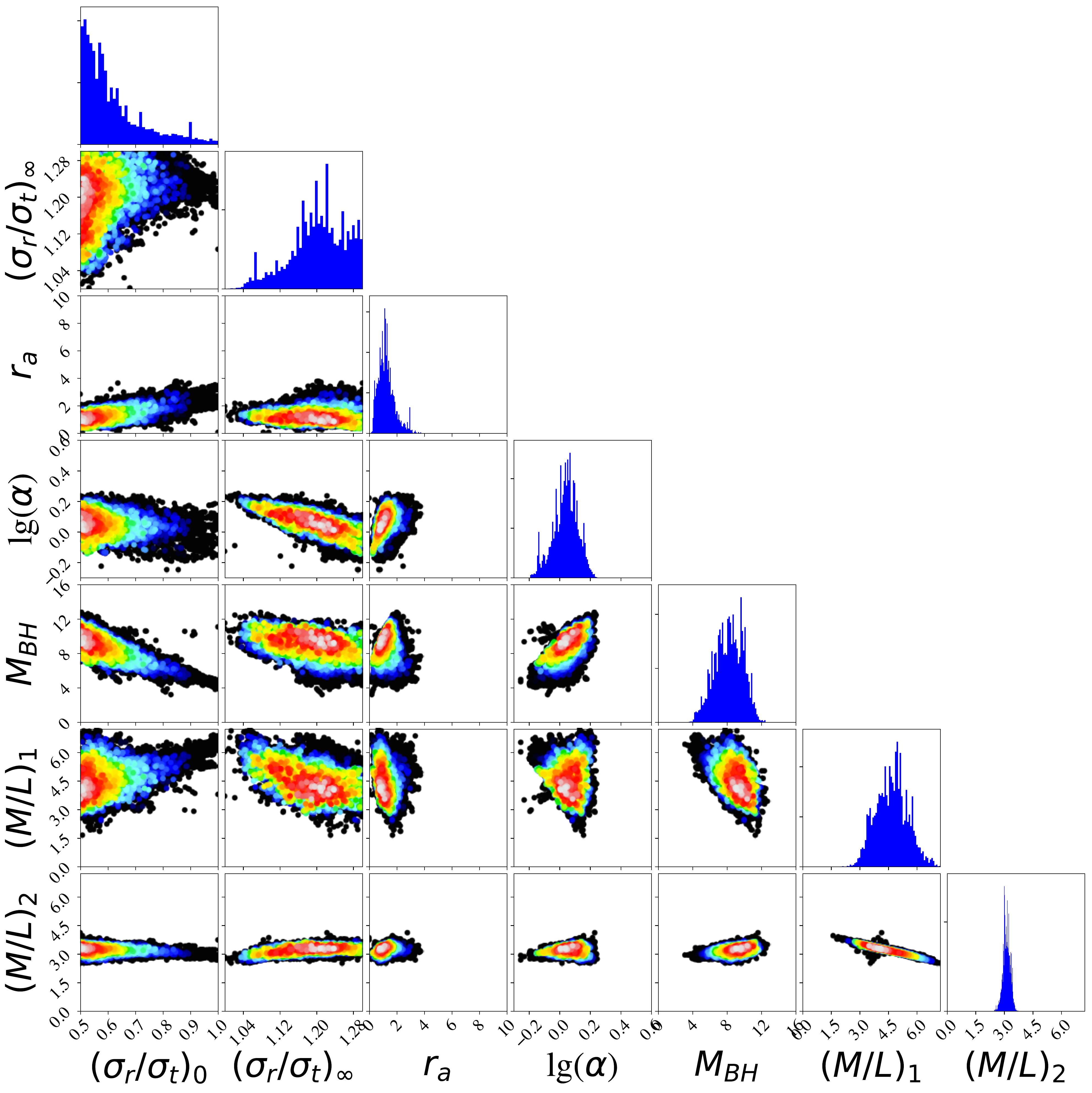}}\\
	\subfloat{\includegraphics[width = 3.0in]{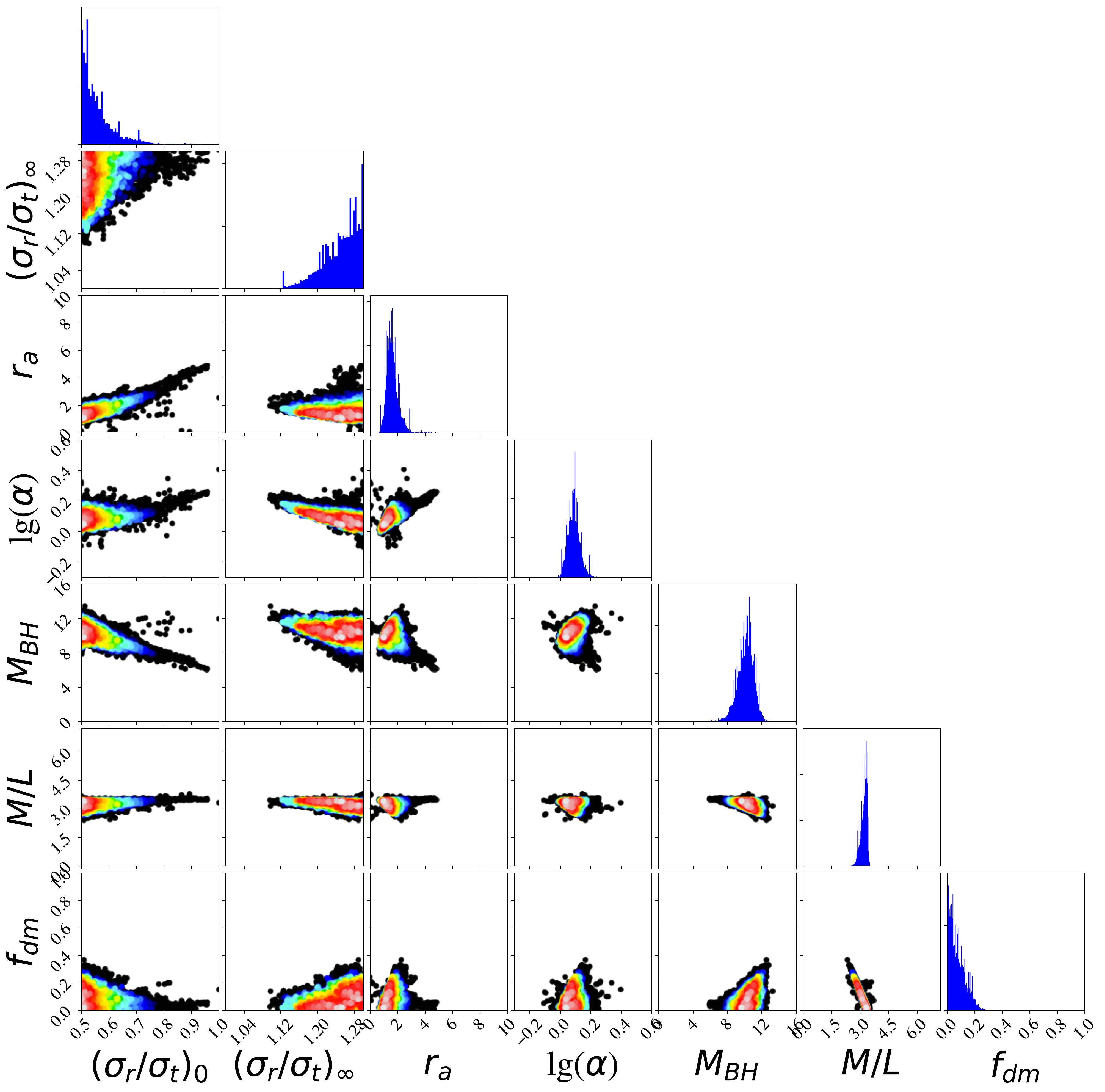}} \hspace{.2cm}
	\subfloat{\includegraphics[width = 3.0in]{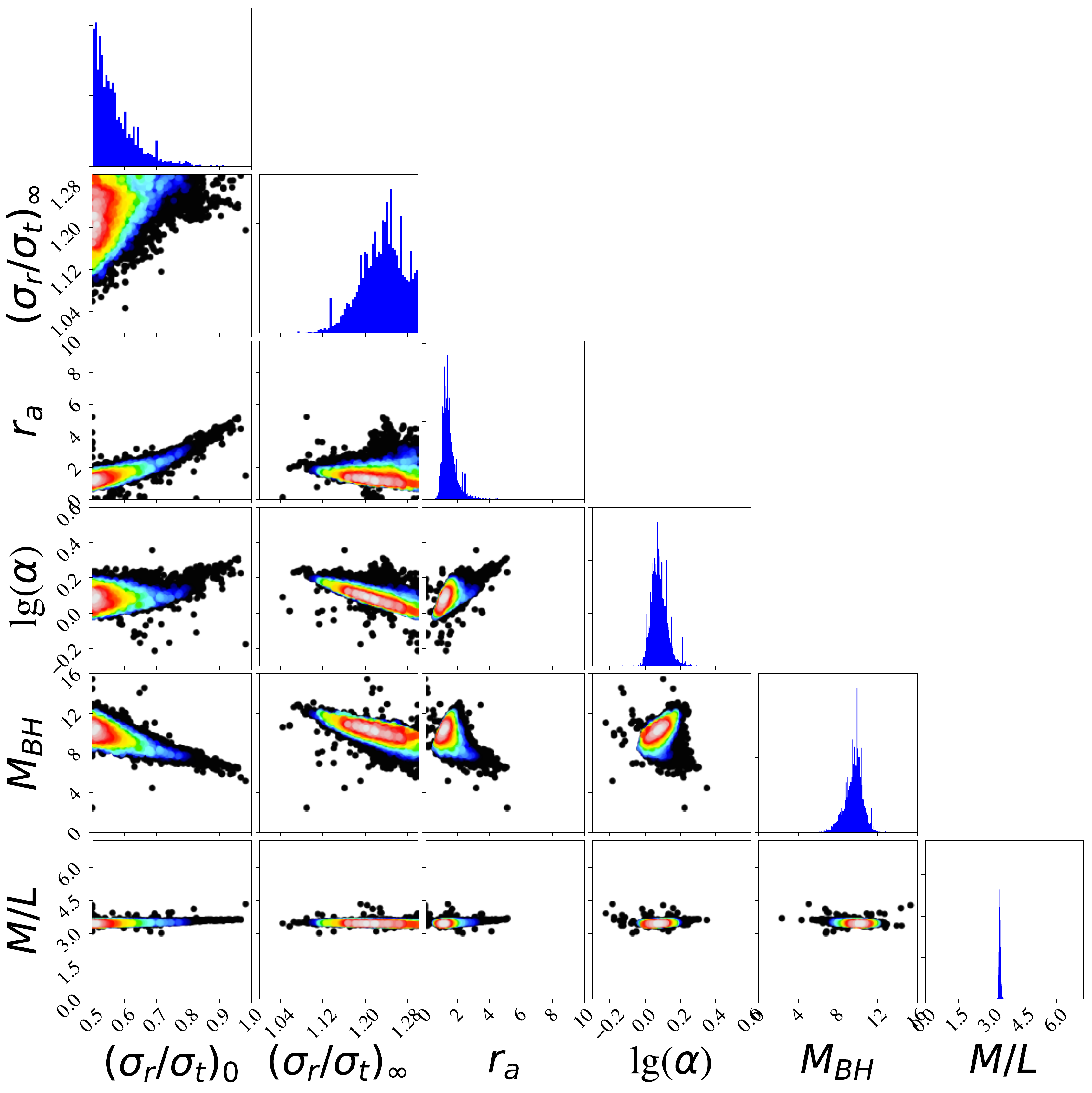}}
	\caption{Corner plots for the joint MUSE RNI + OASIS RNI + SAURON data for each of our four assumed models. Starting from the top left and going clockwise the models are: Varying M/L + NFW DM, Varying M/L without NFW DM, Constant M/L without NFW DM, and constant M/L with NFW DM. }
	\label{fig:MCMC_4mod}
\end{figure*}

One important comment is that these results depend on our use of physically motivated priors. We see in \autoref{fig:MCMC_4mod} that in many cases, the posteriors for $(\sigma_r/\sigma_t)_0$ and $(\sigma_r/\sigma_t)_\infty$ run into the imposed boundaries and thus are unable to explore the full range of parameter space capable of reproducing the data.

\section{Conclusion and Future Prospects}\label{sec:conclusion}
We have studied the galaxy M87 using stellar kinematics from SAURON, OASIS, and MUSE using the code \textsc{JamPy} and our primary conclusions are as follows:

\begin{itemize}
	\item The stellar distribution of M87 can be measured directly within the influence of the AGN. This is done by directly measuring the fraction of the spectral flux due to stars during the kinematic extraction. The shape of the stellar distribution profile used in previous studies \citep{kormendy2009structure} overestimates the the stellar density in the central region of M87 by a factor of 2 (see \autoref{fig:MGE_radial}).
	\item For galaxies with an AGN, the PSF can be accurately measured in integral field spectroscopy by measuring the continuum flux during the kinematic extraction (see \autoref{fig:PSF_slice} and \autoref{tab:PSF}). This is due to the fact that the AGN spectral contribution is thought to be smooth, and hence is well approximated by additive polynomials. We measure the FWHM of the OASIS PSF to be 0\farcs561 and the FWHM of the MUSE PSF to be 0\farcs049. These are consistent with what we expect from the seeing for OASIS and the AO capabilities of MUSE.
	\item We use JAM dynamical models of the kinematics in a Bayesian fashion. We find a preferred black hole mass of $M_{\rm BH}=(8.7\pm1.2\pm 1.3)  \times 10^9 \ M_\odot $ with the second error showing the modelling and kinematic systematic uncertainty. This range is consistent with the EHT measurement and previous stellar dynamical models of M87, though with a distinct preference for a larger black hole mass. Our analysis also highlights the fact that, even with excellent data, the derived black hole mass is sensitive to a variety of assumptions on both the kinematic extraction and the $M/L$ variation. The resulting uncertainties, when accounting for these systematics, are significantly larger than usually adopted.
	\item We find a strong preference for a radially decreasing $M^*/L$ ratio at the lower end of what is found in \cite{sarzi2018muse}. This has the effect of expanding the range of allowed black hole masses to lower values. 
	\item We measure the anisotropy profile of M87 assuming a new flexible analytic parametrization for the anisotropy which is a logistic function of logarithmic radius and find a strong preference for a radial increase. This also clearly shows the mass-anisotropy degeneracy which strongly contributes to the uncertainty in the black hole mass. We conclude that, contrary to what is sometimes assumed, one can obtain stringent constraints on both black hole masses and on the anisotropy profile, using the Jeans equations, by combining priors on the anisotropy, which is now well-understood in galaxy centres, with realistic parametrizations for the total density.
\end{itemize}

There remain many important questions about M87 that are well suited to be studied in the near future. Recent work has suggested a number of different ways that improved modelling of the gas disk is able to bring the supermassive black hole measurements from gas dynamics into agreement with those from stellar kinematics and the EHT \citep{2019ApJ...882...82J,jeter2021reconciling,osorno2023m87}. This, however, relies on resolving details of the gas kinematics within the innermost arcsecond of the galaxy. New and improved datasets could be used to differentiate between these scenarios.

Furthermore, as this work shows, future studies of the supermassive black hole mass of M87 using stellar kinematics will also require detailed studying of systematic effects in order to produce reliable black hole mass results. High quality data for these tasks may not be far off. There is a cycle 1 JWST proposal (2228, PI: Jonelle Walsh) to measure the central supermassive black hole of M87 using NIRSpec. This will cover a wavelength range including the CO bandhead that is similar to the wavelength range covered in \cite{gebhardt2011black} though it will be unaffected by skylines and will have much higher spatial resolution and signal to noise. This will provide the most detailed view of M87's inner kinematics to date.

\section*{Acknowledgements}
This research is based on observations collected at the European Southern Observatory under ESO program 0103.B-0581. This research is based on observations obtained at the Canada-France-Hawaii Telescope (CFHT) which is operated by the National Research Council of Canada, the Institut National des Sciences de l'Univers of the Centre National de la Recherche Scientifique of France, and the University of Hawaii. This research is based on observations made with the NASA/ESA Hubble Space Telescope obtained from the Space Telescope Science Institute, which is operated by the Association of Universities for Research in Astronomy, Inc., under NASA contract NAS 5–26555. These observations are associated with program GO-9401. 

Funding for the SDSS and SDSS-II has been provided by the Alfred P. Sloan Foundation, the Participating Institutions, the National Science Foundation, the U.S. Department of Energy, the National Aeronautics and Space Administration, the Japanese Monbukagakusho, the Max Planck Society, and the Higher Education Funding Council for England. The SDSS Web Site is \url{http://www.sdss.org/}.

The SDSS is managed by the Astrophysical Research Consortium for the Participating Institutions. The Participating Institutions are the American Museum of Natural History, Astrophysical Institute Potsdam, University of Basel, University of Cambridge, Case Western Reserve University, University of Chicago, Drexel University, Fermilab, the Institute for Advanced Study, the Japan Participation Group, Johns Hopkins University, the Joint Institute for Nuclear Astrophysics, the Kavli Institute for Particle Astrophysics and Cosmology, the Korean Scientist Group, the Chinese Academy of Sciences (LAMOST), Los Alamos National Laboratory, the Max-Planck-Institute for Astronomy (MPIA), the Max-Planck-Institute for Astrophysics (MPA), New Mexico State University, Ohio State University, University of Pittsburgh, University of Portsmouth, Princeton University, the United States Naval Observatory, and the University of Washington.

This research has made use of the NASA/IPAC Extragalactic Database (NED), which is funded by the National Aeronautics and Space Administration and operated by the California Institute of Technology. This research has made use of NASA’s Astrophysics Data System Bibliographic Services. This research made use of Montage. It is funded by the National Science Foundation under Grant Number ACI-1440620, and was previously funded by the National Aeronautics and Space Administration's Earth Science Technology Office, Computation Technologies Project, under Cooperative Agreement Number NCC5-626 between NASA and the California Institute of Technology. 

The authors wish to acknowledge CSC – IT Center for Science, Finland, for computational resources.

This research made use of \textsc{Jupyter}
\citep{PER-GRA:2007,Kluyver:2016aa}, \textsc{NumPy} \citep{harris2020array}, \textsc{SciPy} \citep{2020SciPy-NMeth}, \textsc{Matplotlib}
\citep{hunter2007matplotlib}, \textsc{Astropy}, a community-developed core \textsc{Python}
package for astronomy \citep{astropy2013,astropy2018}, \textsc{emcee} \citep{foreman2013emcee}, \textsc{Schwimmbad} \citep{schwimmbad}, \textsc{MgeFit} \cite{cappellari2002efficient}, \textsc{JamPy} \cite{cappellari2008measuring,cappellari2020efficient}, \textsc{pPXF} \cite{cappellari2004parametric,cappellari2017improving,cappellari2022full}.

We thank the anonymous referee for their comments which have improved this manuscript.

D.A.S. is supported by a STFC/UKRI doctoral studentship.

\section*{Data Availability}

The MUSE integral field data underlying this article is publicly available through the ESO science archive facility (\url{http://archive.eso.org/cms.html}). The SAURON integral field data is available on the ATLAS$^{\rm 3D}$ website (\url{https://www-astro.physics.ox.ac.uk/atlas3d/}).

The HST ACS/WFC F850LP image is publicly available through the Mikulski Archive for Space Telescopes (\url{https://archive.stsci.edu/}). The r-band SDSS image is available from \url{https://dr12.sdss.org/}

We also include the OASIS RNI degree 1 and MUSE RNI degree 1 kinematics in the online supplementary materials.




\bibliographystyle{mnras}
\bibliography{M87_BH_bib} 








\bsp	
\label{lastpage}
\end{document}